\documentclass[10pt,journal,compsoc]{IEEEtran}
\usepackage[a-2b]{pdfx}
\usepackage{svg}
\usepackage{balance}  % for  \balance command ON LAST PAGE  (only there!)
\usepackage{graphicx}
\usepackage{array}
\newcolumntype{M}[1]{>{\arraybackslash}m{#1}}
\usepackage{comment}
\usepackage{amsmath}
\usepackage{url}
\usepackage{epsfig}
\usepackage{savesym}
\usepackage{verbatim}
\usepackage{multirow}
\usepackage{epstopdf}
\usepackage{caption}
\usepackage[subrefformat=parens,labelformat=parens]{subcaption}
\usepackage{lipsum}
\usepackage{setspace}
\usepackage{etoolbox}
\usepackage{tikz}
\usepackage{balance}
\usepackage{diagbox}
\usepackage{amsfonts}
\usetikzlibrary{automata, arrows, calc}

\usepackage{enumitem}
\usepackage{mathrsfs}
\DeclareMathAlphabet{\mathcal}{OMS}{cmsy}{m}{n}
\usepackage[linesnumbered,algoruled,boxed,lined,noend]{algorithm2e}
\usepackage{breqn}
\usepackage{makecell} 
\usepackage{float}
\ifCLASSOPTIONcompsoc
  \usepackage[nocompress]{cite}
\else
  \usepackage{cite}
  \fi

\newtheorem{theorem}{Theorem}[section]
\newtheorem{lemma}[theorem]{Lemma}

\newtheorem{example}{Example}[section]
\newtheorem{definition}{Definition}[section]
\newtheorem{proposition}{Proposition}[section]

\SetVlineSkip{0pt}

\SetCommentSty{mycommfont}

\begin{document}
% \newcommand{\cal}[1]{\mathcal{#1}}
%%%%%%%%%%%%%%%%%%%%%
% macros from Byron:%
%%%%%%%%%%%%%%%%%%%%%
\newcommand{\byronnotes}[1]{\textcolor{blue}{\noindent Note: #1}}
\newcommand{\revise}[1]{{#1}}
\newcommand{\choi}[1]{{#1}}
\newcommand{\jiaxin}[1]{{#1}}
\newcommand{\TKDE}[1]{{#1}}
\newcommand{\TKDERF}[1]{{#1}}
\newcommand{\ssize}[1]{\scriptsize{#1}}
\newcommand{\highlight}[1]{\textcolor{red}{#1}}
\newcommand{\vldbjrevise}[1]{{#1}}
\newcommand{\vldbj}[1]{#1}
\newcommand{\pvldb}[1]{{#1}}
\newcommand{\TODO}[1]{\textcolor{BurntOrange}{\textbf{TODO:} #1}}
\newcommand{\Remind}[1]{\textcolor{RubineRed}{\textbf{Reminder:}#1}}
\newcommand{\byronsuggestion}[1]{\textcolor{Green}{\textbf{Reminder:}#1}}
\newcommand{\red}[1]{\textcolor{red}{#1}}
\newcommand{\blue}[1]{\textcolor{blue}{#1}}
\newcommand{\what}[1]{\textcolor{blue}{?#1?}}
\newcommand{\eat}[1]{}
\newcommand{\kw}[1]{{\ensuremath {\textsf{#1}}}\xspace}
\newenvironment{tbi}{\begin{itemize}
		\setlength{\topsep}{0.6ex}\setlength{\itemsep}{0ex}} %\vspace{-0.5ex}}
	{\end{itemize}} %\vspace{-0.5ex}}
\newcommand{\ei}{\end{itemize}}

%%%%%%%%%%%%%%%%%%%%%%%%%
% macros for some terms:%
%%%%%%%%%%%%%%%%%%%%%%%%%
\newcommand{\Gontology}{G_{Ont}}
\newcommand{\PRADS}{\mathsf{PADS}}
\newcommand{\KPADS}{\mathsf{KPADS}}
\newcommand{\BPADS}{\mathsf{BPADS}}
\newcommand{\ADS}{\mathsf{ADS}}
\newcommand{\PKD}{\textsf{\small PKD}}
\newcommand{\FRAMEWORK}{\kw{FRAMEWORK\_NAME}}
\newcommand{\ALGO}{\kw{ALGO\_NAME}}
\newcommand{\PEval}{\mathsf{PEval}}
\newcommand{\IncEval}{\mathsf{IncEval}}
\newcommand{\Assemble}{\mathsf{Assemble}}
\newcommand{\ARef}{\textsf{\small ARefine}}
\newcommand{\ACmpl}{\textsf{\small AComplete}}
\newcommand{\DataGraph}{$G$}
\newcommand{\Ghier}{\mathbb{G}}
\newcommand{\qhier}{\mathbb{Q}}
\newcommand{\subclassOf}{\mathsf{SubClassOf}}
\newcommand{\subtypeOf}{\mathsf{SubTypeOf}}
\newcommand{\answer}{\mathsf{mat}}
\newcommand{\answerb}{\mathsf{mat}^b}
\newcommand{\answerf}{\mathsf{mat}^f}
\newcommand{\Bisim}{\mathsf{Bisim}}
\newcommand{\rank}{\mathsf{rank}}
\newcommand{\dist}{\mathsf{dist}}
\newcommand{\Next}{\mathsf{next}}
\newcommand{\distance}{\mathsf{d}}
\newcommand{\reach}{\mathsf{reach}}
\newcommand{\Generalization}{\mathsf{Gen}}
\newcommand{\Specialization}{\mathsf{Spec}}
\newcommand{\desc}{\mathsf{desc}}
\newcommand{\support}{\mathsf{sup}}
\newcommand{\distort}{\mathsf{DT}}
\newcommand{\degree}{\mathsf{deg}}
\renewcommand{\equiv}{\mathsf{equiv}}
\newcommand{\equivv}[1]{[#1]_\mathsf{equiv}}
\newcommand{\irchi}[2]{\raisebox{\depth}{$#1\chi$}}
\newcommand{\Summarization}{\mathpalette\irchi\relax}
\newcommand{\Configuration}{C}
\newcommand{\radius}{d_{max}}
\newcommand{\Eval}{\mathsf{eval}}
\newcommand{\peval}{\mathsf{peval}}
\newcommand{\F}{\mathcal{F}}
\newcommand{\Score}{\mathsf{scr}}
\newcommand{\ie}{\emph{i.e.,}\xspace}
\newcommand{\eg}{\emph{e.g.,}\xspace}
\newcommand{\wrt}{\emph{w.r.t.}\xspace}
\newcommand{\aka}{\emph{a.k.a.}\xspace}

\newcommand{\equi}{\mathsf{equi}}
\newcommand{\sn}{\mathsf{sn}}
\newcommand{\METIS}{\mathsf{\small METIS}}

\newcommand{\maxsf}{\mathsf{max}}
\newcommand{\PIE}{\mathsf{PIE}}
\newcommand{\PINE}{\mathsf{PINE}}
\newcommand{\PI}{\mathsf{PI}}
\newcommand{\Psf}{\mathsf{P}}
\newcommand{\I}{\mathsf{I}}
\newcommand{\E}{\mathsf{E}}
\newcommand{\model}{\mathsf{model}}
\newcommand{\Algo}{\mathsf{algo}}
\newcommand{\threshold}{\tau}

\newcommand{\cost}{\mathsf{cost}}
\newcommand{\costq}{\mathsf{cost}_\mathsf{q}}
\newcommand{\CR}{\mathsf{compress}}
\newcommand{\DT}{\mathsf{distort}}
\newcommand{\FP}{\mathsf{fp}}
\newcommand{\maxSAT}{\mathsf{maxSAT}}
\newcommand{\True}{\mathsf{T}}
\newcommand{\False}{\mathsf{F}}
\newcommand{\OptGen}{\mathsf{OptGen}}
\newcommand{\freq}{\mathsf{freq}}

\newcommand{\vpd}{V_{pd}}
\newcommand{\vtp}{V_{tbp}}
\newcommand{\BiGindex}{\mathsf{BiG\textnormal{-}index}}
\newcommand{\filter}{\mathsf{filter}}
\newcommand{\ans}{\mathsf{ans\_graph\_gen}}
\newcommand{\Azero}{\mathbb{A}}
\newcommand{\boost}{\mathsf{boost}}
\newcommand{\kws}{\mathsf{kws}}
\newcommand{\bkws}{\mathsf{bkws}}
\newcommand{\fkws}{\mathsf{fkws}}
\newcommand{\rkws}{\mathsf{rkws}}
\newcommand{\dkws}{\mathsf{dkws}}
\newcommand{\knk}{\kw{k-nk}}
\newcommand{\config}{\mathsf{config}}
\newcommand{\content}{\mathsf{isKey}}
\newcommand{\pcnt}{\mathsf{pcnt}}
\newcommand{\private}{\textsf{isPrivate}}

\newcommand{\Path}{{\mathsf{Path}}}
\renewcommand{\P}{{\mathcal P}}
\newcommand{\Partition}{\mathsf{Par}}

\newcommand{\ppkws}{\textsf{\small PPKWS}\xspace}

\newcommand{\DKWS}{\mathsf{DKWS}}
\newcommand{\kDKWS}{\textsf{\small $k$DKWS}\xspace}
\newcommand{\SKWS}{\mathsf{bfkws}}
\newcommand{\KWS}{\textsf{\small KWS}\xspace}
\newcommand{\VU}{\mathbb{V}}
\newcommand{\VI}{\mathcal{V}}
\newcommand{\VM}{\bar{\mathcal{V}}}
\newcommand{\vsf}{\mathsf{v}}
\newcommand{\usf}{\mathsf{u}}
\newcommand{\answerset}{\mathcal{A}}
\newcommand{\candanswerset}{\bar{\mathcal{A}}}
\newcommand{\prune}{S}
\newcommand{\Ud}{\hat{\dist}}
\newcommand{\Ld}{\check{\dist}}
\newcommand{\invert}{\mathcal{I}}
\newcommand{\MB}{u.b}
\newcommand{\MF}{\mathsf{f}}
\newcommand{\Queue}{\mathcal{P}}
\newcommand{\Visit}{V}
\newcommand{\invertV}{V_{\invert}}
\newcommand{\invertE}{E_{\invert}}
\newcommand{\tnormal}[1]{\textnormal{#1}}
\newcommand{\BaselineDKWS}{\mathsf{Baseline}}
\newcommand{\SKWSC}{$\mathsf{BANKS}$-$\mathsf{II}$}
\newcommand{\DKWSBF}{$\mathsf{DKWS}$-$\mathsf{BF}$}
\newcommand{\DKWSNP}{$\mathsf{DKWS}$-$\mathsf{NP}$}
\newcommand{\DKWSPADS}{$\mathsf{DKWS}$-$\mathsf{PADS}$}
\newcommand{\DKWSPINE}{$\mathsf{DKWS}$-$\mathsf{PINE}$}

\newcommand{\Notify}{\mathsf{Notify}}
\newcommand{\Push}{\mathsf{Push}}
\newcommand{\np}{\mathsf{NP}}
\newcommand{\Parameters}{\mathscr{X}}
\newcommand{\grape}{\mathsf{GRAPE}}
\newcommand{\AAP}{\kw{AAP}}
\newcommand{\Buffer}{\mathbb{B}}
\newcommand{\SI}{\kw{SI}}

\newcommand{\SBGindex}{\mathsf{SBGIndex}}

% The index name:
\newcommand{\SGIndex}{\mathsf{SGIndex}} % Summarized generalized index

\newcommand{\Portal}{\mathbb{P}}
\newcommand{\rclique}{\mathsf{r\textnormal{-}clique}}
\newcommand{\Blinks}{\mathsf{Blinks}}
\newcommand{\Rclique}{\mathsf{Rclique}}

\newcommand{\pprclique}{\mathsf{PP\textnormal{-}r\textnormal{-}clique}}
\newcommand{\ppknk}{\mathsf{PP\textnormal{-}knk}}
\newcommand{\ppBlinks}{\mathsf{PP\textnormal{-}Blinks}}

\newcommand{\baselinerclique}{\mathsf{Baseline\textnormal{-}r\textnormal{-}clique}}
\newcommand{\baselineknk}{\mathsf{Baseline\textnormal{-}knk}}
\newcommand{\baselineBlinks}{\mathsf{Baseline\textnormal{-}Blinks}}

\newcommand{\stitle}[1]{\vspace{0.4ex}\noindent{\bf #1}}
\newcommand{\etitle}[1]{\vspace{0.8ex}\noindent{\underline{\em #1}}}
\newcommand{\eetitle}[1]{\vspace{0.6ex}\noindent{{\em #1}}}

%
% Usage \techreport{Text in technical report}{text in the paper}
%\newcommand{\techreport}[2]{#1}
\newcommand{\techreport}[2]{#2}
% The framework name:
\newcommand{\SGFrame}{\mathsf{SGFrame}} % Summarized generalized framework

%%%%%%%%%%%%%%%%%%%%%%%%%
% Auxiliary Macro       %
%%%%%%%%%%%%%%%%%%%%%%%%%
\newcommand{\stab}{\rule{0pt}{8pt}\\[-2.0ex]}
\newcommand{\tab}{\hspace{4ex}}

\newcommand{\Q}{{\cal Q}}

%%%%%%%%%%%%%%%%%%%%%%%%%%
% Environments for IEEE  %
%%%%%%%%%%%%%%%%%%%%%%%%%%

%\newtheorem{proof}{Proof}[section]
%% \renewenvironment{proof}{
%%         \vspace{1ex}
%%         \refstepcounter{proof}
%%         {\noindent\bf Proof \theproof:}}
%%         {\vspace{1ex}}

\newcommand{\eop}{\hspace*{\fill}\mbox{\qed}}

\newcommand*\circled[1]{\tikz[baseline=(char.base)]{
            {\scriptsize \node[shape=circle,draw,inner sep=0.5pt] (char) {#1};}}}

%% spacing %%
%%%%%%%%%%%%%%%%%%%%%%%%%%%% spacing %%%%%%%%%%%%%%%%%%%%%%%%%%%%%%%%%%
% Floating objects style
% \setlength{\floatsep}{-0.3\baselineskip plus 0.1\baselineskip minus 0.1\baselineskip}
% \setlength{\textfloatsep}{-0.3\baselineskip plus 0.1\baselineskip minus 0.1\baselineskip}
% \setlength{\intextsep}{-0.3\baselineskip plus 0.1\baselineskip minus 0.1\baselineskip}
% \setlength{\dbltextfloatsep}{-0.3\baselineskip plus 0.1\baselineskip minus 0.1\baselineskip}
% \setlength{\dblfloatsep}{-0.3\baselineskip plus 0.1\baselineskip minus 0.1\baselineskip}

% \setlength{\belowdisplayskip}{0ex} 
% \setlength{\belowdisplayshortskip}{0ex}
% \setlength{\abovedisplayskip}{0ex} 
% \setlength{\abovedisplayshortskip}{0ex}

% \setlength{\abovecaptionskip}{0em}
% \setlength{\belowcaptionskip}{0em}

\newcommand{\spacecompress}{}

% \titlespacing*{\section}{0pt}{0.35ex}{0.35ex}
% \titlespacing*{\subsection}{0pt}{0.35ex}{0.35ex}

\newtheorem{manuallemmainner}{Lemma}
\newenvironment{manuallemma}[1]{%
  \renewcommand\themanuallemmainner{#1}%
  \manuallemmainner
}{\endmanualtheoreminner}

% \renewcommand{\textfraction}{0.05}
% \renewcommand{\topfraction}{0.95}
% \renewcommand{\bottomfraction}{0.2}
% \renewcommand{\floatpagefraction}{0.99}
%%%%%%%%%%%%%%%%%%%%%%%%%%%%%%%%%%%%%%%%%%%%%%%%%%%%%%%%%%%%%%%%%%%%%%%
% \setlength{\textfloatsep}{0.05cm}
% \setlength{\floatsep}{0.05cm}

%%% Local Variables: 
%%% TeX-master: t
%%% End:

% !TeX root = main.tex

\title{DKWS: A Distributed System for Keyword Search on Massive Graphs (Complete Version)}

\author{Jiaxin Jiang, Byron Choi, Xin Huang, Jianliang Xu and Sourav S Bhowmick
\IEEEcompsocitemizethanks{\IEEEcompsocthanksitem Jiaxin Jiang is with School of Computing, National University of Singapore, Singapore.\protect\\
E-mail: jxjiang@nus.edu.sg
\IEEEcompsocthanksitem  Byron Choi, Xin Huang and Jianliang Xu are with the Department of Computer Science, Hong Kong Baptist University, Hong Kong.  Corresponding author: Byron Choi.\protect\\
  E-mail: \{bchoi, xinhuang, xujl\}@comp.hkbu.edu.hk
  \IEEEcompsocthanksitem Sourav.S. Bhowmick is with School of Computer Engineering, Nanyang Technological University, Singapore. \protect\\
  E-mail: assourav@ntu.edu.sg}
}

\IEEEtitleabstractindextext{
\begin{abstract}
Due to the unstructuredness and the lack of schemas of graphs, such as knowledge graphs, social networks, and RDF graphs, keyword search for querying such graphs has been proposed. As graphs have become voluminous, large-scale distributed processing has attracted much interest from the database research community. While there have been several distributed systems, distributed querying techniques for keyword search are still limited. This paper proposes a novel distributed keyword search system called $\DKWS$. First, we \revise{present} a {\em monotonic} property with keyword search algorithms that guarantees correct parallelization. Second, we present a keyword search algorithm as monotonic backward and forward search phases. Moreover, we propose new tight bounds for pruning nodes being searched. Third, we propose a {\em notify-push} paradigm and $\PINE$ {\em programming model} of $\DKWS$. The notify-push paradigm allows {\em asynchronously} exchanging the upper bounds of matches across the workers and the coordinator in $\DKWS$. The $\PINE$ programming model naturally fits keyword search algorithms, as they have distinguished phases, to allow {\em preemptive} searches to mitigate staleness in a distributed system. Finally, we investigate the performance and effectiveness of $\DKWS$ through experiments using real-world datasets. We find that $\DKWS$ is up to two orders of magnitude faster than related techniques, and its communication costs are $7.6$ times smaller than those of other techniques.
\end{abstract}
}

% !TeX root = main.tex

\maketitle

%\onecolumn

\IEEEraisesectionheading{\section{Introduction}\label{sec:intro}}

\IEEEPARstart{K}{nowledge} graphs, social networks, and RDF graphs have a wide variety of emerging applications, including semantic query processing~\cite{zheng2016semantic}, information summarization~\cite{wu2013summarizing}, community search~\cite{fang2016effective}, collaboration and activity organization~\cite{tian2008efficient}, and user-friendly query facilities~\cite{yi2016autog}. Such graphs often lack useful schema information for users to formulate their queries. To make querying such data easy, {\em keyword search} has been proposed. Users can retrieve information without the knowledge of the schema or query language. In a nutshell, users only specify a set of keywords $Q$ as their query on a data graph $G$. \vldbj{In recent years, there have been well-known projects that build graph-structured databases and allow querying with simply a set of keywords, \eg  BioCyc\footnote{{\tt http://biocyc.org}} and Google's knowledge graph search API.\footnote{{\tt https://developers.google.com/knowledge-graph/}}}

\begin{figure}[tb]
	\begin{center}
	\includegraphics[width=0.45\textwidth]{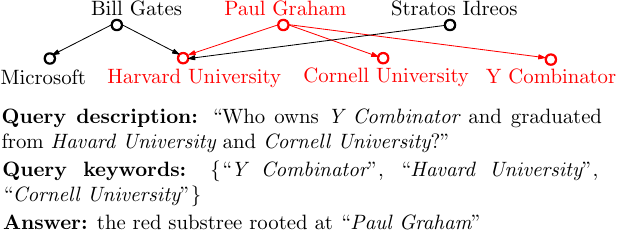}
		\end{center}
	\caption{Example of keyword search on a knowledge graph}
	\label{fig:egkws}
\end{figure}

\eat{For instance, Google's knowledge graph search API\footnote{{\tt https://developers.google.com/knowledge-graph/}} allows users to find matches from their knowledge database, and returns the query answer in the form of subtrees.}

The answer of keyword search semantics (cf. \cite{he2007blinks,DBLP:journals/tkde/YuanLCYWS17,bhalotia2002keyword,kacholia2005bidirectial,fan2017incremental,qin2014scalable,DBLP:conf/www/Shi0K20}) is generally a set of matches, where each match is a rooted subtree of $G$ such that query keywords belong to the labels of leaf vertices. These semantics differ mainly in the score function of the matches. Interested readers may refer to comprehensive surveys on the keyword search semantics for more information~\cite{yang2021keyword,wang2010survey,yu2010keyword}. For example, consider a partial knowledge graph shown in Fig.~\ref{fig:egkws}, where a node is an entity and an edge is a relation between entities. Assume that a user is identifying ``who owns \textit{Y Combinator} and graduated from \textit{Havard University} and \textit{Cornell University}\text{?}''. \vldbj{He/She may simply provide the keywords $Q$=\{\textit{Y Combinator}, \textit{Havard University}, \textit{Cornell University}\} as his/her query.} If users apply the keyword search to the knowledge graph, a substree rooted at \textit{Paul Graham} can be returned as an answer (\eg~\cite{he2007blinks,fan2017incremental,qin2014scalable}).

Nowadays, graphs with billions of vertices or edges are common, and their sizes continue to increase. For example, WebUK~\cite{webuk}, a large Web graph, contains 106 million nodes and 3.7 billion edges. Keyword search often involves numerous traversals of such massive graphs, which are computationally costly. Indexes (\eg for shortest distance computations) on such graphs are often large, \eg $O(|V|^2)$ in the worst case, where $|V|$ is the number of the vertices. Still, it is infeasible to load the index into the main memory, \eg~\cite{he2007blinks,kargar2011keyword}. As a result, distributed graph processing systems are a competitive solution. In this paper, we aim to propose a {\em distributed system} to answer the top-$k$ keyword search on distributed graphs. Intuitively, each worker computes local top-$k$ matches on a graph partition and the global top-$k$ matches are generated from such local matches. However, several major technical challenges of keyword search have not been addressed by existing generic distributed processing systems, \eg~\cite{avery2011giraph,xin2013graphx,fan2017parallel}.

\begin{figure}[tb]
    \centering
    \includegraphics[width=\linewidth]{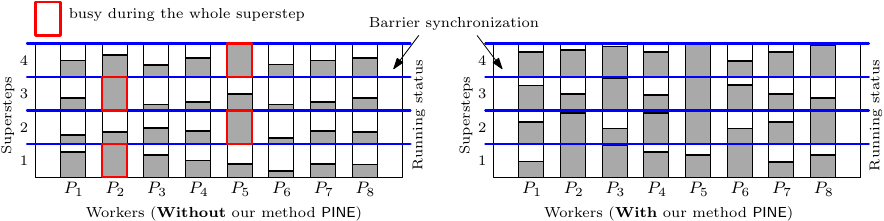}
    \caption{Illustration of stragglers of distributed keyword search \TKDE{(grey denotes the worker $P_i$ is busy, whereas white color denotes the worker $P_i$ is idle.)}}
	\label{fig:straggler}
\end{figure}

\begin{figure}[tb]
	\begin{center}
	\includegraphics[width=0.4\textwidth]{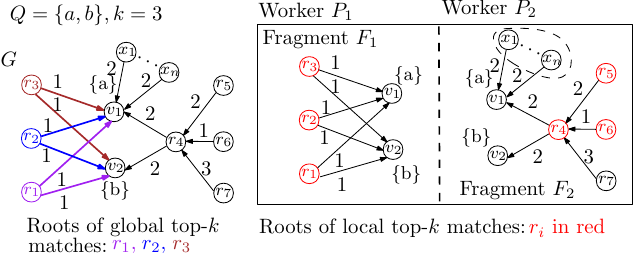}
	\end{center}
	\caption{An example of distributed keyword search}
	\label{fig:intro}
\end{figure}

\stitle{Challenge 1: Straggler problem.} Some workers in a distributed system may take substantially longer than others. We show the working status of supersteps $1$-$4$ of a cluster with $8$ workers \TKDE{($P_1$ to $P_8$)} as shown in Fig.~\ref{fig:straggler}. \TKDE{In each superstep, there are concurrent computation and barrier synchronization. The workers start concurrent computations and become busy (if there is some work) until a barrier synchronization. In the first superstep, worker $P_2$ take longer than the other $7$ workers. They keep waiting idly and the computing power is not used until $P_2$ completes its tasks.} This is known as the {\em straggler problem}. Similarly, $P_2$ is also the straggler in third superstep and $P_5$ is the straggler of second and fourth supersteps. This problem can be caused by either workload imbalance or graph characteristics. \TKDERF{Previous studies have frequently focused on rebalancing partitions~\cite{fan2019dynamic} or predicting machine workloads~\cite{fan2020application} during runtime. Nevertheless, these approaches come with additional costs, including the expenses associated with data transfer. Moreover, setting up the training model for keyword search is a non-trivial task.}

\begin{figure}
		\centering
		\begin{subfigure}[b]{.15\textwidth}
		  \centering
                  %Jiaxin, you may replace it with another figure
                  \includegraphics[width=\textwidth]{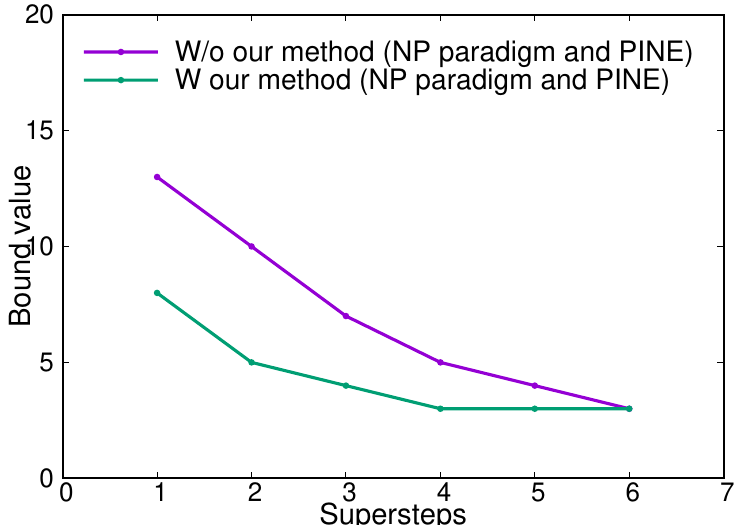}\quad
		\caption{Pruning bounds}	\label{fig:bound}
		\end{subfigure}\quad
		\begin{subfigure}[b]{.15\textwidth}
		\centering
		\includegraphics[width=\textwidth]{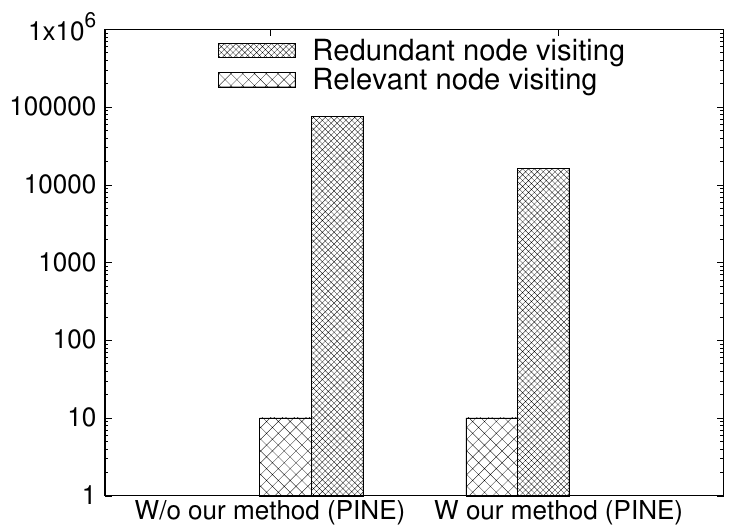}\quad
		\caption{\TKDERF{Visited nodes}}      \label{fig:vnodes}
		\end{subfigure}
		\begin{subfigure}[b]{.15\textwidth}
			\centering
			\includegraphics[width=\textwidth]{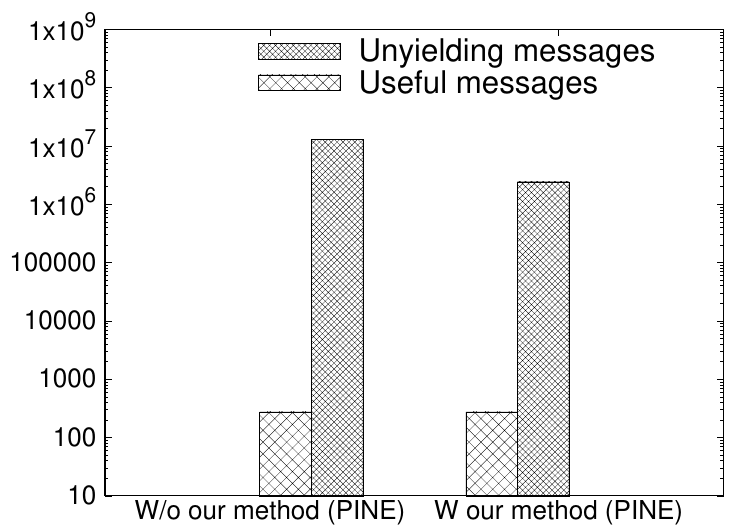}\quad
			\caption{\TKDERF{Messages}}      \label{fig:vmsg}
			\end{subfigure}
		\caption{Illustration of the potential of bound refinements, and messages of distributed keyword search}
	\label{fig:efficiency}
		\end{figure}

\stitle{Challenge 2: Lack of pruning techniques.} Another challenge is that existing sequential keyword search works often utilize the global graph information (\vldbj{\eg the upper bound of the score of the top-$k$ matches}) to develop some pruning techniques, \eg \cite{he2007blinks}, to avoid exhaustive traversals on the graph. Consider a graph $G$ in Fig.~\ref{fig:intro}, the upper bound of the score of top-$k$ matches is $2$ when the subtrees of $G$ rooted at $r_i$ ($i\in \{1,2,3\}$) are retrieved. $r_i$ ($i\in \{4,5,6,7\}$) and $x_i$ ($i\in[1,n]$) are not traversed. However, in a distributed graph system, such pruning techniques can be hardly directly applied since each machine only maintains a graph fragment.
In Fig.~\ref{fig:intro}, the workers $P_1$ and $P_2$ process the graph fragments $F_1$ and $F_2$, respectively. There are two local upper bounds $\prune_1=2$ and $\prune_2 = 8$ generated on $F_1$ and $F_2$, respectively. The search on $F_2$ can only be pruned by $\prune_2$. We show the refinement of bounds after each superstep in Fig.~\ref{fig:bound}. The bound values tighten faster with our techniques. With tighter bounds, unnecessary node visits are significantly reduced, and false matches are pruned early, as shown in Fig.~\ref{fig:vnodes}. \TKDERF{Existing research studies such as \cite{he2007blinks,kargar2011keyword} often rely on indexing distance information for pruning. However, these types of indexing methods are typically designed for single-machine algorithms. Each machine lacks global information, which significantly limits the potential for pruning.}

\eat{
	\begin{figure}
		\centering
		\begin{minipage}[t]{.24\textwidth}
		\includegraphics[width=\linewidth]{./figures/straggler2-eps-converted-to.pdf}
		\end{minipage}
		\begin{minipage}[t]{.24\textwidth}
		\includegraphics[width=\linewidth]{./figures/straggler-eps-converted-to.pdf}
		\end{minipage}

		\caption{Illustration of stragglers of distributed keyword search \TKDE{(grey denotes the worker $P_i$ is busy, whereas white color denotes the worker $P_i$ is idle.)}}
	\label{fig:straggler}
		\end{figure}
}

\stitle{Challenge 3: Message passing.} Since keyword search on graphs often involves numerous traversals, keyword search on distributed graphs might cause massive message passing. For instance, in previous studies ~\cite{DBLP:journals/tkde/YuanLCYWS17} and \cite{michel2005klee}, the local matches rooted at $r_i$ ($i\in [1,2,3]$) (resp. $r_i$ ($i\in [4,5,6]$)) are sent from $F_1$ (resp. $F_2$) to the coordinator for verification. The matches on $F_2$ are not among the final top-$k$ matches, \ie most of the computation on $F_2$ does not lead to matches. As shown in Fig.~\ref{fig:vmsg}, the messages not yielding final matches are significantly reduced in our system. \TKDERF{Previous works such as \cite{karypis1995metis,pacaci2019experimental} have often utilized partitioning strategies to reduce message overhead. However, these studies are designed for general purposes, where the partitioning is primarily based on the graph's structure. In a distributed environment, the message overhead in keyword search often depends on the distribution of the query keywords. This dependency makes these approaches less effective in such scenarios.}

\stitle{Contributions.} In this paper, we propose a system for answering top-$k$ keyword search called $\DKWS$ and show that all three challenges can be addressed. We investigate keyword search algorithms in a distributed environment and the techniques for $\DKWS$ as opposed to individual keyword search semantics (or algorithms).
\setlist{nolistsep}
\begin{enumerate}[wide, labelwidth=!, labelindent=0pt]
\item \revise{We present the {\em monotonic} property with the keyword search algorithm which leads to
  %Popular keyword search algorithms often consist of multiple phases.
correct parallelization.} We show that {\em a sequential keyword search algorithm} can be rewritten into two main phases, (a) backward keyword search ($\bkws$), and (b) forward keyword search ($\fkws$). We propose new lower and upper bounds for pruning in $\fkws$. We prove that $\bkws$ and $\fkws$ implemented in $\DKWS$ are monotonic.   
	\item We propose a {\em notify-push} paradigm for $\DKWS$: (a) each worker {\em asynchronously notifies} the coordinator when the local upper bound is refined; (b) the coordinator maintains a global bound. When it receives the notification from workers, it refines the global upper bound and {\em asynchronously pushes} it to all workers. This incurs a small communication overhead, but the refined global bounds provide global information to workers to prune some search locally. 
    \item We propose a $\PINE$ {\em programming model} that naturally fits the keyword search algorithm that has distinguished search phases. $\DKWS$ launches a {\em preemptive execution} of the searches. Hence, keyword searches are no longer one blocking operation in the distributed environment. We propose staleness indicators and \TKDE{a lightweight cost model} that mitigate the straggler problem.
	\eat{\item We further propose two optimizations to speed up query processing in $\DKWS$, \revise{which can be readily applied to existing graph systems, \eg ~\cite{fan2017parallel,yan2014blogel,avery2011giraph}}. First, we propose a priority queue to maintain the nodes that are needed to be searched in order to \jiaxin{reduce} duplicate traversals. Second, we propose a {\em backtrack graph} to optimize \jiaxin{the propagation of the distance refinement from other fragments}. The use of a backtrack graph lowers the time complexity for the propagation.}
	\item Using real-life graphs, we empirically compare the performance of $\DKWS$ and two baselines. We verify that (a) $\DKWS$ speeds up the query performance of top-$k$ keyword  search up to two orders of magnitude; (b) The communication cost of $\DKWS$ is $7.6$ times smaller than that of baseline; and (c) $\DKWS$ using all optimizations is on average $1.64$ times faster than $\DKWS$ without them.
	\item Due to space limitations, we put the proofs, some optimizations, and more experiments in a technical report~\cite{techreport}.
\end{enumerate}

\stitle{Organization.} Sec.~\ref{sec-pre} provides some background and the problem statement. In Sec.~\ref{sec:skws}, we illustrate an efficient \jiaxin{monotonic} sequential keyword search algorithm. In Sec.~\ref{sec:dkws}, we propose $\DKWS$ and its two novel ideas, namely the notify-push paradigm and the $\PINE$ model. Sec.~\ref{sec:exp} reports experimental results. Sec.~\ref{sec:related}, presents the related work. We conclude the paper and present the future works in Sec.~\ref{sec:conclusions}.

%%%%%% Section 2 %%%%%%%

\section{Preliminaries and problem statement}
\label{sec-pre}

This section presents some background and the problem statement. Some frequently used notations are summarized in Table~\ref{tab-notations}.

% We start with a review of some basic notations.

\begin{figure}[tb]
	\begin{center}
	\includegraphics[width=0.45\textwidth]{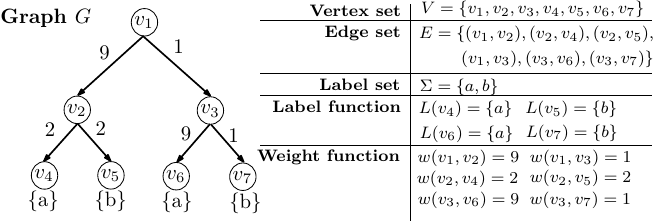}
	\end{center}
	\vldbj{\caption{An example of frequently used graph notations}\label{fig:notations}}
\end{figure}

\stitle{Graphs}. We consider a {\em
	labeled, weighted, directed graph} modeled as $G=(V,E,L,\Sigma,w)$, where (a)~$V$ is a set of
vertices; (b)~$E$ $(\subseteq$ $V$$\times$$V$$)$ is a set of edges; (c)~$\Sigma$
is a set of keywords; (d)~$L$$: $$V$$\rightarrow$$\Sigma$ is a label mapping function such that for each vertex $v\in V$, $L(v)$ maps $v$ to a subset of \TKDERF{labels/keywords} in $\Sigma$; and (e)~$w(e)$ is a positive weight of an edge $e=(u,v)\in E$. For simplicity, we may omit $L$, $\Sigma$ and $w$ when they are irrelevant to the discussions. The size of the graph is denoted by $|G|=|V|+|E|$.

\begin{example}
Consider a graph $G$ in Fig.~\ref{fig:notations}, a) $V=\{v_1,v_2,v_3,v_4,v_5,v_6,v_7\}$ is the vertex set, b) $E$ is a set of edges, \eg $(v_1,v_2)\in E$ is an edge, c) $\Sigma = \{a,b\}$ is a set of \TKDERF{keywords}, d) $L$ maps each vertex in $V$ to a subset of \TKDERF{keywords} in $\Sigma$, \eg $L(v_4) = \{a\}\subseteq \Sigma$, and e) $w$ maps each edge in $E$ to a positive weight, \eg $w(v_1,v_2) = 9$.
\end{example}

\stitle{Partition strategy}. 
%This is to facilitate data-partitioned parallelism.
Given a number $m$, a strategy $\Partition$
partitions a graph $G$ into {\em fragments}
$\mathcal{F}$ = $\{F_1, \ldots, F_m\}$ such that each 
$F_i = (V_i, E_i, L_i)$ is a subgraph of $G$, 
$E = \bigcup_{i \in [1, m]} E_i$, $V = \bigcup_{i \in [1, m]} V_i$ and \vldbj{$L_i = L$}, 
and $F_i$ resides at worker 
$P_i$, where $i\in [1,m]$ is the fragment id. There are two special sets of nodes for each fragment. 
\begin{itemize}[leftmargin=*]
\item $F_i.I\subset V_i$: the set of nodes $v \in V_i$ such that there is
an edge $(v', v)$ {\em incoming} from a node $v'$ in $F_j$ ($i \neq j$); and
\item  $F_i.O$: the set of nodes $v'$ such that there exists an {\em outgoing} edge $(v, v')$ in $E$, $v\in V_i$ and $v'$ is in some $F_j$ ($i \neq j$).

%we refer to $v$ as a {\em border node}
%and $e$ as a {\em cross edge};
%another fragment $F_j$, \ie $v$ is a border node in $F_j$; and
\end{itemize}
  
In addition, we denote $\F.O$ = $\bigcup_{i \in [1, m]} F_i.O$, and $\F.I$ = $\bigcup_{i \in [1, m]} F_i.I$. We refer
	to the nodes in $F_i.I \cup F_i.O$ as the {\em border nodes} (\aka~~{\em portal nodes})
	of $F_i$ \wrt~$\Partition$. Partition strategies (\eg~\cite{karypis1995metis}) are orthogonal to our work. \TKDERF{In this paper, we utilize the edge-cut partitioning approach, where vertices are assigned to different partitions. As a result, edges may span across two partitions.}

\noindent\TKDERF{
\stitle{Platform.} In this work, we propose our system, $\DKWS$, built on top of the code-base of  $\grape$~\cite{fan2017parallel}. $\grape$ exemplifies a generic approach to parallel computations through a programming model that consists of three functions for implementing user-defined algorithms - $\PEval$, $\IncEval$, and $\Assemble$. These functions together form the $\PIE$ program paradigm. $\grape$ parallelizes the sequential algorithms (and minor revisions are required). $\grape$ inherits all optimization strategies available for sequential algorithms and graphs, such as indexing. $\DKWS$ inherits the strengths of $\grape$ while introducing a novel efficient paradigm $\PINE$ (detailed in Sec.~\ref{sec:dkws}) and novel optimizing such as indexing techniques for keyword search.
}

%%%%%%%%%%%%%%%%%%%%%%%%%%%%%%%
\begin{table}[tb!]
	\begin{scriptsize}
		\begin{center}
  \caption{\TKDE{Frequently used notations}}\label{tab-notations}
			\begin{tabular}
				{|c|p{0.68\linewidth}|} \hline Notation & Meaning \\
				\hline
                                \hline
                $Q$ & a set of query keywords $Q=\{q_1,q_2,\ldots q_l\}$\\ \hline
                $\tau$ & the \jiaxin{threshold} of the distance between a distinct root and its leaf nodes\\ \hline	
                $T$ & a match to query $(Q, \tau)$\\ \hline
				$\Score(u)$ & the score of a match $T$ rooted at $u$ \\ \hline
				$\SKWS/\bkws/\fkws$ & Sequential keyword search/backward search/forward search\\ \hline
				$\answer_u/\answerb_u/\answerf_u$ & the (partial) match found by $\SKWS/\bkws/\fkws$  \\ \hline
				$\dist(u, v)$ & the shortest distance between $u$ and $v$ \\ \hline
%				$G$ & graph, directed or undirected \\ \hline
				$\answerset$ & the answer, which contains top-$k$ matches \\ \hline 
				$P_0$, $P_i$ & $P_0$: the coordinator; $P_i$: workers, where $i \in [1, m]$ \\ \hline
				$\Partition$ & graph partition strategy \\ \hline
				\eat{$G_\P$ & the fragmentation graph of $G$ via $\P$ \\ \hline}
				$\F$ & fragmentation (\aka partition) $\{F_1, \ldots, F_m\}$ \\ \hline
                $M_i$ & messages designated to worker $P_i$ \\ \hline
			\end{tabular}
		\end{center}
	  \setlength{\tabcolsep}{0.5em}
	\end{scriptsize}
\end{table}

%%%%%%%%%%%%%%%%%%%%%%%%%%%%%%%
\begin{table}[tb!]
	\caption{\TKDE{Representative keyword search involved traversals and shortest distance computation or estimation}}
  \label{tab-kws}
  \begin{scriptsize}
	  \begin{center}
		  \setlength{\tabcolsep}{0.4em}
		  \begin{tabular}{|c|c|c|c|c|}
			\hline
			\multirow{2}{*}{\makecell[c]{Keyword search \\semantics}} & \multirow{2}{*}{$\bkws$} & \multirow{2}{*}{$\fkws$} & \multicolumn{2}{c|}{Indexing techniques} \\ \cline{4-5} 
							 &                       &                       & Exact index                 & Apx. index                 \\ \hline
		
								 Distinct root trees & \cite{he2007blinks,DBLP:journals/tkde/YuanLCYWS17}, \cite{fan2017incremental,qin2014scalable}  & \cite{DBLP:journals/tkde/YuanLCYWS17} & \cite{he2007blinks} & \cite{DBLP:journals/tkde/YuanLCYWS17,Jiang2020PPKWSAE}\\ \hline
								 Group Steiner trees & \cite{bhalotia2002keyword,kacholia2005bidirectial} & \cite{bhalotia2002keyword,DBLP:conf/www/Shi0K20,kacholia2005bidirectial} & \cite{DBLP:conf/www/Shi0K20}  &  $-$ \\ \hline
								 Other semantics & \cite{yang2019efficient,jiang2015exact} &  \cite{shi2016top,jiang2015exact} & \cite{jiang2015exact,kargar2011keyword} & \cite{qiao2013top} \\ \hline
			\end{tabular}
	  \end{center}
  \end{scriptsize}
\end{table}

\stitle{Semantics of keyword search ($\kws$) for graphs.} Several keyword query semantics have been proposed,~\eg~\cite{he2007blinks,kargar2011keyword,DBLP:journals/tkde/YuanLCYWS17}. \revise{They are driven by various interesting applications. We list some representative works of keyword search and their characteristics in Tab.~\ref{tab-kws}. Many of them involve backward search ($\bkws$) and/or forward search ($\fkws$).} We consider the same query semantic of \cite{he2007blinks,fan2017incremental,qin2014scalable}, which is the most popular semantic among the others.\footnote{According to Google scholar in \jiaxin{Jun} 2023, the total number of citations of the query semantic \cite{he2007blinks} received $718$ citations.} A keyword query is a binary tuple ($Q$, $\threshold$) which contains a set of keywords $Q$=\{$q_1$,$\ldots$,$q_l$\} and a distance threshold $\threshold$. Given a graph $G=(V,E)$, a match of $Q$ in $G$ is a subgraph of $G$, denoted by $T=\{u,\langle v_1,\ldots v_l\rangle\}$, such that  (i) $T$ is a tree rooted at $u$; (ii) $\forall i \in [1,l]$, $v_i$ is a leaf vertex of $T$ and $q_i\in L(v_i)$; and (iii) $\dist(u,v_i) \leq \threshold$, where $\dist(u,v_i)$ is the shortest distance between $u$ and $v_i$. Existing works design indexes for distance estimation/computations. However, the indexes for computing exact matches are large on massive graphs and also non-trivial to be adapted in a distributed environment.  On the other hand, the indexes for approximate match computation return bounds for pruning false matches. This work proposes new bounds for pruning false matches and adopts a lightweight index.

\TKDERF{Keyword searches can yield numerous matches, particularly within a massive graph. However, users are often concerned with interpreting most compact matches. As such, our focus is on the top-$k$ query that determines the \textit{top-$k$ matches} as the query answers. To facilitate this approach for top-$k$ queries, each vertex can serve as the root match only once. The more compact, the higher the rank. Accordingly, we augment the query structure from $(Q,\tau)$ to $(Q,\tau,k)$, where $\tau$ is the distance threshold between the root vertex and the leaf vertices.}

\TKDERF{It is well-received that an ideal match is a compact structure that contains all keywords. Hence, existing studies assign a score to each match $T$, using the root $u$ as a basis. This score is denoted as $\Score(u)$. In this context, a lower score for $T$ signifies a more compact match, considered preferable. Specifically, we employ the same score function as presented in \cite{he2007blinks,fan2017incremental,qin2014scalable}. This function is defined as follows.}

% A keyword search may have numerous matches in a massive graph. Users may only be interested in interpreting the most relevant matches. Therefore, we focus on finding the {\em top-k matches} as the query answer. To enable the top-$k$ queries, each vertex as the root match can occur only once. We extend the query $(Q,\tau)$ to $(Q,\tau,k)$.  Intuitively, a match should contain all keywords in a small structure. A score is therefore assigned to a match $T$ rooted at $u$. We denote the score of a match $T$ by $\Score(u)$. The lower the score of $T$ is, the more compact and better the match is. Specifically, we adopt the same score function of \cite{he2007blinks,fan2017incremental,qin2014scalable}, as defined below.

\spacecompress
\begin{definition}[Score function $\Score(u)$]\label{def:score}
	  Given a match, $T=\{u,\langle v_1,\ldots v_l\rangle\}$, to the query $(Q,\tau, k)$, the {\em score of $T$} is denoted by $\Score(u) = \sum_{i\in [1,l]} \dist(u,v_i)$, where $\dist(u, v_i)$ is the shortest distance between $u$ and $v_i$. 
\end{definition}
\spacecompress

\stitle{Problem statement.} Given a graph $G$, a keyword query $(Q,\tau,k)$, we investigate a distributed system to compute the top-$k$ matches $\answerset$ (\ie the answer) of the query on $G$.

\eat{
\stitle{Remarks.} 
The definitions of keyword search semantics often involve the shortest distances of nodes, \eg Definition~\ref{def:score}. The query algorithms ofter require numerous shortest distance computations. Such computations in a distributed environment are inefficient as the shortest might span across multiple fragments. We aim to optimize two major traversals, namely forward expansion and backward expansion, in a distributed environment. They can be extended/adapted to implement traversals of other keyword search algorithms such as \cite{he2007blinks,bhalotia2002keyword,jiang2015exact}.
}

\section{Backward and forward keyword search}\label{sec:skws}

In this section, we discuss \revise{the monotonic property} of keyword search ($\kws$) algorithms, which is crucial for its correct parallelization~\cite{fan2017parallel}.  Specifically, backward and forward keyword search ($\SKWS$) consists of two phases, namely, backward keyword search ($\bkws$) and forward keyword search ($\fkws$). \TKDERF{Intuitively, $\bkws$ starts from the vertices that contain the query keywords and performs a backward search to identify potential vertices that might serve as the roots of a match. $\fkws$ initiates its search from these identified roots and proceeds forward. The objective of $\fkws$ is to discover any missing keywords within the subtrees that consider these vertices as roots.} We prove that both $\bkws$ and $\fkws$ have the monotonic property (detailed at the end of Sec.~\ref{sec:bkws} and ~\ref{sec:fkws}, respectively). The monotonic property of a few popular keyword search algorithms, such as \cite{he2007blinks,qin2014scalable,fan2017incremental}, can be analyzed similarly, which is omitted.

\subsection{Monotonic algorithms for keyword search}

This subsection presents how the keyword search algorithm has the monotonic property. More specifically, the monotonic property is defined with a {\em partial order} of {\em match variables} from a {\em finite domain}. Intuitively, the shortest distance between the root $u$ and a query keyword $q\in Q$, denoted as $\dist(u,q)$ (\ie $\dist(u,q) = \min\{\dist(u,v)| q\in L(v) \}$), is of a finite domain. When the monotonic property holds, its value decreases or remains unchanged during query processing and converges to the \textit{exact} shortest distance after query processing ends. Before providing further details, we present the structure of the match variable, $\answer_u$, which maintains the substree rooted at $u$.

\TKDERF{
\begin{definition}[Match $\answer_u$]
For a given graph $G$ and a query $(Q,\tau, k)$, a match $\answer_u$ with its root at $u$ represents a \textit{map}. $\forall q\in Q$, if $\dist(u,q) \leq \tau$, $\answer_u[q]$ is set to $\dist(u,q)$. Otherwise, $\answer_u[q]$ is set to $\mathsf{null}$.
\end{definition}
}

\noindent\TKDERF{
\stitle{Complete matches and partial matches.} $\answer_u[q]$  is initialized to ${\mathsf{null}}$. Throughout the search process, certain keywords for all $u$ may be discovered within $\answer_u$ and $\answer_u[q]$ is set to $\dist(u,q)$, while others may remain $\mathsf{null}$. Formally, a match $\answer_u$ is referred to as a {\em partial match} if and only if $\exists q\in Q$, $\answer_u[q]$ is ${\mathsf{null}}$ (\ie some keyword is not matched). Otherwise, $\answer_u$ is a {\em complete match}.
}

\begin{definition}[Monotonic $\kws$ algorithm]\label{def:monotone}
	 Given a graph $G=(V,E)$, where each node $u\in V$ is associated with $\answer_u$. A {\em monotonic keyword search algorithm} $\kws$ satisfies the following conditions:
	\begin{enumerate}[leftmargin=*]
		\item $\answer_u$ of all vertices are in a finite domain; and
		\item there exists a partial order $\preceq$ on $\answer_u$ such that, $\forall u\in V$, $\kws$ updates $\answer_u$ in the order of $\preceq$. 
	\end{enumerate}
\end{definition}

We next illustrate the details of the monotonic property of $\kws$ in relation to a finite domain and a partial order on the matches.

\stitle{(1) Finite domain of $\kws$}. To illustrate a finite domain of \choi{match variables}, we encode ${\mathsf{null}}$ with a constant large value $+\infty$ larger than $\sum_{e_i\in E} w(e_i)$. Consider the value of $\answer_u[q]$. $\answer_u[q] \in \{\sum_{e_i\in E'} w(e_i) | E'\subseteq E\}\cup \{+\infty\}$.

\noindent\TKDERF{
\stitle{(2) Partial order of $\kws$}. We propose the partial order $\preceq$ on $\answer_u$ which is defined as follows. Suppose $\kws$ updates the (partial or complete) matches by following an order $\preceq$. If $\answer_u' \preceq \answer_u$, $\answer'_{u}[q]\leq \answer_{u}[q]$ or $\answer_{u}[q] = \mathsf{null}$, then $\preceq$ is a partial order of $\kws$. Intuitively, $\kws$ follows the partial order and keeps \textit{refining} the distances between the roots of the matches and the query keywords to obtain the top-$k$ complete matches.
}

\stitle{Remarks.} A keyword search algorithm $\kws$ can be parallelized and terminated with the correct answer (\aka the top-$k$ matches) in a distributed environment if $\kws$ \revise{is correct for the query $Q$ on a single machine and has a monotonic property}. We follow the proof pipeline of \textbf{Theorem 1} in \cite{fan2017parallel}.

\noindent{\em (i) Termination.} In each superstep, at least one $\answer_u$ has to be updated. Given a graph $G$, the number of distinct values to update $\answer_u$ is bounded since all $\answer_u$ are in a finite domain and updates follow the partial order $\preceq$. Therefore, the number of supersteps is bounded.

\noindent\TKDERF{
{\em (ii) Correctness.} Given that $\kws$ is correct for query $Q$, at the superstep $R=1$, $\kws$ returns a set of correct local matches with roots in each fragment $F_i$. Matches with roots on portal nodes are passed to their copies in other fragments (if any) at the end of each superstep. At the superstep $R=s$, each node $u$ contains its local match $\answer_u$ from the superstep $R=(s-1)$ and the matches $\answer'_u$ which are rooted at its copies in other fragments. Therefore, $\kws$ can compute the correct match for the current superstep for each node. The correctness of the final matches is thus established by induction on the supersteps.
}

%We analyze the monotonic property of these algorithms at the end of  Sec.~\ref{sec:bkws} and \ref{sec:fkws}, respectively. These are used to analyze the correctness of $\SKWS$ on $\DKWS$ in Sec.~\ref{sec:analysis}.
 
\subsection{Monotonic backward search ($\bkws$)}\label{sec:bkws}

\TKDE{We next present the major steps of backward search for keyword search ($\bkws$), which is essential to many previous studies, \eg ~\cite{he2007blinks,DBLP:journals/tkde/YuanLCYWS17,fan2017incremental,qin2014scalable}. The detailed pseudocode is illustrated with {\em Lines~\ref{peval-bkws-init-ans}-\ref{peval-bkws-end}} of Algo.~\ref{algo:pevalkws}, to be discussed with $\PINE$ in Sec.~\ref{sec:dkwspine}. Given a keyword query, $\bkws$ goes through three key steps. First, $\bkws$ initializes a set of search origins. Second, $\bkws$ expands the search origins backward. Third, a complete match is found once the node $u$ is expanded by all search origins. We elaborate the details below.}

% \stitle{Overview of $\bkws$.} 

\stitle{Answer $\answerset$.} The answer to the query is a set of the top-$k$ matches at the end of the query algorithm. If $\answer_u$ is a top-$k$ match, $\answer_u\in \answerset$. We use $\prune$ to denote the score of the current $k$-th match in $\answerset$. Hence, $\prune$ is the {\em upper bound} of the score of any match in $\answerset$. Given a complete match $\answer_u$, if $\Score(u) > \prune$, we say that $\answer_u$ is a {\em candidate match}, \ie it is not among the current top-$k$ matches. Candidate matches may be refined and added to $\answerset$ by traversing adjacent fragments.

\stitle{Maintenance of answer.} $\SKWS$ maintains the top-$k$ matches $\answerset$ in a priority queue of a fixed size $k$ and is ordered in descending order according to the scores of the matches. The match at the head of $\answerset$ has the highest priority to be removed as it has the least compact structure. $\prune$ is initialized to $+\infty$. It remains unchanged when $|\answerset| < k$. Otherwise, it is always set to the score of the match at the head of $\answerset$. $\prune$ will be refined once there is a match found with a score which is smaller than $\prune$. Formally, when a candidate match $\answer_u$ is refined, the following are checked:
\begin{enumerate}[leftmargin=*]
	\item if $|\answerset| < k$, $\answer_u$ is inserted into $\answerset$ directly; and
	\item if $|\answerset| = k$ and $\Score(u)<\prune$, the match at the \jiaxin{head} of $\answerset$ is removed and $\answer_u$ is inserted into $\answerset$.
	\eat{\item otherwise, $\answer_u$ is skipped and it will be checked again once it is further refined. $\prune$ is set to the score of the \jiaxin{head} of $\answerset$.} 
\end{enumerate}

\etitle{(Step 1) Initialization.} Consider a set of query keywords $Q=\{q_1,q_2,$ $\ldots,q_l\}$. We denote the set of vertices that contain the keyword $q\in Q$ as $O_{q}$ (\aka search origin), and the set of vertices that could reach $q$ (\ie one of the vertices in $O_{q}$) as $V_{q}$.

\etitle{(Step 2) Backward expansion.} $\bkws$ expands the vertex set $O_{q}$ backwardly. In each search step, $\bkws$ compares the next vertex to be expanded for each query keyword, and the vertex $u$ with the smallest distance to the search origin is selected. In the expansion, $u$ is added to $V_{q}$ and $\answer_u$ is checked whether it is a complete match, where $(u,v)$ is an incoming edge of $v$. If (a) $\sum_{q\in Q} \dist(u,q)$ $>\prune$, where \vldbj{$u$ is the nearest vertex of $O_{q}$} and has not been expanded by query keyword $q$ (\ie $u_i=\arg\min_u\dist(u,v)$, where $u\not\in V_{q}$ and $v\in O_{q}$) or (b) all adjacent vertices of $V_{q}$s are expanded, the expansion stops. Otherwise, the backward expansion continues.

\etitle{(Step 3) Match discovery.} It discovers a complete match rooted at $u$ such that $u$ can reach at least one node that contains $q$, for each $q \in Q$, \ie $u\in \bigcap_{q\in Q} V_{q}$.

\begin{figure}[tb]
	\begin{center}
		\includegraphics[width=0.5\textwidth]{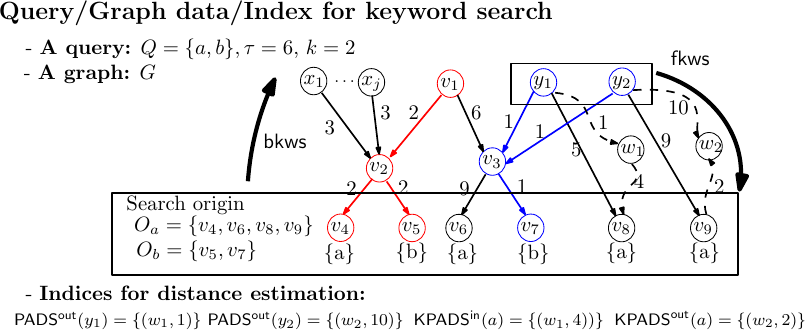}
	\end{center}
	\caption{A query, a data graph (top) and indexes (bottom) for the illustration of the key steps of $\bkws$ and $\fkws$}\label{fig:dataeg}
\end{figure}

\begin{figure*}[tb]
	\begin{center}
	\includegraphics[width=\textwidth]{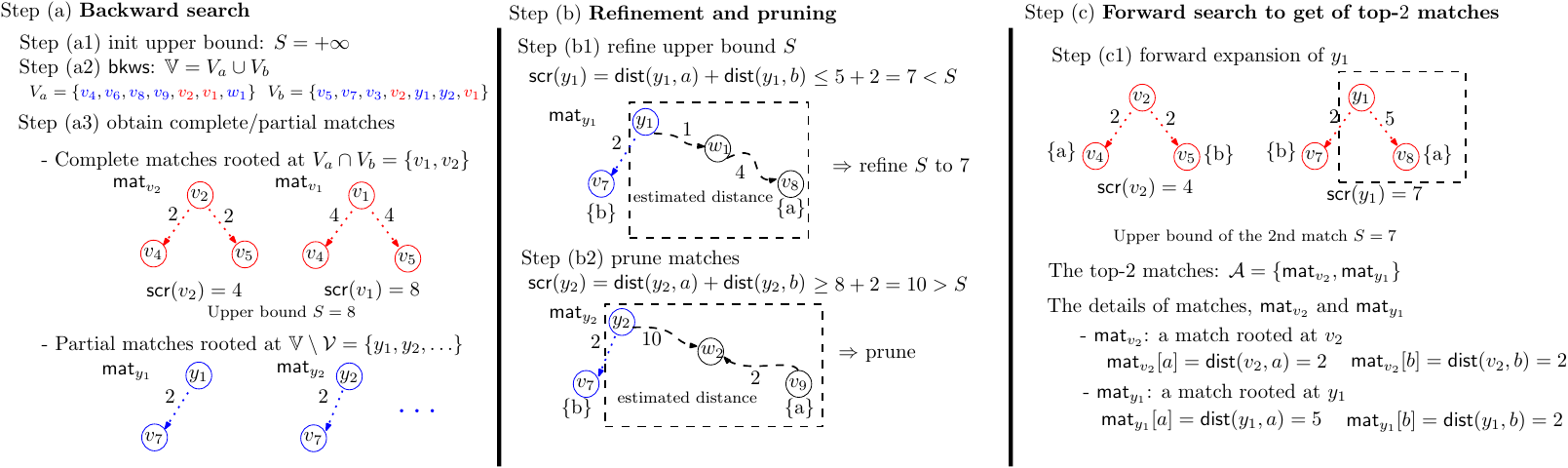}
		\end{center}
	\caption{Key steps: backward search (Sec.~\ref{sec:bkws}), refinement and pruning (Example~\ref{eg:refine} and~\ref{eg:prune}), and forward search (Sec.~\ref{sec:fkws})}
	\label{fig:skws}
\end{figure*}

\eat{
\begin{example}
      The backward search algorithm finds the top-$2$ matches rooted at $v_2$ and $v_1$, respectively. The match rooted at $v_2$ is discovered by the backward expansion from $v_4$ and $v_5$, while that rooted at $v_1$ is discovered by the backward expansion from $v_6$ and $v_7$.
\end{example}
}

\begin{example}
	Consider the graph in \vldbj{Fig.~\ref{fig:dataeg}}. Given a keyword query $(Q,\tau, k)$, where $Q=\{q_1,q_2\}$ that is $q_1=a$ and $q_2=b$, $\tau=6$ and $k=2$. For brevity, Fig.~\ref{fig:dataeg} only shows the vertex labels relevant to $Q$. \vldbj{We illustrate the backward search with Step (a) in Fig.~\ref{fig:skws}.} Initially, $O_a=\{v_4, v_6, v_8, v_9\}$ and $O_b=\{v_5, v_7\}$. The backward expansion iterates over $V_a$. The first seven vertices are $[v_4, v_6, v_8, v_9, v_2, v_1, w_1]$, which are ordered by the first time the vertices expanded. Similarly, the vertices of $V_b$ can be expanded as follows: $[v_5, v_7, v_3, v_2, y_1, y_2, v_1]$. Two complete matches rooted at $v_1$ and $v_2$ are discovered. The score of $\answer_{v_1}$ (resp. $\answer_{v_2}$) is $\Score(v_1)=8$ (resp. $\Score(v_2)=4$). Hence, the upper bound $\prune = 8$. The next vertex to expand for $V_a$ is $x_1$. The next vertex to expand for $V_b$ is $x_1$, too. $\dist(x_1, a) + \dist(x_1,b) = 5 + 5 = 10 > \prune$. The subsequent backward expansions, such as $x_i$ ($i\in [1,j]$), are skipped since the termination condition is met.
\end{example}

\eat{
\begin{figure}[tb]
	\begin{center}
	\includegraphics[width=0.5\textwidth]{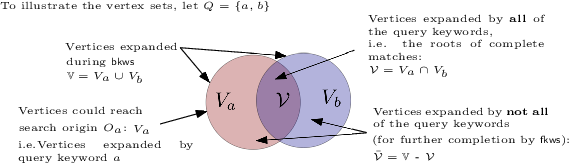}
		\end{center}
	\caption{Illustration of the relationship of the vertex sets in $\bkws$ and $\fkws$ (cf. Tab~\ref{tab-notations-ans})}
	\label{fig:venn}
\end{figure}
}

\stitle{Analysis of $\bkws$.} We show that $\bkws$ identifies all the partial and complete matches. We denote the union (resp. intersection) of $V_{q}$ by $\VU=\bigcup_{q\in Q} V_{q}$ (resp. $\VI=\bigcap_{q\in Q} V_{q}$). We note that $u\in \VU \setminus \VI$ reaches some of the query keywords but not all of them, \ie $\answer_u$ is a partial match. We denote the set of roots by $\VM = \VU \setminus \VI$ and have the following proposition.

\begin{proposition}\label{prop:s_completed}
	The node set visited by $\bkws$, $\VU$, has the following properties:\\
	(1) $\forall u\not\in \VU$, $\answer_u\not\in \answerset$; and (2) $\forall \answer_u\in \answerset$, $u\in \VU$.
\end{proposition}
\begin{IEEEproof}
The proof is presented in Appx.~\ref{appendix-skws-completeness} of \cite{techreport}.
\end{IEEEproof}

\TKDERF{Intuitively, if a vertex is not visited during the backward expansion of any query keyword, it cannot serve as the roots of the top-$k$ matches. Prop.~\ref{prop:s_completed} ensures that the roots of the top-k matches are in $\VU$. Some vertices in $\VU$ that are not roots of the top-k matches will be further pruned in $\fkws$ (Sec.~\ref{sec:fkws}).}

\begin{example}
	We illustrate the key steps of $\bkws$ with the graph in Fig.~\ref{fig:skws}(a). $V_a$ = $\{v_4,v_6,v_8,v_9,v_2,v_1,w_1\}$ and $V_b$ = $\{v_5,v_7,v_3,v_2,y_1,y_2,v_1\}$ are expanded, after $\bkws$. $\VI$ = $V_a\cap V_b$ = $\{v_1,v_2\}$ are the roots of the complete matches, \ie they can reach the vertices containing the query keywords $\{a,b\}$. $\VU$ = $\{v_4,v_6,v_8,v_9,v_2,v_1,w_1,v_5,v_7,v_3,y_1,y_2\}$ are the vertices which are traversed during $\bkws$. $\VM$ = $\VU\setminus\VI$ = $\{v_4,v_6,v_8,v_9,w_1,v_5,v_7,v_3,y_1,y_2\}$ are the vertices that are not backward traversed by either keyword $a$ or $b$.
\end{example}

\jiaxin{\stitle{Correctness.} \revise{By using match refinement, $\bkws$ is monotonic. Since $\answer_{u}[q]$ is refined when a shorter path between $u$ and $q$ in a complete match $\answer_u$ or a new path between $u$ and a missing keyword $q$ in a partial match $\answer_{u}$ is identified, the refinement follows the partial order $\preceq$ on $\answer_u$.} We recall that $\answer_{u}[q]$ is from a finite domain $\{\sum_{e_i\in E’}$ $ w(e_i)\}\cup \{\vldbj{+\infty}\}$, where $E’$ is any subset of $E$. By Def.~\ref{def:monotone}, $\bkws$ can be parallelized and terminated correctly.}

\stitle{Complexity.} $\bkws$ takes $O(|Q|(|E|+|V|\log |V|))$, where $|Q|$ is the number of query keywords. For simplicity, we provide the analysis in Appx.~\ref{sec:complexity_bkws} of \cite{techreport}. \vldbj{The size of a match, $\answer_u$, is bounded by $O(|Q|)$. Hence, the space complexity is bounded by $O(|Q||V|)$.}

\subsection{Monotonic forward search ($\fkws$)}\label{sec:fkws}

%\stitle{\ref{sec:skws}.(II) Forward search ($\fkws$)}

The main purpose of $\fkws$ is to retrieve the missing keywords of the partial matches via forward expansion. \revise{$\fkws$ is also widely used in existing keyword search algorithms, \eg \cite{he2007blinks,DBLP:journals/tkde/YuanLCYWS17,bhalotia2002keyword,DBLP:conf/www/Shi0K20,kacholia2005bidirectial}}. Due to a potentially large number of partial matches, forward expansion for the vertices $\VM$ could be costly. Existing studies can be space-consuming~\cite{he2007blinks} or do not guarantee exact matches~\cite{kacholia2005bidirectial}. We describe the forward expansion and propose \TKDE{\textit{new bounds}} for pruning in $\fkws$.
%, which make forward expansion efficient and is less space-consuming. Firstly, we describe forward expansion below.

\eat{Hence, Kacholia \cite{kacholia2005bidirectial} took the advantage of the vertex prestige to avoid some redundant or unnecessary search. However, it highly relies on the graph structure and cannot return the exact top-$k$ answers. He et al. \cite{he2007blinks} proposed a vertex-keyword index to avoid forward expansion. However, such an index is space-consuming.}

\stitle{Forward expansion.} Consider a partial match $\answer_u$. Suppose a query keyword $q\in Q$ is missing in $\answer_u$. $\fkws$ forward expands from $u$ by using Dijkstra's algorithm to retrieve the nearest node that contains $q$.

\stitle{Pruning in $\fkws$.} Some forward expansions do not lead to complete matches and can be pruned as shown in Prop.~\ref{prop:pruning}.

\begin{proposition}\label{prop:pruning}
Consider the forward expansion for vertex $u$. Suppose the next vertex to be expanded by Dijkstra's algorithm is $v$, the forward expansion is terminated when any of the following conditions holds.
\begin{enumerate}[label=(\roman*), leftmargin=*]
	\item $q\in L(v)$, \ie the keyword $q$ is found;
	\item $\dist(u,v) > \tau$, the vertex containing keyword $q$ is farther than $\tau$ from $u$ or does not exist (\ie $ \dist(u,q) > \tau$); or \label{condition:c2}
	\item $\Score(u) + \dist(u,v) > \prune \Rightarrow \Score(u) + \dist(u,q) > \prune$. \label{condition:c3}
\end{enumerate}
\end{proposition}

% motivations

\TKDERF{As indicated by Condition~\ref{condition:c2}, if $\dist(u,q)$ has been indexed, early termination can be determined if $\dist(u,q) > \tau$. Furthermore, Condition~\ref{condition:c3} posits that the current top $k$-th match score, denoted as $\prune$, serves as an upper bound. If $\dist(u,q)$ is indexed, we can employ a tightly estimated upper bound of $\prune$ to facilitate decisions on early termination. Thus, we engage state-of-the-art indexing techniques — PageRank-based All-distances Sketches ($\PRADS$) and PageRank-based Keyword Distance Sketches ($\KPADS$)~\cite{Jiang2020PPKWSAE}. Specifically, $\PRADS(u)$ is a sketch for $u$, which indexes the shortest distance between $u$ and the sketch's centers (some vertices in the graph). Given that $\PRADS(v_i)$ and $\PRADS(v_j)$ may share common centers where $q\in L(v_i) \cap L(v_j)$, these shared centers can be merged. In the process of merging, only the smallest distance is retained. $\KPADS(q)$ sketch is constructed through such merges and is used to index the shortest distance between the keyword $q$ and the centers. These sketches assist in estimating both the upper and lower bounds of the shortest distance between $u$ and $q$, where $u$ belongs to $\VM$ and $q$ is a missing keyword in $\answer_u$.}

\stitle{Indexing.} $\PRADS{}$ and $\KPADS{}$ have been shown to be both space- and time-efficient in practice with theoretical guarantees on the accuracy of the shortest distance which can be readily distributed.
However, we remark that \cite{Jiang2020PPKWSAE} considered undirected graphs. To support {\em directed graphs}, we make a modification to $\PRADS$ as follows. The sketch of a node $u$ is two sets of vertices and their corresponding shortest distances from (resp. to) $u$, denoted by $\PRADS^{\mathsf{out}}(u)=\{(w,d)\}$ (resp. $\PRADS^{\mathsf{in}}(u)=\{(w,d)\}$), where $w\in V$ and $d=\dist(u,w)$ (resp. $d=\dist(w,u)$). Similarly, the sketch of a keyword $q$ is denoted by $\KPADS^{\mathsf{out}}(q)=\{(w,d)\}$ (resp. $\KPADS^{\mathsf{in}}(q)=\{(w,d)\}$), where $w\in V$ and $d=\dist(q,w)$ (resp. $d=\dist(w,q)$). For brevity, we leave the construction pseudo-code of $\PRADS$ and $\KPADS$ in~\cite{techreport}. 

Since $\PRADS$ yields estimated bounds, $\fkws$ needs to handle both approximate and exact matches. Specifically, $\fkws$ computes the {\em upper} bound of the score for any $u\in \VM$, $\Score(u)$, by estimating the shortest distance between $u$ and the missing keywords, \ie $\Sigma \dist(u,q_i) + \Sigma \answer_u[q_j]$, where $q_i,q_j\in Q$, $q_i$ is missing from $\answer_u$ whereas $q_j$ has been found in $\answer_u$. If the upper bound is smaller than $\prune$, $\answer_u$ is inserted into $\answerset$ and $\prune$ is refined accordingly. To avoid ambiguity, we denote the $\answerset$ that may consist of exact matches and approximate matches by $\hat{\answerset}$. The approximate matches in $\hat{\answerset}$ are further refined during forward expansion. $\hat{\answerset}$ is eventually refined to yield $\answerset$.

Next, we present the upper and lower bounds of $\dist(u,q)$ for the termination of forward expansion from $u$. These bounds can be applied to other keyword search semantics as they involve numerous distance computations, such as ~\cite{kargar2011keyword,efficient2020kargar,he2007blinks}.

\stitle{(1) Upper bound of the shortest distance.} Given a shortest distance query $(u,q)$, the upper bound is computed by $\PRADS^{\mathsf{out}}(u)$ and $\KPADS^{\mathsf{in}}(q)$ as follows:
\begin{equation}\label{equ:estimateUpper}
\dist(u,q) \leq \dist(u,w) + \dist(w,q),
\end{equation}
where $(w, \dist(u,w)) \in \PRADS^{\mathsf{out}}(u), (w, \dist(w,q)) \in \KPADS^{\mathsf{in}}(q)$, and $w$ is a common center in $\PRADS^{\mathsf{out}}(u)$ and $\KPADS^{\mathsf{in}}(q)$.
%}

\begin{example}\label{eg:refine}
	Consider the graph in Fig.~\ref{fig:skws}(b1). $\PRADS^{\mathsf{out}}(y_1)=\{(w_1, 1)\}$ and $\KPADS^{\mathsf{in}}(a) = \{(w_1,4)\}$. The common center of $\PRADS^{\mathsf{out}}(y_1)$ and $\KPADS^{\mathsf{in}}(a)$ is $w_1$. Hence, the upper bound of the shortest distance between $y_1$ and keyword $a$ is derived by $\dist(y_1,w_1) + \dist(w_1, a) = 5$. Then, the upper bound of the score of the match rooted at $y_1$ is $7$. Since the upper bound is smaller than $\prune$, the approximate match $\answer_{y_1}$ is inserted into $\answerset$ to yield $\hat{\answerset}$. $\prune$ is refined accordingly, $\prune = \Score(y_1)$. 
\end{example}

\stitle{(2) Lower bound of the shortest distance.} We also derive a lower bound of the shortest distance between $u$ and $q$ by exploiting $\PRADS^{\mathsf{out}}(u)$ and $\KPADS^{\mathsf{out}}(q)$ to prune unnecessary traversals in an early stage of forward expansion. We have the following inequality.
\begin{equation}\label{equ:lowerbound}
\dist(u,q) \vldbj{\geq} \dist(u,w) - \dist(q,w),
\end{equation}
where $(w, \dist(u,w)) \in \PRADS^{\mathsf{out}}(u), (w, \dist(q,w)) \in \KPADS^{\mathsf{out}}(q)$, and $w$ is a common center in $\PRADS^{\mathsf{out}}(u)$ and $\KPADS^{\mathsf{out}}(q)$. Therefore, the minimum of $\dist(u,w) - \dist(q,w)$ is the lower bound of $\dist(u,q)$.

\eat{
\begin{theorem}\label{theorem:lower_bound}
	Given a vertex $u\in V$ and a keyword $q\in Q$, the following inequality holds:
	\begin{equation}\label{equ:estimateLower}
		\dist(u,q) \geq \dist(u,w) - \dist(q,w),
		\end{equation}
		where $(w, \dist(u,w)) \in \PRADS^{\mathsf{out}}(u)$, and $(w, \dist(q,w)) \in \KPADS^{\mathsf{out}}(q)$.
\end{theorem}
\begin{proof}
\eat{According to the definition of $\PRADS$ and $\KPADS$, $d_u = \dist(u,w)$ and $d_{q} = \dist(q,w)$.}
By the triangle inequality, we have
\begin{equation}
	\dist(u,q) + \dist(q,w) \geq \dist(u,w)	
\end{equation} 
\begin{equation}
 \dist(u,q) \geq \dist(u,w) - \dist(q,w)
\end{equation}
\end{proof}
}

If the lower bound is larger than $\tau$, the forward expansion from $u$ is simply skipped, since $\dist(u,q) > \tau$, and Prop.~\ref{prop:pruning}-Condition \ref{condition:c2} is already satisfied. Similarly, if the lower bound of the score of the match rooted at $u$ is larger than $\prune$, Prop.~\ref{prop:pruning}-Condition \ref{condition:c3} is met.

\begin{example}\label{eg:prune}
	Consider the graph in Fig.~\ref{fig:skws}(b2). Suppose $\PRADS^{\mathsf{out}}(y_2)=\{(w_2, 10)\}$ and $\KPADS^{\mathsf{out}}(a) = \{(w_2,2)\}$. The common center of $\PRADS^{\mathsf{out}}(y_2)$ and $\KPADS^{\mathsf{out}}(a)$ is $w_2$. The lower bound of the shortest distance between $y_2$ and keyword $a$, $\dist(y_2, a)$, is derived by $\dist(y_2,w_2) - \dist(a, w_2) = 8$. The lower bound of the score of the match rooted at $y_2$ is $10 > \prune$. The forward expansion of $y_2$ is pruned.
\end{example}

%Eq.~\ref{equ:estimateUpper} and Eq.~\ref{equ:estimateLower}  are true because of the triangle inequality. It is worth noting that any 2-hop label technique could be adopted. We adopt \PRADS{} and \KPADS{} because their accuracies have been verified while they are both space- and time-efficient in practice with theoretical guarantees on the accuracy of the shortest distance \cite{Jiang2020PPKWSAE}.

\stitle{Correctness.} The analyses of partial order and the finite domain of $\fkws$ are similar to those of $\bkws$. Hence, $\fkws$ can be parallelized correctly since it has the monotonic property.

\stitle{Complexity.} In the worst case, $\fkws$ performs a single source shortest path computation for each vertex $u\in \VM$. Therefore, the time complexity of $\fkws$ is bounded by $O(|\VM|$ $(|E|+|V|\log |V|))$. \TKDE{The space complexity $\fkws$ is identical to that of $\bkws$, which is bounded by $O(|Q||V|)$, whereas the space for $\PRADS{}$ and $\KPADS{}$ is $O(|V|\log |V|)$~\cite{Jiang2020PPKWSAE}.}

\section{Distributed keyword search ($\DKWS$)}\label{sec:dkws}

We illustrate $\PIE$~\cite{fan2017grape} with a keyword search algorithm, denoted as $\kws$. (a) $\PEval$ is {\em partial evaluation} of $\kws$. Partial results are passed to the next function. (b) $\IncEval$ is {\em incremental evaluation} of $\kws$ that takes partial results and computes the changes. $\IncEval$ is repeated until no more changes are computed. (c) $\Assemble$ collects local matches from workers. These functions are evaluated in a non-preemptive manner and defined formally as follows.

\TKDERF{
The partial evaluation ($\PEval$) utilizes a query $Q$ and a fragment $F_i$ of the graph $G$ as inputs. $\PEval$ then concurrently computes partial answers, represented as $Q(F_i)$, consisting of current $\answer_u$ for all $u\in V$ at each worker $P_i$.
}

\TKDERF{
The incremental evaluation ($\IncEval{}$) takes four inputs: a query $Q$, a fragment $F_i$ of graph $G$, partial results derived from the application of the query to the fragment $Q(F_i)$, and a message $M_i$. The function then incrementally computes $Q(F_i\oplus M_i)$, optimizing the computation of $Q(F_i)$ from the previous superstep to maximize efficiency. After every execution of $\IncEval{}$, $\DKWS$ updates its state by considering $F_i \oplus M_i$ and $Q(F_i \oplus M_i)$ as the new $F_i$ and $Q(F_i)$, respectively, forming the input for the incremental computation in the next superstep.
}

\TKDERF{$\Assemble$ starts its computation when $M_i$ is empty for any worker $P_i$. $\Assemble$ accepts $Q(F_i \oplus M_i)$ as inputs. It consolidates for all $i\in [1,m]$, $Q(F_i \oplus M_i)$, to compute the final answer $Q(G)$.}

\stitle{Architecture of $\DKWS$ (Fig.~\ref{fig:workflow}).} The coordinator $P_0$ is responsible for receiving and transmitting the query to all workers. Workers $P_1$ to $P_n$ are in charge of computing the query on their fragments $F_1$ to $F_n$. When receiving the query, the workers perform $\PEval$. During each superstep ($\IncEval$) of query computation, a selector of each $P_i$ decides to perform either $\bkws$ or $\fkws$ on $F_i$. When all workers meet the termination condition, the coordinators assemble the (local) top-$k$ matches and select the (global) top-$k$ matches from the local ones.

\stitle{Programming model of $\DKWS$.} $\DKWS$ differs from previous studies in two major ways: (1) $\DKWS$ is the first to introduce a {\em notify-push} paradigm into a distributed programming model. \jiaxin{The notify-push paradigm allows the coordinator and workers to asynchronously exchange refined bounds (Sec.~\ref{sec:np-paradigm}) at runtime; \eat{While $\DKWS$ has synchronous evaluation, the bounds can be exchanged before the end of each superstep, which provides a global scope for each worker;}} and (2) $\PINE$ consists of \underline{P}Eval,  \underline{I}ncEval (\underline{$n$} subtasks), and one Ass\underline{e}mble functions (Sec.~\ref{sec:dkwspine}) that the users can use to solve their problems by composing several $\mathsf{PI}$ algorithms and assembling the matches at the end, rather than only one $\PIE$ algorithm. $\DKWS$ runs the $\mathsf{PI}$ algorithms in a {\em preemptive} manner and therefore interactions, such as exchange of tighter bounds (presented in Sec.~\ref{sec:skws}) between them are possible.

\begin{figure*}
\begin{minipage}[t]{0.34\textwidth}
\includegraphics[width=\textwidth]{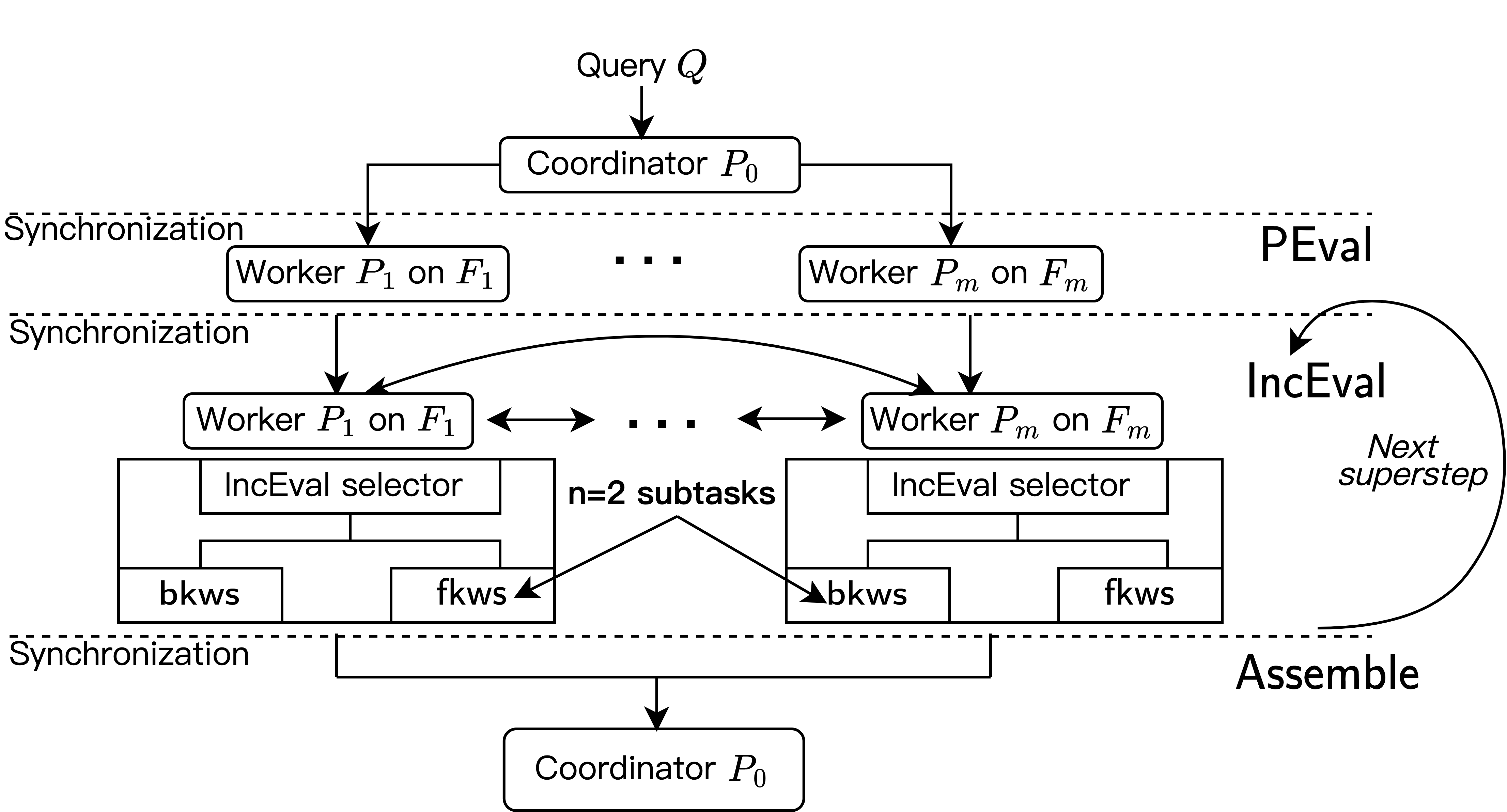}
	\caption{Workflow of $\DKWS$}
	\label{fig:workflow}
\end{minipage}\hspace{1em}
    \begin{minipage}[t]{0.63\textwidth} 
    \includegraphics[width=\textwidth]{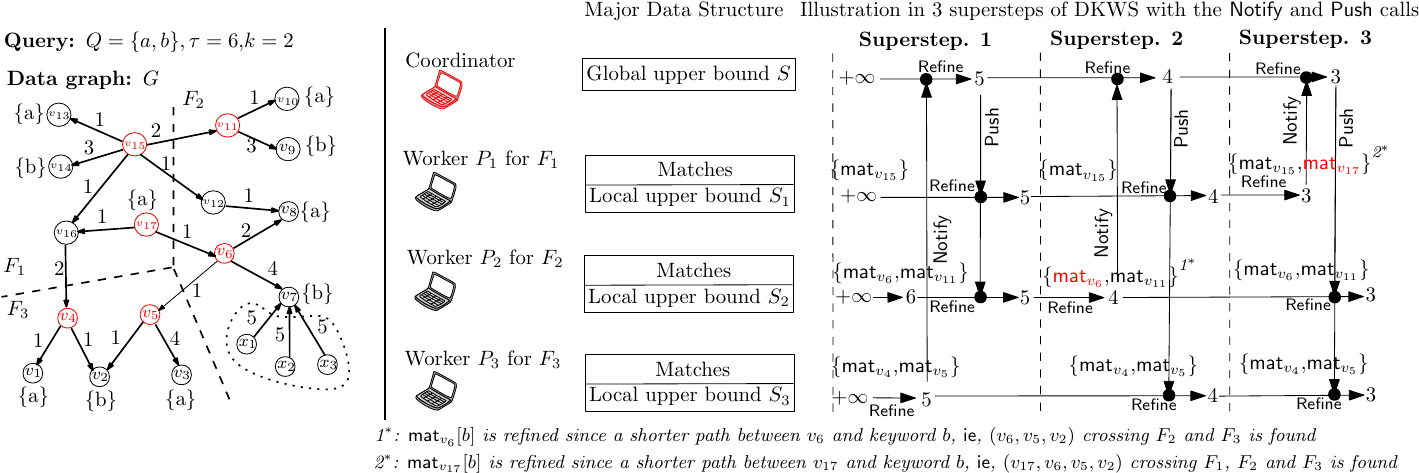}
    \caption{Illustration of the change of bounds during query processing of $\DKWS$}
	\label{fig:dkws-example}
    \end{minipage}
\end{figure*}

%\jiaxin{\DKWS extends the $\PIE$ programming model ($\mathsf{\underline{P}Eval}$, $\mathsf{\underline{I}ncEval}$ and $\mathsf{Ass\underline{e}mble}$) of \cite{fan2017parallel} to the $\PINE$ programming model }

% Byron: the above is some old text. Maybe useful later.
%A sequential algorithm often has a series of sub-algorithms which can be interleaved. However, $\PIE$ cannot fully takes the advantages of the effectiveness of each sub-algorithm since differenct sub-algorithms might have some outdated messages which can be pruned by the other sub-algorithms.

\eat{
\begin{figure}[tb]
	\begin{center}
% 	\includesvg{./figures/dkwsframework.svg}
	\includegraphics[width=0.4\textwidth]{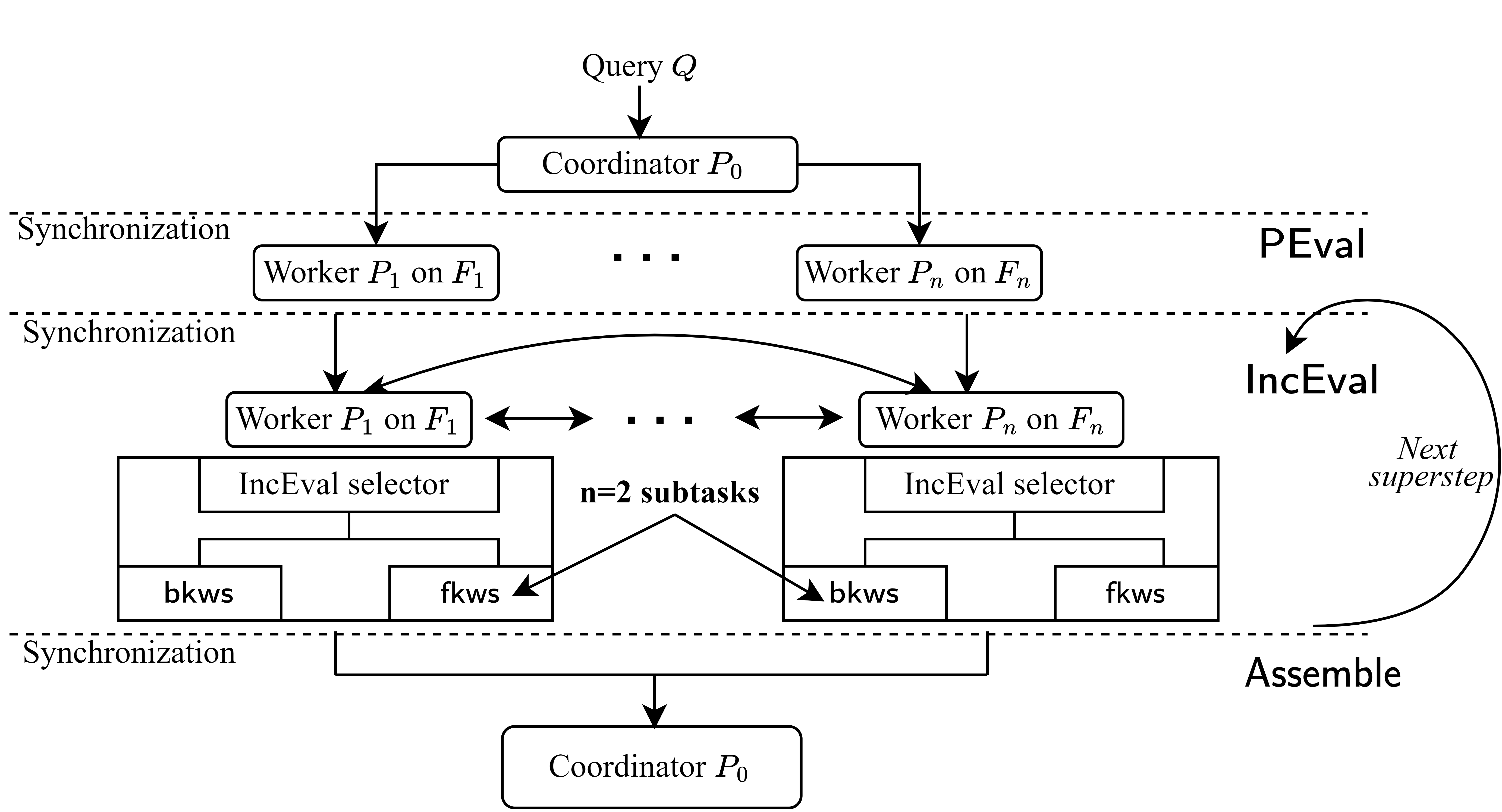}
		\end{center}
	\caption{Workflow of the $\PINE$ programming model in $\DKWS$}
	\label{fig:workflow}
\end{figure}
}

%out-dated
% Next, we illustrate how \DKWS implements the backward search and forward search using PINE (detailed in Sec.~\ref{sec:np-paradigm}). In particular, \DKWS implements $\SKWS$ by two $\PIE$ algorithms, namely $\bkws$ and $\fkws$, respectively. In $\bkws$ (Fig.~\ref{fig:notifypush}, Step 1.1), \DKWS performs the backward expansion starting from the vertices which contain a given query keyword (detailed in Sec.~\ref{sec:dkwspine}.1). In $\fkws$ (Fig.~\ref{fig:notifypush}, Step 1.2), \DKWS starts the forward search from the roots of partial answers to retrieve the missing query keywords. (detailed in Sec.~\ref{sec:dkwspine}.2).

\subsection{Notify-Push (NP) paradigm}\label{sec:np-paradigm}
%\stitle{\ref{sec:dkws}.(I) Notify-Push paradigm}

\eat{
\begin{figure}[tb]
  %\begin{center}
	  \includegraphics[width=0.25\textwidth]{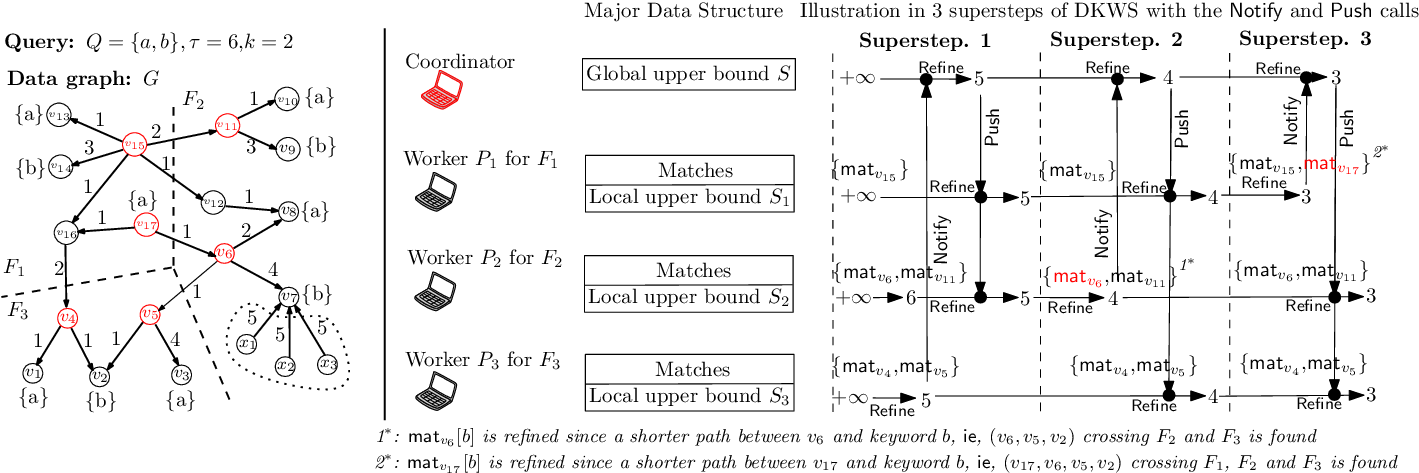}
		%\end{center}
	\caption{Example of query processing of $\DKWS$}
	\label{fig:dkws}
\end{figure}
}

%Workers may perform unnecessary computations as they do not have more global information.

With the Notify-Push paradigm, the bounds can be exchanged at run-time and provide a global scope for each worker. The pruning techniques are more efficiently on the workers with the tighter bounds.

%%%%%%%%Answer Expanding
\begin{algorithm}[tb]
    \caption{API of $\DKWS$}\label{algo:dkwsapi}
    \footnotesize
    \SetKwProg{Fn}{Function}{}{}
    \Fn{$\mathsf{Notify}$ \textnormal{(Worker id $i$, Local upper bound $\prune_i$):}}{
        Worker $P_i$ sends $\prune_i$ to notify  the coordinator $P_0$ \\
        Coordinator refines the global upper bound $\prune$ by $\min\{\prune, \prune_i\}$\\
    }
    \Fn{$\mathsf{Push}$ \textnormal{(Worker id $i$, Global upper bound $\prune$):}}{
        Coordinator $P_0$ pushes $\prune$ to  worker $P_i$ \\
        Worker $P_i$ refines the local upper bound $\prune_i$ by $\min\{\prune, \prune_i\}$\\
    }
    
    \end{algorithm}

\begin{definition}[$\Notify$ API]
$\Notify(i, \prune_i)$ is an API that a worker refines the global upper bound $\prune$ with the local upper bound $\prune_i$. $\Notify(i, \prune_i)$ takes a worker's id $i$ and a local upper bound $\prune_i$ as input. $\Notify$ must be invoked by a worker $P_i$ to notify the coordinator with its worker id $i$ and the local upper bound $\prune_i$.
\end{definition}

\stitle{Local upper bound $S_i$.} For fragment $F_i$, $\DKWS$ maintains a local upper bound $\prune_{i}$ to prune false matches locally.  $\DKWS$ maintains a priority queue $\answerset_i$ with a fixed size $k$ to store the local top-$k$ matches, which are ordered in descending order of the score of the matches for each fragment $F_i$. Once a better match is inserted into $\answerset_i$, $\prune_{i}$ is refined locally. The worker $P_i$ sends the refined local upper bound $\prune_{i}$ to the coordinator $P_0$ and {\em notifies} the coordinator to refine the global upper bound by calling function $\Notify(i, \prune_i)$.

\begin{definition}[$\Push$ API]
$\Push(i,\prune)$ is an API that the coordinator $P_0$ broadcasts the global upper bound $\prune$ to all the workers and refines the local upper bounds. $\Push(i, \prune)$ takes a worker's id $i$ and the global upper bound $\prune$ in the coordinator $P_0$ as input. $\Push$ is invoked by the coordinator $P_0$ and pushes the global upper bound $\prune$ to worker $P_i$.
\end{definition}

\stitle{Global upper bound $\prune$}. When the coordinator receives a local upper bound from a worker, it refines its local upper bound table which records the local upper bounds from all the workers. The global upper bound $\prune$ is the smallest among the local upper bounds. To avoid excessive refinements, the coordinator maintains a notification counter $N_i$ for each fragment $F_i$. Consider any $N_i$. If $\max\{N_j | j\in [1, m]\} - N_i$ is larger than a threshold, and $S_i > \prune$, this implies $P_i$ may be doing unnecessary computation on the fragment $F_i$ for a long time. The coordinator {\em push}es the global upper bound to $F_i$ by calling  $\Push$($i, \prune$). Once $P_i$ receives the global upper bound $\prune$, it  refines the local upper bound $\prune_i$ with $\prune$.

\eat{
\begin{figure*}[tb]
	\begin{center}
	\includegraphics[width=0.55\textwidth]{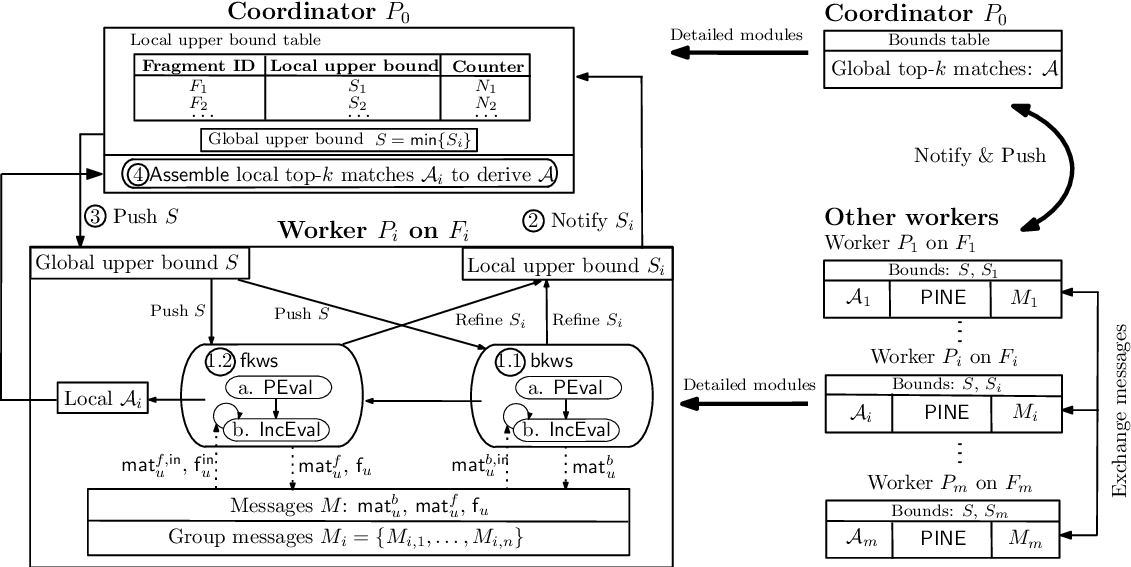}
		\end{center}
	\caption{Illustration of the Notify-Push paradigm with Coordinator $P_0$ and Workers $P_i$, and other workers (sketches)}
	\label{fig:notifypush}
\end{figure*}
}

\TKDERF{Note that the notify-push paradigm is established on the fact that the local upper bound $\prune_i$ is the upper bound of the global upper bound $\prune$. We formalize this as follows.}

\TKDERF{
\begin{lemma}
    $\forall i\in [1,m]$, $\prune_i \leq \prune$.
\end{lemma}
\begin{IEEEproof}
    We can prove this assertion by contradiction. Let's suppose that $\prune_i > \prune$. By definition, $\prune_i$ (resp. $\prune$) denotes the score of the local (resp. global) $k$-th match, represented by $\answer_{u_k}$ (resp. $\answer_{u_k'}$). Considering any local top-$k$ match $\answer_{u_j}$ with $j\in[1,k]$, $\Score(\answer_{u_j}) \leq \Score(\answer_{u_k}) < \Score(\answer_{u_k'})$. This implies that $\answer_{u_k'}$ is not included among the global top-$k$ matches, as there are $k$ matches with lower scores on $F_i$. Hence, we deduced that $\prune_i \leq \prune$.
\end{IEEEproof}
}

\begin{example}
        Given a distributed graph $G$, which has been partitioned into three fragments ($F_1$, $F_2$, and $F_3$), shown in Fig.~\ref{fig:dkws-example}, assume that the query keywords are $Q=\{a,b\}$ and $k=2$. In the first superstep of $\DKWS$, $P_3$ finishes the computation earlier since the size of $F_3$ is smaller. The local upper bound $\prune_2$ on $F_2$ is $ 6$. Without the $\np$ paradigm, $P_2$ does not terminate until all vertices of $F_2$ are traversed. The vertices $x_1$, $x_2$, and $x_3$, are not pruned by $\prune_2$ since $\dist(x_1,b) = 5 < \prune_2$ and the termination condition of backward expansion is not met as presented in Sec.~\ref{sec:bkws}. With the paradigm, $P_3$ sends $\prune_3=5$ to the coordinator by $\Notify(3,\prune_3)$. Then, the coordinator refines the global upper bound $\prune$ with $\prune_3$ and pushes the global upper bound to all the workers, \eg $P_2$, by $\Push(2, \prune)$. Once $P_2$ receives the global upper $\prune = 5$, it refines the local upper bound $\prune_2$ (denoted by $6\xrightarrow[]{R}5$) accordingly. Since the paradigm allows exchanging the bounds during a superstep, if $x_1$, $x_2$ and $x_3$ have not been visited, they are pruned, since $\dist(x_i, b) = 5 \geq \prune_2$. Then, the termination condition of backward search is met. Similarly, in Superstep 2, the backward expansion at $v_{16}$ is skipped.
\end{example}

\stitle{\it Remarks.} $\DKWS$ is efficient for several reasons: (a) $\Notify$ API provides a way for each worker to send refined bounds which help to prune more false matches on stragglers; and (b) The communication cost is small since $\DKWS$ only exchanges the local upper bounds rather than intermediate matches during distributed query evaluation.

\subsection{$\PINE$ programming model}\label{sec:dkwspine}

\stitle{\ref{sec:dkwspine}.1 Overview of $\PINE$}

\noindent\TKDERF{
The overview of $\PINE$ is illustrated with Fig.~\ref{fig:workflow}. $\PINE$ consists of \underline{P}Eval and \underline{I}ncEval of \underline{$n$} subtasks, along with one Ass\underline{e}mble. In the first superstep, $\PEval$ of all subtasks are executed in each worker $P_i$. In subsequent supersteps, each worker $P_i$ features an $\IncEval{}$ selector that decides which subtask's $\IncEval{}$ to execute. This granular level of execution is designed to address the straggler problem \TKDE{(refer to the Challenge 1 in Sec.~\ref{sec:intro})}.
}

We next illustrate the $\PINE$ programming model with an efficient implementation of $\SKWS$. There are two subtasks, $\bkws{}$ (Sec.~\ref{sec:dkwspine}.2) and $\fkws{}$ (Sec.~\ref{sec:dkwspine}.3), respectively. For each subtask, we only need to declare its messages, $\PEval$ and $\IncEval{}$. We propose preemptive execution of $\IncEval$s in $\DKWS$ (shown in Fig.~\ref{fig:workflow}). We use $\answerb_u$ (resp. $\answerf_u$) to denote the partial match found by $\bkws$ (resp. $\fkws$) rooted at $u$. Finally, we implement $\Assemble$ by collecting the local top-$k$ matches from all fragments to yield the global top-$k$ matches after both the $\IncEval$s terminate.

\stitle{\ref{sec:dkwspine}.2 $\PI$ for $\bkws$}

\etitle{Message declaration.} $\DKWS$ declares a variable $\answerb_u$ for each vertex $u$, where $\answerb_u$ is a map such that $\answerb_u[q]=\langle \vsf, \distance\rangle$ is used to denote the shortest distance between $u$ and a query keyword $q\in L(\vsf)\cap Q$, \ie $\distance = \dist(u,q)$. \TKDERF{Intuitively, $u$ is considered as the root of a match, while $\vsf$ is a leaf vertex of the match, the labels of which contain a query keyword, $q$.}

%%%%%%%%Answer Expanding
\begin{algorithm}[tb]
  \caption{{$\PEval$} for $\bkws$}\label{algo:pevalkws}
  \footnotesize
  \SetKwProg{Fn}{Function}{}{}
  \KwIn{$F_i(V, E, L)$, $Q=\{q_1,\ldots, q_l\}$, $\threshold$}
  \KwOut{$Q(F_i)$ consisting of current $\answerb_u$ for all $u\in V$}
  \blue{init a local upper bound $\prune_i$ with a large value\label{peval-bkws-init}} \\
  \blue{For each node $u\in V$, init a match variable $\answerb_u$ to $\mathsf{null}$\label{peval-bkws-init-ans}}\\
  \ForEach(\tcp*[h]{\blue{init the searching origin}}){$q\in Q$}{ \label{peval-bkws-init-start}
	\blue{init search priority queue $\Queue_q = \emptyset$} \eat{\tcp*[h]{$\langle u,d \rangle\in \Queue_q$ is a vertex-distance pair and ordered in the ascending order of $d=\dist(u, q)$}}\\
	\blue{init visited vertices set $\Visit_{q}=\emptyset$} \\
	%\blue{init visited vertices set $\Visit_{j}=\emptyset$}\\
  	\ForEach{\blue{$u\in O_{q}$}}{
		  \blue{$\answerb_u[q] = 0$} \\
		  \blue{$\Queue_q.\mathsf{push}(\langle u, 0 \rangle)$} \label{peval-bkws-init-end}
  	}
  }
  $\mathsf{BackwardExpand}(\Queue)$\tcp*[h]{$\Queue=\{\Queue_q | q\in Q\}$} \\

  \Fn{$\mathsf{BackwardExpand}(\Queue)$}{
 	\While{$\exists \Queue_{q}$ \tnormal{is not empty and} $\prune_i > \Sigma \Queue_{q}.\mathsf{top}().\distance$}{\label{peval-bkws-whole-start}
		pick $\Queue_{q}$ \tnormal{from all the search queues with minimal} $\Queue_{q}.\mathsf{top}().\distance$\\
		$\langle u, d\rangle = \Queue_{q}.\mathsf{top}()$ \\
		% $d = \Queue_{j}.\mathsf{top}().\distance$ \\
		$\Visit_q.\mathsf{add}(u)$\\
 		\ForEach{$e = (u', u)\in E$ \tnormal{and} $u' \not\in \Visit_q$}{
 			$d' = w(e) + d$ \\
 			\If{$d' < \tau$ \tnormal{and} $d' < \answerb_{u'}[q]$}{
 				$\answerb_{u'}[q] = d'$ \\
				 $\Queue_q.\mathsf{push}(\langle u', d'\rangle)$\\
				 \If{$\answerb_u$ is a complete match \tnormal{and} $\Score(u) < \prune_i$}{
					 $\answerset_i.\mathsf{push}(\langle u, \answerb_u\rangle)$ \\
					 $\prune_i = \Score(\answerset_i.\mathsf{top}().\usf)$\label{peval-bkws-end} \\ 
					 \highlight{$\Notify$($i$, $\prune_i$)} \label{peval-bkws-np}\\ 
				 }
 			}
 		}
 	}
  }
 \etitle{Message segment:} \blue{$M_i = \{\answerb_u | u\in F_i.I\}$ \label{peval-bkws-msg}}
 
  \end{algorithm}

\etitle{(1) Partial evaluation} ($\PEval$) for $\bkws$ (Algo.~\ref{algo:pevalkws}). \jiaxin{Upon receiving a query $Q$, $\PEval$ computes the partial matches of $\bkws$, $\answerb_u$ on $F_i$ locally, for all $i\in [1,m]$ in parallel.} $P_i$ initializes its local upper bound $\prune_i$ with a large constant value and initializes a match variable $\answerb_u$ for each vertex (Lines~\ref{peval-bkws-init}-\ref{peval-bkws-init-ans}). Lines~\ref{peval-bkws-init-start}-\ref{peval-bkws-init-end} initialize the search origins and the priority queue for the search. Lines~\ref{peval-bkws-whole-start}-\ref{peval-bkws-end} present the pseudo-code of $\bkws$ (described in Sec.~\ref{sec:bkws}).  In addition, in the $\np$ paradigm, at runtime, $\PEval$ sends the local upper bound $\prune_i$ to the coordinator and notifies it to refine the global upper bound when $\prune_i$ is refined (Line~\ref{peval-bkws-np}). In Line~\ref{peval-bkws-msg}, the messages are grouped into $M_i$ at the incoming portal nodes on fragment $F_i$. Partial matches that are relevant to $F_j$ ($M_{i,j}=\{\answerb_u | u\in F_i.I \cap F_j.O\}\in M_i$) are transmitted to worker $P_j$.  

%%%%%%%%Answer Expanding
\begin{algorithm}[tb]
  \caption{{$\IncEval$} for $\bkws$}\label{algo:inckws}
  \SetKwProg{Fn}{Function}{}{}
  \footnotesize
  \KwIn{$F_i(V, E, L)$, $Q=\{q_1,\ldots, q_l\}$, $\threshold$, $Q(F_i)$, message $M_i$}
  \KwOut{$Q(F_i \oplus M_i)$ consisting of current $\answerb_u \in \answerset_i$, where $u\in V$}
  init $\Visit_q$, $\Queue_q$ for each query keyword $q\in Q$\\
  \ForEach{\blue{$\answer_u^{b,\mathsf{in}}  \in M_i$}}{ \label{msg:start}
  	\ForEach{\blue{$q\in Q$ \tnormal{and} $\answerb_u[q] > \answer_u^{b,\mathsf{in}}[q]$}}{
  		\blue{$\answerb_{u}[q] = \answer_u^{b,\mathsf{in}}[q]$} \\
  		\blue{$\Queue_q.\mathsf{push}(\langle u, \answerb_u[q] \rangle)$} \label{msg:end}
  	}
  }
  $\mathsf{BackwardExpand}(\Queue)$ \tcp*[h]{$\Queue=\{\Queue_q | q\in Q\}$}\\
 
 \etitle{Message segment:} \blue{$M_i = $\{$\answerb_u | u\in F_i.I$\}}
  \end{algorithm}

\etitle{(2) Incremental computation} ($\IncEval$) for $\bkws$ (Algo.~\ref{algo:inckws}). Upon receiving messages $M_i$, $\IncEval$ iteratively computes the partial matches, $\answerb_u$, on $F_i$ with the updates (messages) $M_i$. Specifically, if the distance between $u$ and $q\in Q$, \ie  $\distance = \answerb_u[q]$, is refined by using message $M_i$, $u$ is pushed into the priority queue $\Queue_q$ with the refined  distance. Then, $\IncEval$ propagates the distance refinement to the affected area by $\bkws$. Worker $P_i$ notifies the coordinator $P_0$ once the local upper bound $\prune_i$ is refined by invoking $\Notify(i,\prune_i)$. At the end of $\IncEval$, the messages are grouped into $M_i$ at the incoming portal nodes and sent to the relevant workers, similar to $\PEval$.

\stitle{Completeness.} We assume that $\DKWS$ takes $R$ supersteps to finish the evaluation of a keyword query. We denote the vertices that have been visited on $F_i$ at the $s$-th ($s\leq R$)  superstep for a query keyword $q_j\in Q$ by $\Visit_{q_j,i}^s$.  We denote the union set of all the visited vertices by $\VU$. Hence, $\VU = \bigcup_{i\in [1,m], j\in [1,l], s\in [1,R]}\Visit_{q_j,i}^s$, where $m$ is the number of workers and $l=|Q|$. We have the following proposition.

\begin{proposition}\label{prop:d_completed}
    Suppose the top-$k$ matches of a keyword query is $\answerset$ and all the visited vertices $\VU$, the following hold:\\
    \;\;\;\;\;\;(1) $\forall u\not\in \VU$, $\answer_u\not\in \answerset$; and (2) $\forall \answer_u\in \answerset$, $u\in \VU$.
\end{proposition}

\begin{IEEEproof}
    The proof is presented in Appx.~\ref{appendix-completeness} of \cite{techreport}.
\end{IEEEproof}

\begin{figure}
\includegraphics[width=1.05\linewidth]{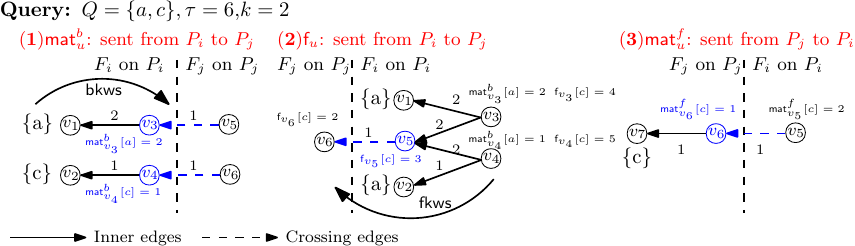}
    \caption{\TKDE{Message exchange during query processing ($\answer_u^b$ (resp. $\answer_u^f$): keep track the shortest distance between $u$ and a query keyword by $\bkws$ (resp. $\fkws$); and $\MF_u$: the longest distance needed to be forward expanded starting from $u$)}}
	\label{fig:dkwsmsg}
\end{figure}

\begin{example}
    As shown in Fig.~\ref{fig:dkwsmsg}.(1), when $\bkws{}$ expands from $v_1$ to $v_3$, $\answerb_{v_3}[a]=2$ is sent from $F_i$ to $F_j$ since $v_3\in F_i.I$. Similarly, $\answerb_{v_4}[c]=1$ is sent from $F_i$ to $F_j$. $\IncEval{}$ of $\bkws$ is invoked in $F_j$ to search for matches.
\end{example}

%\subsection{$\PIE$ for forward search}\label{sec:pie-fkws}
\stitle{\ref{sec:dkwspine}.3 $\PI$  for $\fkws$}

\etitle{Message declaration.} $\DKWS$ declares a variable $\answerf_u$, where $\answerf_u$ is a map, $\answerf_u[q] = \langle \vsf, \distance \rangle$, where $\distance$ is the shortest distance between vertex $u$ and a query keyword $q\in L(\vsf) \cap Q$, \ie $\distance = \dist(u,q)$. $\answerf_u$ is to keep track of the updates to $u$ during the forward expansion. $\DKWS$ also declares a variable $\MF_u$  for each vertex $u$ to indicate the distances of the longest forward expansion of retrieving missing keywords starting from $u$. Formally, $\MF_u$ is a map $(q,\distance)$, where $q\in Q$ is a query keyword and $\distance$ is the longest distance needed to be forward expanded starting from $u$ to retrieve the query keyword $q$. 

%%%%%%%%Answer Expanding
\begin{algorithm}[!t]
  \caption{{$\PEval$} for $\fkws$}\label{algo:pevalfkws}
  \footnotesize
  \SetKwProg{Fn}{Function}{}{}
  \KwIn{$F_i(V, E, L)$, $Q=\{q_1,\ldots, q_l\}$, $\threshold$}
  \KwOut{$Q(F_i)$ consisting of current $\answer_u \in \answerset_i$ for all $u\in V$}
  load the indexes $\PRADS{}$ and $\KPADS$\label{algo:fkws-peval-load-index}\\
  \blue{maintain the vertices to be forward expanded  in $\VM$, \ie roots of partial matches\\
  for $u\in \VM$, init a forward match $\answerf_u$ and a forward distance $\MF_u$ \\  
  }
  \ForEach{$u\in \VM$}{ \label{algo:pevalfkws_forward_start}
  $\mathsf{forwardExpand}(u,Q,\prune_i, \answerset_i)$ \label{algo:pevalfkws_forward_end_f}
  }
\Fn{$\mathsf{forwardExpand}(u,Q,\prune_i, \answerset_i)$}{
	\ForEach{$v$ \textnormal{in the Dijkstra's traversal of $u$}}{
		\If{\tnormal{not} $\mathsf{isCandidate}(u)$ \tnormal{or all the} $q\in\MF_u$ \tnormal{are found}}{
			break 
		}
		
	\ForEach{$q\in \MF_u$}{
		\If(\tcp*[h]{missing keyword is found}){$q\in L(v)$}{ \label{algo:visitstart} 
			$\answerf_u[q] = \dist(u,v)$\\
			marks that $q\in \MF_u$ is found \\ 
			\eat{propagate the refinement of $\answerf_v[q]$ by $\invert_v$ backwardly \\}
		}\ElseIf{$\dist(u,v) > \answerf_u[q]$ \tnormal{\textbf{or}} $\dist(u,v) + \Score(u) > \prune_i$}{
			marks that $q\in \MF_u$ is found \label{algo:visitend} \tcp*[h]{Prop.\ref{prop:pruning} Condition \ref{condition:c3} is met
			} \\
		}\ElseIf{$\dist(u,v) +  \answer_v[q] < \tau$ \label{algo:found}}{
			\tcp*[h]{Refine $\answerf_u$ by a found match $\answer_v$}\\
			$\answerf_u[q] = \min\{\answerf_u[q], \dist(u,v)+ \answer_v[q]\}$ \label{algo:refine} \\ 
		}\ElseIf{$v\in F_i.O$}{
			\tcp*[h]{foward expansion on other fragments}\\
			$\MF_v[q] = \max\{\MF_v[q], \MF_u[q] - \dist(u,v)\}$ \\ 
			
		}
	}
}
	\If{$\answer_u$ is a complete match \tnormal{\textbf{and}} $\Score(u) < \prune_i$}{
			$\answerset_i.\mathsf{push}(\langle u, \answer_u\rangle)$ \\
			$\prune_i = \Score(\answerset_i.\mathsf{top}().\usf)$ \\ 
			\highlight{$\Notify$($i$, $\prune_i$)} \label{algo:pevalfkws_forward_end}
		}
}
\Fn{$\mathsf{isCandidate}(u)$}{
	$\Ld(u,F_i.O)\leftarrow$ estimate the lower bound between $u$ and $F_i.O$ by Eq~\ref{equ:lowerbound}\label{algo:checkup} \\
	\ForEach{$q\in \MF_u$}{
		$\Ld(u,q)\leftarrow$ estimate the lower bound between $u$ and $q$ by Eq~\ref{equ:lowerbound} \label{algo:checkuq}\\
		\If{$\Ld(u,q) > \MF_u[q]$ \textnormal{\textbf{and}} $\Ld(u,F_i.O)> \MF_u[q]$}{
			\Return $\mathsf{False}$\\
		}
	}
	\Return $\mathsf{True}$
}
 
 \etitle{Message segment:} \blue{$M^1_{i} = \{\answerf_u | u\in F_i.I\}$ and $M^2_{i} = \{\MF_u | u\in F_i.O\}$}\label{algo:peval:fkws:end}
 
\end{algorithm}

\etitle{(1) Partial evaluation} ($\PEval$) for $\fkws$ (Algo.~\ref{algo:pevalfkws}). $\fkws$ mainly conducts the forward expansion to complete the partial matches. Lines~2-3 are the initialization of the vertices for the expansion and match variables. In the forward expansion starting from $u$, if any condition(s) in Prop.~\ref{prop:pruning} is met, the expansion is terminated (Lines~\ref{algo:visitstart}-\ref{algo:visitend}). Suppose $u$ is expanded to vertex $v$ and the missing keyword $q$ is found in $\answer_v = \answerb_v\cup \answerf_v$, $\answerf_u[q]$ is refined (Lines~\ref{algo:found}-\ref{algo:refine}). If $v\in F_i.O$, the remaining distance of the forward expansion to retrieve the query keyword $q$ on other fragments is stored in $\MF_v[q]$ (Lines~\ref{algo:pevalfkws_forward_end}). 

At the end of $\PEval$ (Line~\ref{algo:peval:fkws:end}), messages $\answerf_u$ (resp. $\MF_u$) are grouped into $M_i^1$ (resp. $M_i^2$) in worker $P_i$. $M_{i,j}\in M_i$ is sent to worker $P_j$. Formally, $M^1_{i,j}=\{\answer_u | u\in F_i.I \cap F_j.O\}$ and $M^2_{i,j}=\{\MF_u | u \in F_i.O \cap F_j.I\}$ are sent from worker $P_i$ to worker $P_j$. Moreover, $\PEval$ sends the refined $\prune_i$ to the coordinator and notifies it to refine the global upper bound $\prune$ once the local upper bound is refined.

\etitle{(2) Incremental computation} ($\IncEval$) for $\fkws$ is derived with the following two modifications.

\etitle{(2.1) Refinement propagation.} Firstly, $P_i$ receives the partial matches, $\answerf_u$ in previous supersteps from other fragments through the portal nodes. If a shorter path between $u$ and $q$ is found crossing multiple fragments, the forward match $\answerf_u[q]$ is refined. $\IncEval$ propagates the distance refinement to the ancestor vertices.

\etitle{(2.2) Incremental forward expansion.} Secondly, upon receiving some forward expansion requests from other fragments, worker $P_j$ further forward expands to retrieve missing keywords on $F_j$ through the incoming portal nodes, $F_j.I$. Specifically, if $\MF_u^{\mathsf{in}}\in M_j^2$ is received and $u\not\in \VM$, $u$ is added to $\VM$. Since search requests come from different fragments, $\MF_u$ keeps the largest one for each keyword. If $u$ is forward expanded in previous iterations for query keyword $q$ or $\MF_u^{\mathsf{in}}[q]$ is smaller than $\MF_u[q]$, $\MF_u^{\mathsf{in}}[q]$ is skipped.

At the end of $\IncEval$, the partial matches found by forward expansions are grouped into $M_i^1$ and the remaining forward expansion requests are grouped into $M_i^2$, respectively, for fragment $F_i$ and sent to the corresponding fragments, which is the same as that of $\PEval$.

\begin{example}
    As shown in Fig.~\ref{fig:dkwsmsg}.(2), when $\fkws{}$ expands from $v_4$ to $v_5$ to search for the missing keyword $c$, $\MF_{v_5}[c]=3$ is sent from $F_i$ to $F_j$ since $v_5\in F_i.O$. $\IncEval{}$ of $\fkws$ is invoked in $F_j$ to search on $F_j$ forwardly.  Once the keyword $c$ is retrieved in $F_j$ as recorded in $\answerf_{v_6}[c]=1$ (shown in Fig.~\ref{fig:dkwsmsg}.(3)), $\answerf_{v_6}[c]=1$ is sent to $F_i$ via the portal node $v_6\in F_j.I$.
\end{example}

\stitle{\ref{sec:dkwspine}.4 Preemptive execution of $\IncEval$s in $\PINE$}

Even if the complexity of $\bkws$ (analyzed in Sec.~\ref{sec:bkws}) is smaller than that of $\fkws$ (analyzed in Sec.~\ref{sec:fkws}), running $\bkws$ first and then $\fkws$ may not exhibit the best query performance in practice. In particular, we provide three insights: (a) $\bkws$ increases the size of $\VM$ but $k$ of top-$k$ is fixed. Relatively more vertices of $\VM$ may not be  backward expanded to final matches; (b) some early messages from $\bkws$ may not effectively refine into tight upper bounds for $\fkws$; and (c) some workers running $\bkws$ can be stragglers, as $\fkws$ is blocked by them, \ie it cannot yet start. Hence, $\PINE$ \TKDE{provides a lightweight selector (as shown in Fig.~\ref{fig:workflow}) and} allows the computation of $\bkws$ and $\fkws$ in a preemptive manner. At runtime, each worker $P_i$ determines to {\em execute either $\bkws$ or $\fkws$, independently}. Each of them maintains a set of {\em status parameters} to estimate the performance improvement of executing either $\bkws$ or $\fkws$.

\etitle{Message buffers.} Each worker $P_i$ maintains two message buffers $\Buffer_i^b$ and $\Buffer_i^f$ to keep track of backward and forward messages from other workers. The more messages are accumulated in $\Buffer_i^b$ (resp. $\Buffer_i^f$), the earlier $P_i$ should start $\bkws$ (resp. $\fkws$) computation, and vice versa. 

\etitle{Expansion distance.} Denote $\answer_u^{b,\mathsf{in}}$ as a message generated by $\bkws$ and maintained in $\Buffer_i^b$. If $\answerb_u[q]$ is larger than $\answer_u^{b,\mathsf{in}}[q]$, $\DKWS$ needs to backward expand starting from $u$. We call $d_q^b =\min\{ \prune - \answer_u^{b,\mathsf{in}}[q], \tau\}$ {\em backward expansion distance starting from $u$ for query keyword $q$}.  Without prior statistics, if $d_q^b$ is larger, the backward expansion is more costly and the messages are less likely to finally yield one of the top-$k$ matches. Worker $P_i$ may stop expanding $\answerb_u[q]$ by postponing the execution of $\bkws$, but start $\fkws$ to prune some unyielding messages. Similarly, we define $d_q^f =\min\{\MF_u^{\mathsf{in}}[q], \tau\}$, the {\em forward expansion distance}.  

\etitle{Staleness indicators.} Inspired by the complexities of $\bkws$ and $\fkws$, we propose the staleness indicators of the accumulated backward messages and forward messages for each worker $P_i$, denoted by $\SI_i^b$ and $\SI_i^f$. $\SI_i^b$ and $\SI_i^f$ are formally defined below:

\noindent\begin{minipage}{.5\linewidth}
    \begin{equation}
        \begin{footnotesize}
          \SI_i^b =
          \begin{cases}
              +\infty, & \text{if  $\Buffer_i^b$ is empty} \\
              \frac{\mathlarger{\sum}_{u\in F_i.O}\mathlarger{\sum}_{q\in Q}d^b_q}{|\Buffer^b_i|}, & \text{otherwise}
          \end{cases}
        \end{footnotesize}
      \end{equation}
      \end{minipage}%
    \begin{minipage}{.5\linewidth}
        \begin{equation}
            \begin{footnotesize}
              \SI_i^f =
              \begin{cases}
                  +\infty, & \text{if  $\Buffer_i^f$ is empty} \\
                  \frac{\mathlarger{\sum}_{u\in F_i.I}\mathlarger{\sum}_{q\in Q}^{|Q|}d^f_q}{|\Buffer^f_i|}, & \text{otherwise}
              \end{cases}
            \end{footnotesize}
          \end{equation}
    \end{minipage}
    
\noindent where $d^b_j$ (resp. $d^f_j$) is the average backward (resp. forward) searching distance for query keywords and $|\Buffer^b_i|$ (resp. $|\Buffer^f_i|$) is the size of backward (resp. forward) messages buffer.

If $\SI_i^b < \SI_i^f$, worker $P_i$ conducts $\bkws$. Otherwise, worker $P_i$ conducts $\fkws$. $\PINE$ is able to simulate $\PIE$ by enforcing $\SI_i^b$ to $+\infty$ at the even supersteps and $\SI_i^b$ to $+\infty$ at the odd supersteps.

\begin{figure}[tb]
	\begin{center}
		\includegraphics[width=0.5\textwidth]{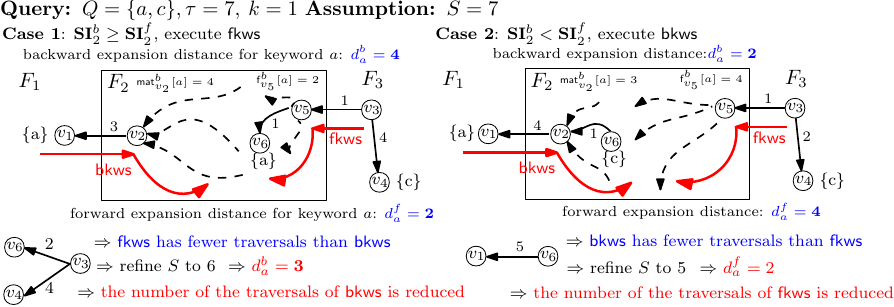}
	\end{center}
 \vspace{-3mm}
	\caption{\TKDERF{Illustration of the preemptive execution}}\label{fig:Preemptive}
\end{figure}

\TKDERF{
\begin{example}
Consider the two cases in Fig.~\ref{fig:Preemptive}. In \textbf{Case 1}, when the backward expansion from $v_1$ is performed via $v_2$, the backward expansion distance starting from $v_2$ is $4$ and $\SI^b_2=4$. When the forward expansion from $v_3$ is performed via $v_5$, the forward expansion distance from $v_5$ is $2$ and $\SI^f_2=2$. Since $\SI^f_2 < \SI^b_2$, $\fkws$ has a higher priority to be executed. We can observe that an answer rooted at $v_3$ is returned, and the upper bound $\prune$ is refined to $6$. Consequently, $d_a^b$ is refined to $3$, and fewer traversals are required in the next iteration. Similarly, in \textbf{Case 2}, $\bkws$ has a higher priority and produces the matches earlier, which reduces the number of traversals of $\fkws$ after the upper bound $\prune$ is refined.
\end{example}
}

\stitle{\ref{sec:dkwspine}.5 $\Assemble$ for $\SKWS$}

$\DKWS$\eat{differs from ~\cite{DBLP:journals/tkde/YuanLCYWS17} in the following aspects. The coordinator of $\DKWS$ does not collect the local candidate matches at runtime. Instead, it} only collects the local top-$k$ matches to yield the global top-$k$ matches $\answerset$ by selecting the top-$k$ matches from $\bigcup_{i\in [1,m]} \answerset_i$ after the executions of $\IncEval$s of $\bkws$ and $\fkws$ have terminated. \jiaxin{Hence, the cost of collecting local matches from all the workers is bounded by $O(km)$.}

\eat{ (Step \circled{2} of Fig.~\ref{fig:notifypush})}

%\revisethesis{
    \begin{example}
        Consider the graph and query in Fig.~\ref{fig:dkws-example}. Local matches of $F_1$ are rooted at $v_{15}$ and $v_{17}$ and $\Score(v_{15}) = 4$ and $\Score(v_{17})=3$. Similarly, we have two local matches on $F_2$ with $\Score(v_{11}) = 4$ and $\Score(v_{6})=4$ and two local matches on $F_3$ with $\Score(v_{4}) = 2$ and $\Score(v_{5})=5$. Hence, the coordinator collects all the $6$ local matches. The matches rooted at $v_4$ and $v_{17}$ are returned since they are the top-$2$ among the $6$ matches.
    \end{example}
%}

\subsection{Analysis of $\SKWS$ on $\DKWS$}\label{sec:analysis}

In this section, we present an analysis of the correctness of $\PINE$. Following \cite{valiant1990general} and \cite{fan2020adaptive}, a parallel model $\model_1$ can be optimally simulated by another one $\model_2$ if there exists a compilation algorithm that transforms any program \jiaxin{on} $\model_1$ with a constant cost $C$ to a program \jiaxin{on} $\model_2$ with a cost $O(C)$.

\eat{$\DKWS$ is a variation of $\grape$. However, $\DKWS$ is finer-grained and more flexible for users. There have been abundant algorithms that consist of a set of distinguishable subtasks. $\DKWS$ parallelizes such algorithms automatically and efficiently.}

\spacecompress
\begin{proposition}\label{prop:simulate}
    A $\PINE$ algorithm  can be compiled into a $\PIE$ algorithm with a cost $O(C)$. 
\end{proposition}
\spacecompress
\spacecompress
\begin{IEEEproof}
    Any $\PINE$ algorithms developed on $\DKWS$ can be compiled into a $\PIE$ algorithm. Given a $\PINE$ algorithm $\Algo$ that consists of $n$ $\PEval$s (denoted by $\Psf_i$, where $i\in [1,n]$), $n$ $\IncEval$s (denoted by $\I_i$, where $i\in [1,n]$), and one $\Assemble$ (denoted by $\E$). $\Algo$ is compiled into $\grape$ by a $\PIE$ algorithm as follows. (a) $\PEval$ of $\grape$ runs $\Psf_i$s sequentially over the workers. The messages are exchanged by $\PEval$ after $\Psf_n$ is executed. (b) $\IncEval$ of $\grape$ introduces a selection control mechanism by a switch statement. $\grape$ plugs $\I_i$ into the $i$-th branch of the switch statement. The control flow of $\IncEval$ execution is determined by staleness indicators provided by users. The messages are exchanged at the end of each round of $\IncEval$. (c)~$\Assemble$ of $\grape$ is identical to $\E$. %We show that the $\PIE$ algorithm incurs no more cost than $\Algo$.
\end{IEEEproof}

\spacecompress
Due to Prop.~\ref{prop:simulate}, $\DKWS$ inherits all properties of $\grape$ (Theorem $1$ of~\cite{fan2017parallel}), including convergence and correctness theorems. 

\begin{theorem}\label{theorem:pine}
  The general form of a $\PINE$ algorithm consists of the following:
  \begin{enumerate}
  \item $n$ $\PEval$s (denoted by $\Psf_i$, where $i\in [1,n]$), 
  \item $n$ $\IncEval$s (denoted by $\I_i$, where $i\in [1,n]$), and
  \item one $\Assemble$ (denoted by $\E$), and any partition strategy $\Partition$.
  \end{enumerate}
  The $\PINE$ algorithm on $\DKWS$ terminates correctly if
  \begin{enumerate}[label={(\alph*)}]
  \item $\I_i$ satisfies the monotonic condition,\footnote{There exists a partial order on the variables attached on the vertices such that $\IncEval$ updates the variables in the partial order~\cite{fan2017parallel}.} for all $i\in [1,n]$; and
  \item $\Psf_i$, $\I_i$ and $\E$ are correct \wrt $\Partition$.\footnote{$\Psf_i$ is correct if it returns correct answer on an input graph $G$ for any queries. $\I_i$ is correct if it returns correct answer on an input graph $G$ and a set of messages for any queries. $\E$ is correct if it yields the answer on the input graph $G$ by assembling all the local matches.} 
  \end{enumerate}
%    , then $\DKWS$ with $\Psf_i$, $\I_i$ and $\E$ guarantee to terminate correctly.
\end{theorem}

\spacecompress
\begin{IEEEproof}
    The proof is presented in \ref{appendix-pine} of \cite{techreport}.
\end{IEEEproof}
\spacecompress

The correctness of $\SKWS$ implemented using $\PINE$ is assured by the correctness of $\bkws$ and $\fkws$ (Sec.~\ref{sec:bkws} and \ref{sec:fkws}) and Theorem~\ref{theorem:pine}.

\noindent\vldbj{\stitle{Complexities.} The time complexity of $\bkws$ (resp. $\fkws$) is $O(|Q|(|E| + |V|\log |V|))$ (resp. $O(|\VM|(|E| + |V|\log |V|))$). The space complexity of $\bkws$ and $\fkws$ is bounded by $O(|Q||V|)$. The size of $\mathsf{PADS}(u)$ is bounded by $O(\log |V|)$. Hence, the overall index size of $\mathsf{PADS}$ is bounded by $O(V\log |V|)$ (cf.~\cite{Jiang2020PPKWSAE}).}

\section{EXPERIMENTAL STUDY}\label{sec:exp}
We experimentally evaluate (1) efficiency, (2) performance under different settings, and (3) communication costs on massive graphs with competitors ~\cite{qin2014scalable} and ~\cite{fan2017incremental}.

\subsection{Experimental setup} 
\stitle{Software and hardware.} Our experiments were run on a cluster with eight machines. Each machine had one Xeon X5650 CPU, 128GB memory and was running CentOS 7.4. The implementation was made memory-resident. \revise{We used METIS~\cite{karypis1995metis} as the graph partition strategy.}

\stitle{Algorithms.} We implemented all algorithms in C++. The settings followed ~\cite{qin2014scalable} and ~\cite{fan2017incremental} whenever appropriate. Our implementation of $\PINE$ was done by modifying the $\PIE$ model running on the platform of $\grape$~\cite{fan2017parallel}. We used the following implementations for algorithms.

\begin{enumerate}[wide, labelwidth=!, labelindent=0pt]
	\itemsep0em 
    \item \etitle{\DKWSBF.} We implemented $\SKWS$ using the $\PIE$ programming model (detailed in Sec.~\ref{sec:dkws}). 
    \item \etitle{\DKWSPADS.} \revise{We applied $\PRADS$ and $\KPADS$ to \DKWSBF{} for deriving a lower bound between a vertex and a query keyword for pruning the forward expansion as proposed (detailed in Sec.~\ref{sec:skws}).} 
    \item \etitle{\DKWSNP.} We applied $\np$ paradigm to \DKWSPADS.
    \item \etitle{\DKWSPINE.} We applied $\PINE$ model to \DKWSNP.
    \item  \etitle{$\BaselineDKWS$.} \TKDERF{We implemented the distributed algorithms proposed in ~\cite{qin2014scalable} and \cite{fan2017incremental}, both of which share the same keyword semantics as ours. These were established on $\grape$\cite{fan2017parallel}, serving as our baseline algorithms.} We did not compare $\DKWS$ with \cite{DBLP:journals/tkde/YuanLCYWS17} since their algorithm (a) returns a set of approximate matches, and (b) proposes a different subtree semantic.
    \item \revise{\etitle{\SKWSC.}  \SKWSC~\cite{kacholia2005bidirectial} is the only sequential algorithm we could run on a single machine. In particular, \SKWSC{} does not require massive indexes. \SKWSC{} is widely used in the experimental comparison of existing works, such as~\cite{he2007blinks,yang2019efficient}.}
\end{enumerate}

\begin{table}[!t]
%\setlength{\abovecaptionskip}{10pt}
%\setlength{\belowcaptionskip}{10pt}
%\captionsetup{font={scriptsize}}
% \setlength{\tabcolsep}{0.3em}
\caption{Statistics of real-world datasets}\label{table:Statistics}
\centering
\scriptsize{}
\begin{tabular}{|c|c|c|c|}
  \hline
  {\bf Datasets} & {\bf $|V|$} & {\bf $|E|$}  & avg. \# of keywords per node \\
  \hline
  YAGO3 & 2,635,317 & 5,260,573 & 3.79  \\
  \hline
  DBpedia & 5,795,123 & 15,752,299 & 3.72  \\
  \hline
  DBLP & 2,221,139 & 5,432,667  & 10  \\
  \hline
  WebUK & 133,633,040  &  5,507,679,822  & 1  \\
  \hline
\end{tabular}
\end{table}

\stitle{Datasets.} We used four popular real-world graphs:  (a) YAGO3~\cite{mahdisoltani2014yago3}, a large knowledge base with 2.6 million entities and 5.26 million factors; (b) WebUK~\cite{webuk}, a large Web graph with 106 million nodes and 3.7 billion edges; (c) DBLP~\cite{dblp} is a social network with 2.2 million authors and 5.4 million collaboration relationships; and (d) DBpedia~\cite{dbpedia} is a knowledge base with 5.8 million entities and 15.7 million factors. \revise{These datasets are widely used in previous keyword search works such as ~\cite{he2007blinks,kargar2011keyword,efficient2020kargar,DBLP:conf/www/Shi0K20} or used to test the scalability of distributed graph evaluation systems, such as ~\cite{fan2017parallel,yan2014blogel}.}

\stitle{Queries.} We followed~\cite{DBLP:journals/tkde/YuanLCYWS17} to generate the queries by varying the number of query keywords $|Q|$. The number ranged from $2$ to $6$. The average query time is stable when the number of queries is $50$. Hence, we generated $50$ random synthetic keyword queries for each query size in our experiments and reported the average evaluation time.

\stitle{Default settings.} We fixed $k=10$, the number of query keywords $|Q|$ to $4$, the number of workers to $8$, and the $\tau$ to $3$. Each worker was assigned one fragment.  Unless specified otherwise, we conducted experiments with default values of the parameters and varied values of a specific parameter.

\subsection{Experimental results}

\begin{figure*}[tp]
  %\vspace{-0.2cm}
%   \hspace*{-10mm}
\begin{center}
\includegraphics[width=0.75\textwidth]{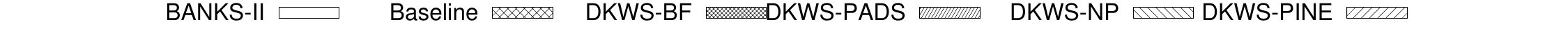}
\end{center}
		\begin{minipage}{\linewidth}	
			\begin{tabular}{cc}
				% 1.1
			\begin{minipage}[t]{0.2\textwidth}
		   	    \includegraphics[width=\textwidth]{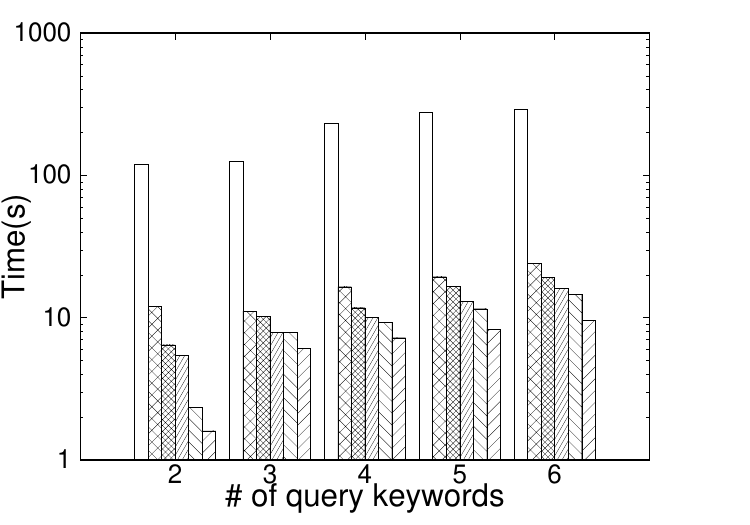}
          \vspace{-1.8em}
				\subcaption{\ssize{Vary $|Q|$ (YAGO3)}}
				\label{fig:dkws-baseline-yago}
			\end{minipage}
				% 1.2
				\begin{minipage}[t]{0.2\textwidth}
				\includegraphics[width=\textwidth]{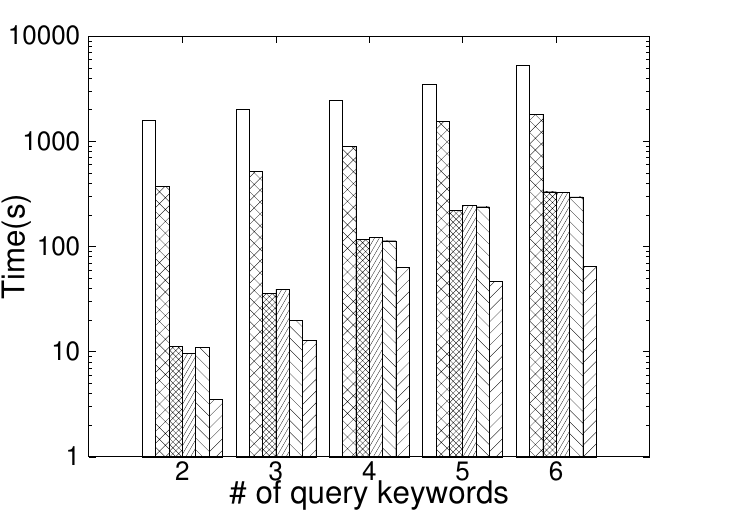}
    \vspace{-1.8em}
				\subcaption{\ssize{Vary $|Q|$ (WebUK)}}
				\label{fig:dkws-baseline-webuk}
				\end{minipage}
				% 1.3
				\begin{minipage}[t]{0.2\textwidth}
					\includegraphics[width=\textwidth]{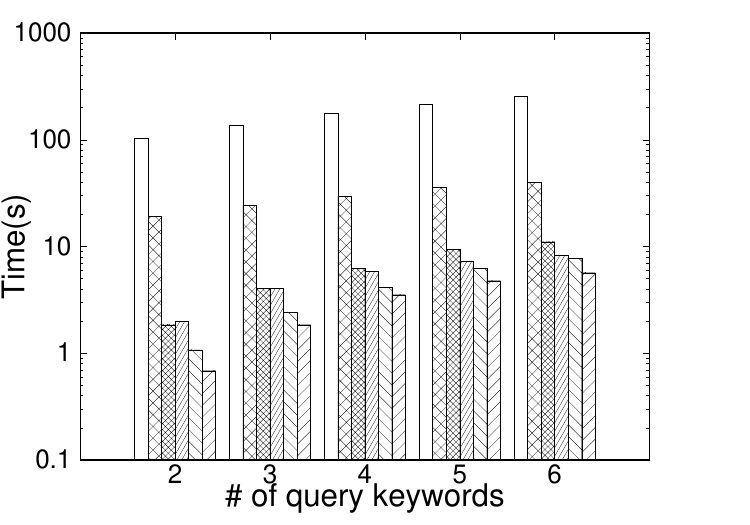}
					\vspace{-1.8em}
					\subcaption{\ssize{Vary $|Q|$ (DBLP)}}
					\label{fig:dkws-baseline-dblp}
					\end{minipage}
					% 1.4
					\begin{minipage}[t]{0.2\textwidth}
					\includegraphics[width=\textwidth]{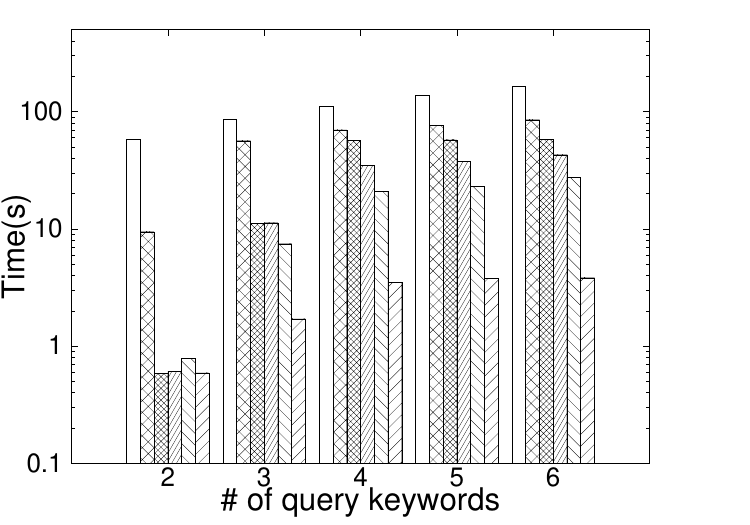}
					\vspace{-1.8em}
					\subcaption{\ssize{Vary $|Q|$ (DBpedia)}}
					\label{fig:dkws-baseline-dbpedia}
					\end{minipage}
		    \begin{minipage}[t]{0.2\textwidth} 
			\includegraphics[width=\textwidth]{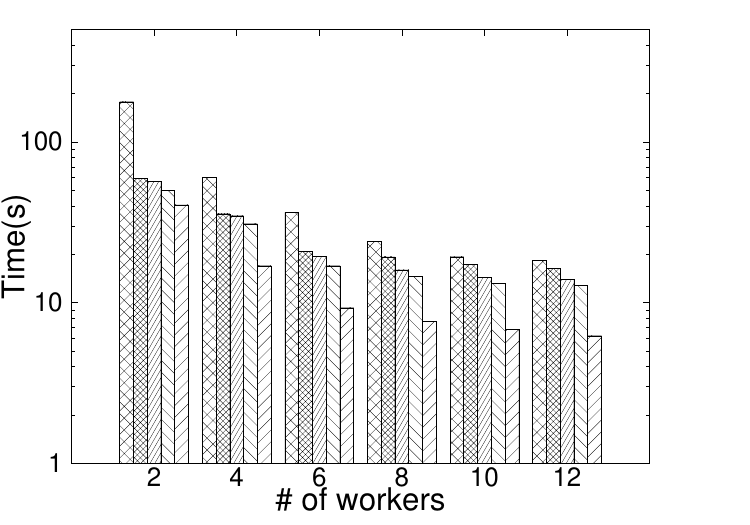}
			\vspace{-1.8em}
			\subcaption{\TKDERF{\ssize{Vary workers (YAGO3)}}}
			\label{fig:dkws-vary-workers-yago-appdix}
			\end{minipage}			
				\end{tabular}

		\begin{tabular}{cc}			
			% 2.1
		
		\begin{minipage}[t]{0.2\textwidth} 
			\includegraphics[width=\textwidth]{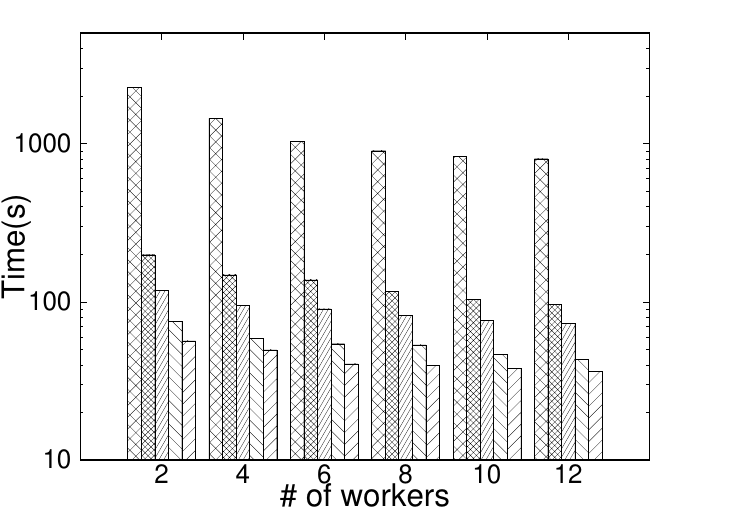}
			\vspace{-1.8em}
			\subcaption{\TKDERF{\ssize{Vary workers (WebUK)}}}
			\label{fig:dkws-vary-workers-webuk}
			\end{minipage}
	 % 2.2
	 \begin{minipage}[t]{0.2\textwidth} 
		\includegraphics[width=\textwidth]{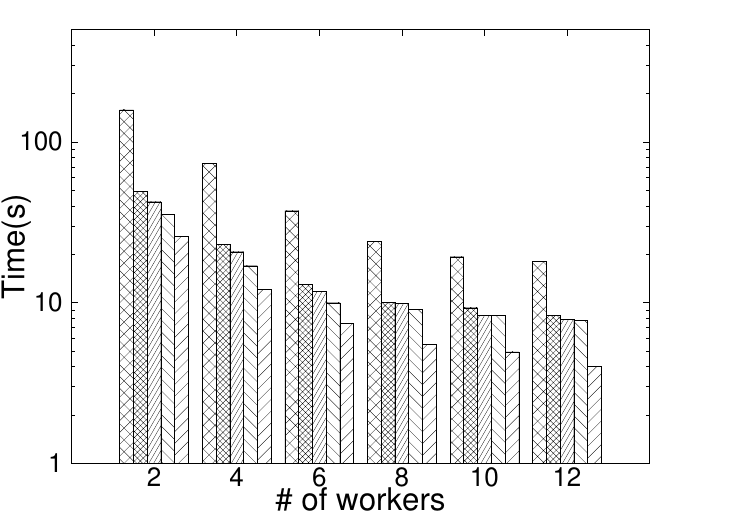}
		\vspace{-1.8em}
		\subcaption{\TKDERF{\ssize{Vary workers (DBLP)}}}
		\label{fig:dkws-vary-workers-dblp}
		\end{minipage}
			% 2.1

	% 2.4
	\begin{minipage}[t]{0.2\textwidth} 
		\includegraphics[width=\textwidth]{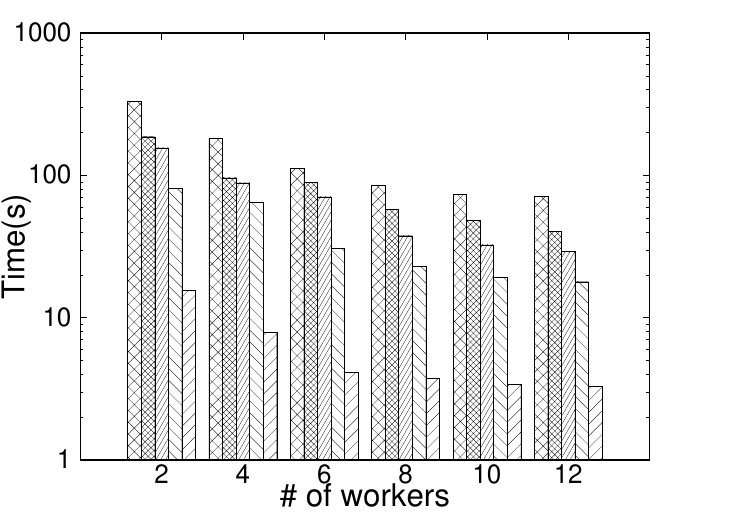}
		\vspace{-1.8em}
		\subcaption{\TKDERF{\ssize{Vary workers (DBpedia)}}}
		\label{fig:dkws-vary-workers-dbpedia-appdix}
		\end{minipage}
% 2.3
\begin{minipage}[t]{0.2\textwidth} 
			\includegraphics[width=\textwidth]{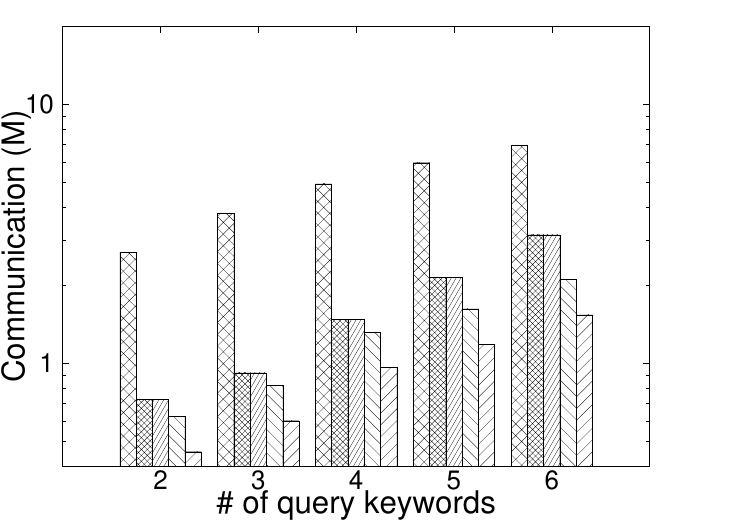}
			\vspace{-1.8em}
			\subcaption{\ssize{Communication (YAGO3)}}
			\label{fig:dkws-baseline-msg-yago}
	\end{minipage}
     \begin{minipage}[t]{0.2\textwidth} 
	    \includegraphics[width=\textwidth]{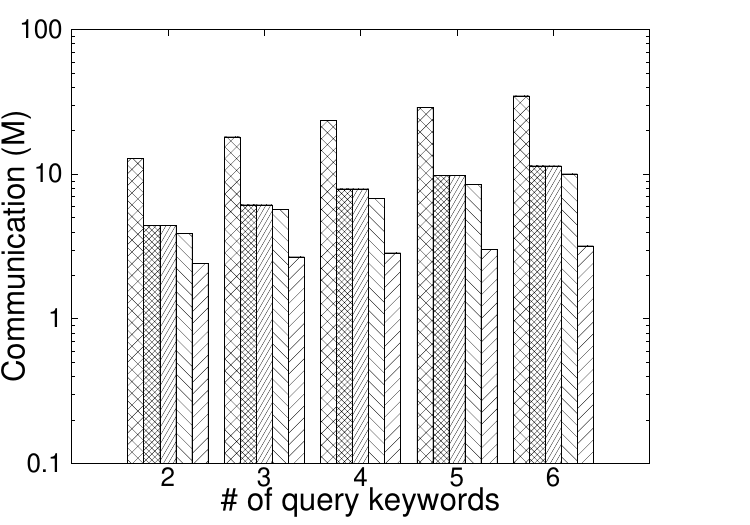}
	    \vspace{-1.8em}
	    \subcaption{\ssize{Communication (WebUK)}}
	    \label{fig:comm-webuk}
    \end{minipage}
		\end{tabular}

\begin{tabular}{cc}
    
% 2.4
    \begin{minipage}[t]{0.2\textwidth} 
	    \includegraphics[width=\textwidth]{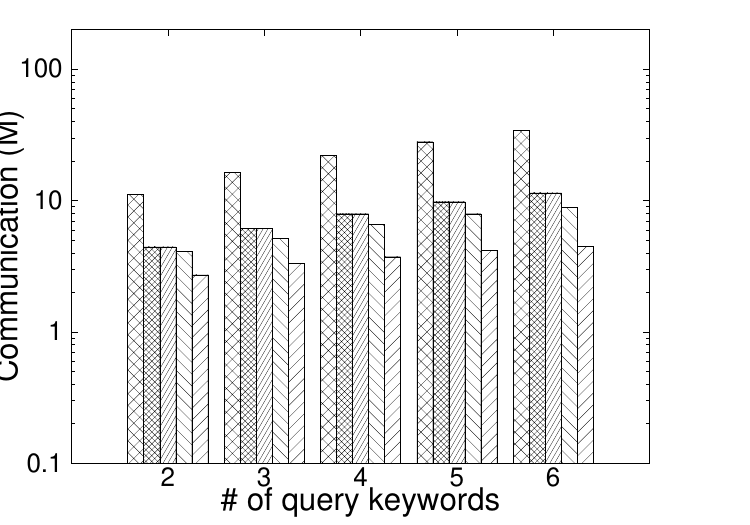}
	    \vspace{-1.8em}
	    \subcaption{\ssize{Communication (DBLP)}}
	    \label{fig:comm-DBLP}
	\end{minipage}
		\begin{minipage}[t]{0.2\textwidth} 
			\includegraphics[width=\textwidth]{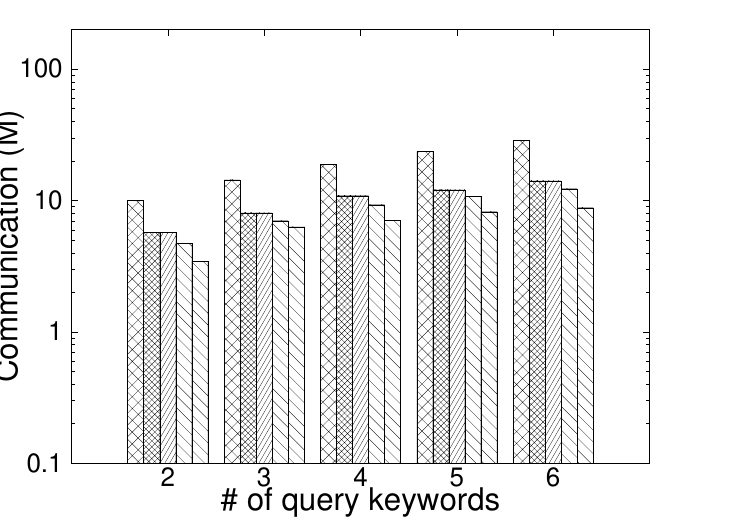}
			\vspace{-1.8em}
			\subcaption{\ssize{Communication (DBpedia)}}
			\label{fig:dkws-baseline-msg-dbpedia}
		\end{minipage}
		\begin{minipage}[t]{0.2\textwidth} 			\includegraphics[width=\textwidth]{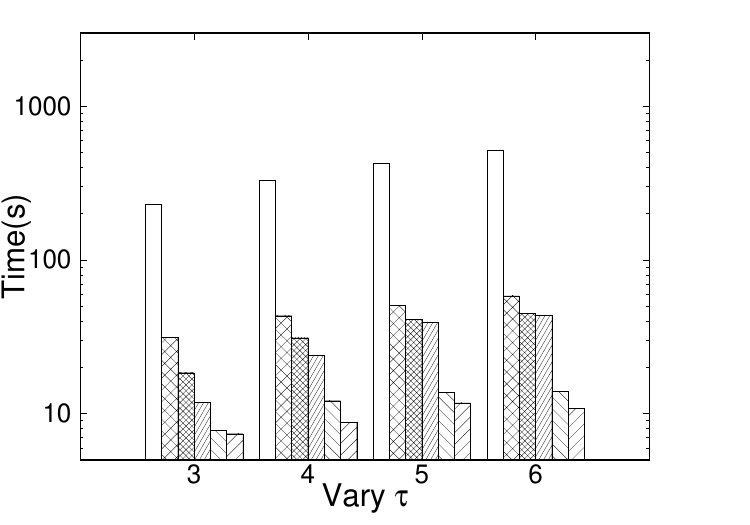}
		\vspace{-1.8em}
				\subcaption{\ssize{Vary $\tau$ (YAGO3)}}
				\label{fig:dkws-vary-tau-yago}
			\end{minipage}
			
			% 3.2
			\begin{minipage}[t]{0.2\textwidth} 
				\includegraphics[width=\textwidth]{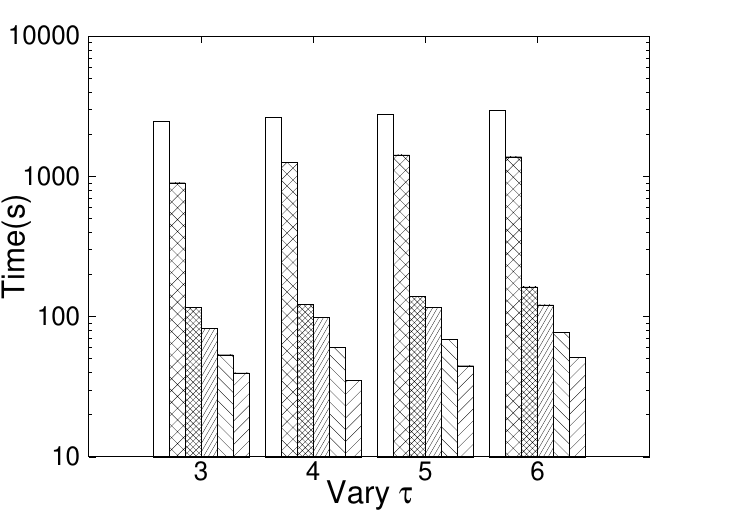}
				\vspace{-1.8em}
				\subcaption{\ssize{Vary $\tau$ (WebUK)}}
				\label{fig:dkws-vary-tau-webuk}
				\end{minipage}
		 % 3.3
		 \begin{minipage}[t]{0.2\textwidth} 
			\includegraphics[width=\textwidth]{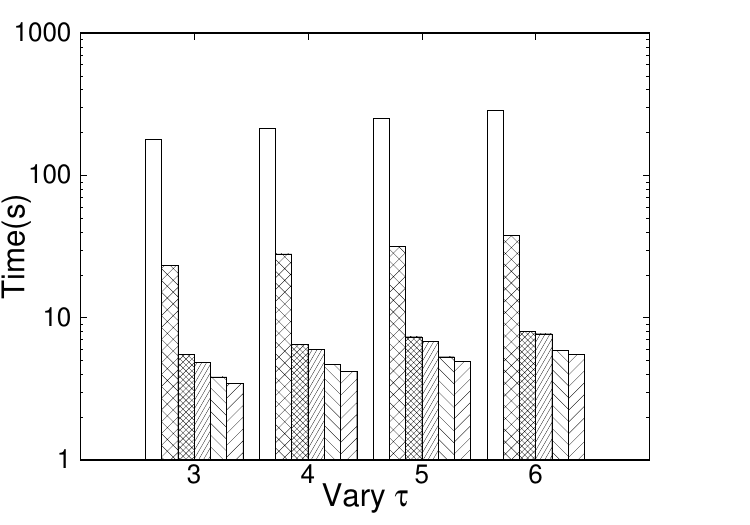}
			\vspace{-1.8em}
			\subcaption{\ssize{Vary $\tau$ (DBLP)}}
			\label{fig:dkws-vary-tau-dblp}
			\end{minipage}
\end{tabular}

		\begin{tabular}{cc}
				% 3.1

		% 3.4
		\begin{minipage}[t]{0.2\textwidth} 
			\includegraphics[width=\textwidth]{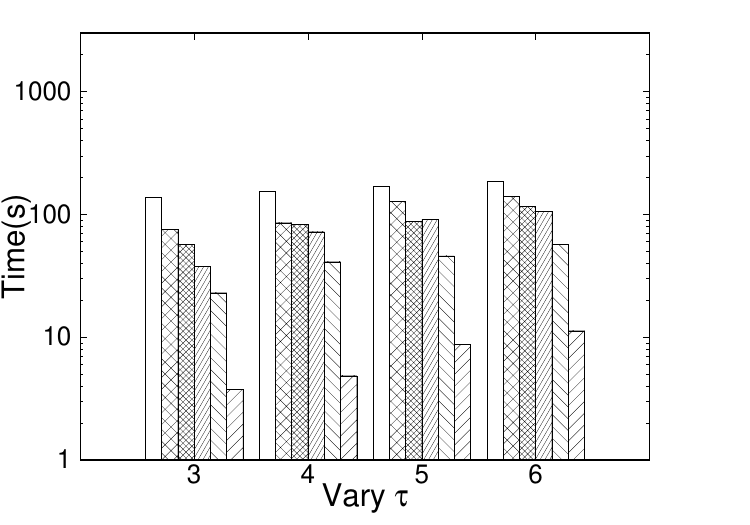}
			\vspace{-1.8em}
			\subcaption{\ssize{Vary $\tau$  (DBpedia)}}
			\label{fig:dkws-vary-tau-dbpedia}
			\end{minipage}
		\begin{minipage}[t]{0.2\textwidth} 
			\includegraphics[width=\textwidth]{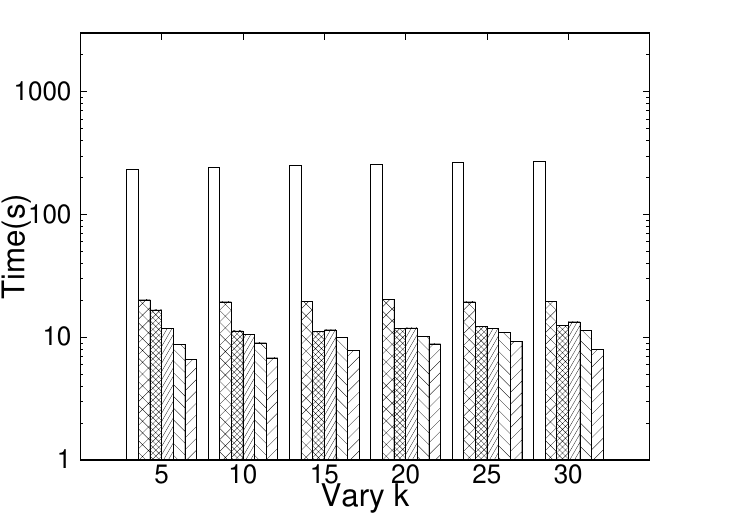}
			\vspace{-1.8em}
			\subcaption{\ssize{Vary $k$ (YAGO3)}}
			\label{fig:dkws-vary-topk-yago-appdix}
			\end{minipage}
             % 4.2
			\begin{minipage}[t]{0.2\textwidth} 
				\includegraphics[width=\textwidth]{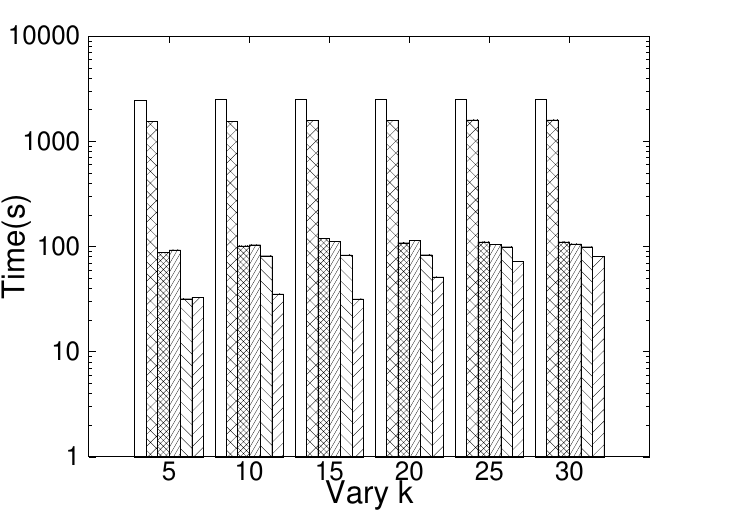}
				\vspace{-1.8em}
				\subcaption{\ssize{Vary $k$ (WebUK)}}
				\label{fig:dkws-vary-topk-webuk}
				\end{minipage}
		 % 4.3
		 	\begin{minipage}[t]{0.2\textwidth} 
			\includegraphics[width=\textwidth]{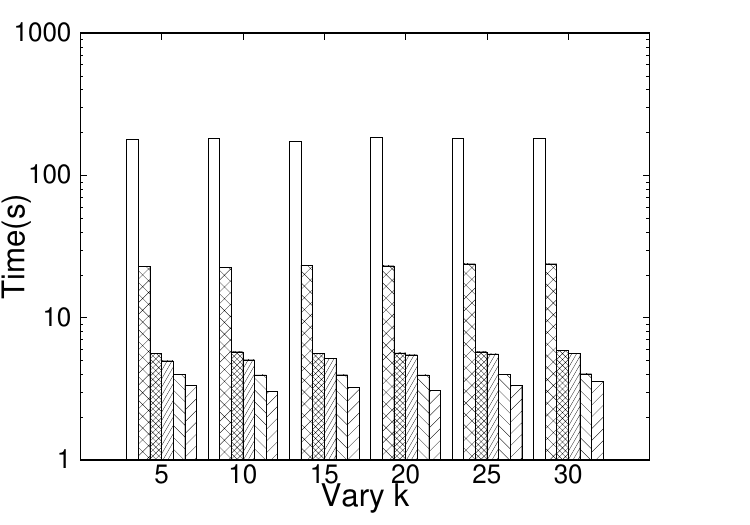}
			\vspace{-1.8em}
			\subcaption{\ssize{Vary $k$ (DBLP)}}
			\label{fig:dkws-vary-topk-dblp}
			\end{minipage}
	% 4.4
	\begin{minipage}[t]{0.2\textwidth} 
		\includegraphics[width=\textwidth]{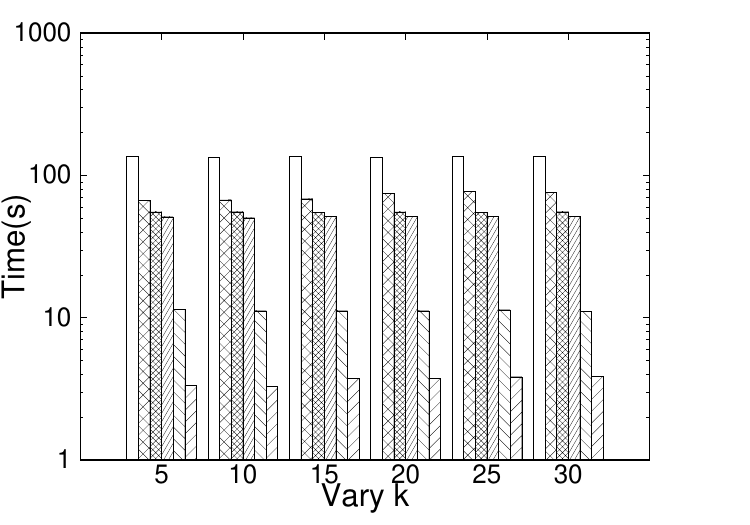}
		\vspace{-1.8em}
		\subcaption{\ssize{Vary $k$  (DBpedia)}}
		\label{fig:dkws-vary-topk-dbpedia-appdix}
		\end{minipage}	
			\end{tabular}

%%%%%%%% Followings are messages. For backup. 
\eat{
		\begin{tabular}{cc}
			% 3.1
		\begin{minipage}[t]{0.23\textwidth} 
			\includegraphics[width=\textwidth]{./figures/dkws-baseline-msg-yago-eps-converted-to.pdf}
			\subcaption{Vary $\#$ of query keywords (YAGO3)}
			\label{fig:dkws-baseline-msg-yago}
		\end{minipage}
		% 3.2
		\begin{minipage}[t]{0.23\textwidth} 
			\includegraphics[width=\textwidth]{./figures/dkws-baseline-msg-webuk-eps-converted-to.pdf}
			\subcaption{\scriptsize Vary $\#$ of query keywords (WebUK)}
			\label{fig:dkws-baseline-msg-webuk}
		\end{minipage}
		% 3.3
		\begin{minipage}[t]{0.23\textwidth} 
			\includegraphics[width=\textwidth]{./figures/dkws-baseline-msg-dblp-eps-converted-to.pdf}
			\subcaption{\scriptsize Vary $\#$ of query keywords (DBLP)}
			\label{fig:dkws-baseline-msg-dblp}
		\end{minipage}
			% 3.4
		\begin{minipage}[t]{0.23\textwidth} 
			\includegraphics[width=\textwidth]{./figures/dkws-baseline-msg-dbpedia-eps-converted-to.pdf}
			\subcaption{\scriptsize Vary $\#$ of query keywords (DBpedia)}
			\label{fig:dkws-baseline-msg-dbpedia}
		\end{minipage}
		\end{tabular}
		
		\begin{tabular}{cc}
			% 4.1 
			\begin{minipage}[t]{0.23\textwidth} 
				\includegraphics[width=\textwidth]{./figures/dkws-vary-workers-msg-yago-eps-converted-to.pdf}
				\subcaption{\scriptsize Vary $\#$ of the workers (YAGO3)}
				\label{fig:dkws-vary-workers-msg-yago}
			\end{minipage}
				  % 4.2
			\begin{minipage}[t]{0.23\textwidth} 
				\includegraphics[width=\textwidth]{./figures/dkws-vary-workers-msg-webuk-eps-converted-to.pdf}
				\subcaption{\scriptsize Vary $\#$ of the workers (WebUK)}
				\label{fig:dkws-vary-workers-msg-webuk}
			\end{minipage}
				  % 4.3
				  \begin{minipage}[t]{0.23\textwidth} 
					\includegraphics[width=\textwidth]{./figures/dkws-vary-workers-msg-dblp-eps-converted-to.pdf}
					\subcaption{\scriptsize Vary $\#$ of the workers (DBLP)}
					\label{fig:dkws-vary-workers-msg-webuk}
				\end{minipage}
				% 4.4
				\begin{minipage}[t]{0.23\textwidth} 
					\includegraphics[width=\textwidth]{./figures/dkws-vary-workers-msg-dbpedia-eps-converted-to.pdf}
					\subcaption{\scriptsize Vary $\#$ of the workers (DBpedia)}	\label{fig:dkws-vary-workers-msg-dbpedia}
				\end{minipage}
		\end{tabular}
}

	\caption{Query performance on the four real-life datasets}
	\label{fig:performance}
		\end{minipage}
	%        \vspace{-1cm}
	\end{figure*}

\stitle{Exp-1: Efficiency.} We firstly evaluated the efficiency of $\DKWS$ by varying the number of query keywords $|Q|$ from $2$ to $6$. \vldbj{All algorithms take longer when $|Q|$ gets larger since the size of search space increases.} The results are shown in Fig.~\ref{fig:performance}\subref{fig:dkws-baseline-yago} to Fig.~\ref{fig:performance}\subref{fig:dkws-baseline-dbpedia}.

\eat{
\todo{extracting the data for straggler}
}

\noindent(a) On YAGO3, \DKWSPADS{} is on average $1.24$ times faster than \DKWSBF{}. The main reason is that most forward expansions are pruned by $\PRADS$. \DKWSNP{} is $2.32$ times faster than $\BaselineDKWS$ as \DKWSNP{} avoids the straggler problem. The slower workers are terminated early by using the global upper bound. \DKWSPINE{} is $3.3$ times faster than $\BaselineDKWS$ since the computing tasks of \DKWSPINE{} are finer-grained, avoiding the straggler problem and tighter bounds are retrieved by taking advantage of both $\bkws$ and $\fkws$.

\noindent(b) On WebUK, \DKWSBF{} is on average $14.6$ times faster than $\BaselineDKWS$. The reason for such a significant speedup is that a tight local upper bound on WebUK is derived early, since WebUK is denser than the other three datasets. Hence, the vertices, which require forward expansion, are few. \DKWSPADS{} (resp. \DKWSNP{} and \DKWSPINE{}) is $17.05$ (resp. $21.45$ and $46.8$) times faster than $\BaselineDKWS$. 

\noindent(c) On DBLP, the query time of \DKWSBF{} is $5.73$ times faster than $\BaselineDKWS$. Since the diameter of DBLP is small, the forward expansion distance is not far. \DKWSPADS{} is $6.12$ times faster than $\BaselineDKWS$. Pruning by $\PRADS$ on DBLP is not as obvious as that on the other three datasets due to the small graph diameter. \DKWSNP{} (resp. \DKWSPINE) is on average $9.22$ (resp. $12.86$) times faster than $\BaselineDKWS$. 

\noindent(d) The query performance improvement on DBpedia is similar to that on YAGO3. \DKWSBF{} (resp. \DKWSPADS, \DKWSNP, and \DKWSPINE) is $5.06$ (resp. $5.31$, $5.83$, and $22.32$) times faster than $\BaselineDKWS$.

In a nutshell, the performance improvement is due to the following reasons: (a) \DKWSBF{} avoids the exhaustive explorations by the backward search and forward search; (b) \DKWSPADS{} prunes the redundant forward search by computing the tight lower bound of the shortest distance between a vertex and a query keyword, which prunes some unnecessary forward search at an early stage; (c) \DKWSNP{} improves the query performance by exchanging the local upper bounds to yield a global upper bound, which reduces the stragglers' computation; and (d) \DKWSPINE{} further improves the query performance since it is finer-grained.

\noindent\TKDERF{
\stitle{Exp-2: \vldbj{Scalability}.} We next investigated the \vldbj{scalability} of $\DKWS$ over real-life graphs by varying the number of workers ($m$) from $2$ to $12$. \vldbj{a) All algorithms take a shorter time when the number of workers becomes larger, as expected. b) All algorithms scale reasonably well with the increase of $m$. When $m$ increases from $2$ to $12$, the running time of $\BaselineDKWS$ (resp. \DKWSBF{}, \DKWSPADS{}, \DKWSPINE{} and \DKWSNP{}) decreases to $21.63\%$ (resp. $23.18\%$, $28.99\%$, $31.62\%$, and $25.75\%$) on average. c) \DKWSNP{} consistently outperforms $\BaselineDKWS$, \DKWSBF{}, \DKWSPADS{} and \DKWSPINE{} for all queries. Specifically,} the results are shown in Fig.~\ref{fig:performance}\subref{fig:dkws-vary-workers-webuk} to Fig.~\ref{fig:performance}\subref{fig:dkws-vary-workers-dbpedia-appdix}. \DKWSBF{} and \DKWSPADS{} take less time when the number of workers increases. More specifically, \DKWSBF{} is on average $1.65$ (resp. $8.81$, $2.65$, and $1.60$) times faster than $\BaselineDKWS$ on YAGO3 (resp. WebUK, DBLP, and DBpedia), when the number of workers varies from $2$ to $12$. \DKWSPADS{} is on average $1.81$ (resp. $13.11$, $2.92$, and $2.11$) times faster than $\BaselineDKWS$ on YAGO3 (resp. WebUK, DBLP, and DBpedia). The reason is that  \DKWSBF{}, and \DKWSPADS{} avoid exhaustive search and prune some redundant stale computations. By exchanging the local upper bounds, \DKWSNP{} exploits parallelism, since it reduces the straggler problem. \DKWSNP{} is on average $2.03$ (resp. $21.19$, $3.31$, and $3.67$) times faster than $\BaselineDKWS$ on YAGO3 (resp. WebUK, DBLP, and DBpedia). \DKWSPINE{} is the most efficient since the computing tasks are finer-grained. On average, \DKWSPINE{} is $3.47$ (resp. $26.94$, $5.00$, and $22.82$) times faster than $\BaselineDKWS$ on YAGO3 (resp. WebUK, DBLP, and DBpedia). It is also worth noting that in a single-machine environment, there is no difference in the performance of \DKWSPINE{}, \DKWSNP{}, and \DKWSPADS{}. This is because the fine-grained execution of $\PINE$ and notify-push paradigm are not activated in a single-machine setting.
}

\begin{figure}[tp]
	\centering
	\begin{subfigure}[b]{0.24\textwidth}
		\centering
		\includegraphics[width=\textwidth]{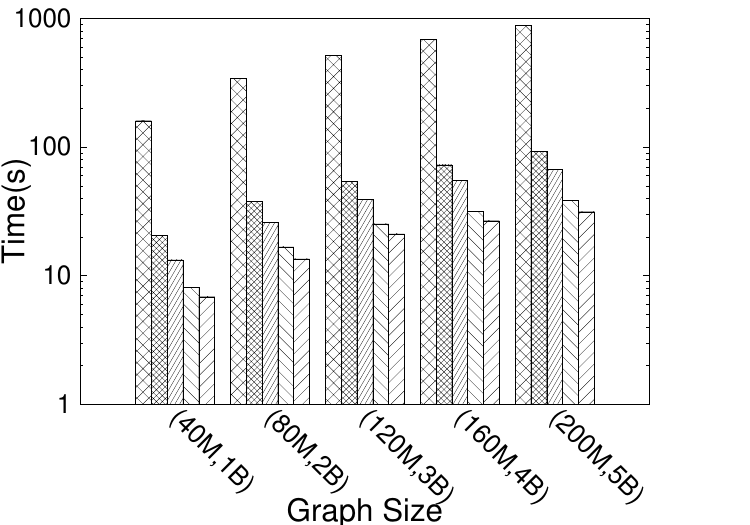}
	\end{subfigure}
	\begin{subfigure}[b]{0.24\textwidth}
		\centering
		\includegraphics[width=\textwidth]{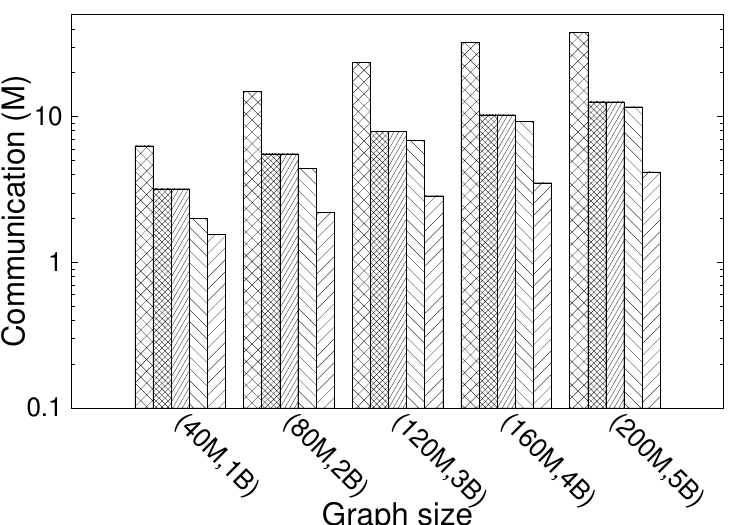}
	\end{subfigure}
	\caption{\TKDERF{Scalability on synthetic graphs}}
	\label{fig:scalability}
\end{figure}

\noindent\TKDERF{\etitle{Impact of the graph size $|G|$.} We also evaluated the scalability of $\DKWS$ over larger synthetic graphs. We use the graph generator of ~\cite{fan2017parallel} to produce graphs $G=(V,E,L)$ with $L$ drawn from an alphabet $\mathcal{L}$ of $50$ labels. It is controlled by the numbers of nodes $|V|$ and edges $|E|$, up to $200$ million and $5$ billion, respectively. Fixing $n=12$, we varied $|G|$ from $(40$M$,1$B$)$ to $(200$M$,5$B$)$. As reported in Fig.~\ref{fig:scalability}, the results are consistent with Fig.~\ref{fig:performance} over real-life graphs. (a) All algorithms take a longer time when the $G$ gets larger, as expected. (b) $\DKWS$ scales reasonably well with the increase of $|G|$. When $G$ increased by $5$ times, the running time of $\BaselineDKWS$ (resp. \DKWSBF{}, \DKWSPADS{}, \DKWSPINE{} and \DKWSNP{}) increases by $6.7$ (resp. $6.3$, $6.8$, $6.3$ and $6.1$) times. \DKWSPINE{} consistently outperforms $\BaselineDKWS$, \DKWSBF{}, \DKWSPADS{} and \DKWSNP{}.}

\stitle{Exp-3: Impact of parameters.} The elapsed time of keyword search is relevant to the threshold, $\tau$ and the number of matches, $k$. We next present the impact of these parameters. 

\etitle{Impact of threshold $\tau$.} $\tau$ has been a crucial parameter of keyword search. According to the findings of \cite{coffman2012empirical,DBLP:journals/tkde/YuanLCYWS17}, $\tau=5$ is large enough to obtain satisfactory matches in real applications. Hence, we next evaluated the \vldbj{scalability} of $\DKWS$ by varying $\tau$ from $3$ to $6$. a) All algorithms take longer when $\tau$ becomes larger, as expected, since there are more candidate answers generated during the backward expansion and forward expansion. b) All algorithms scale reasonably well with the increase of $\tau$. When $\tau$ increases from $3$ to $6$, the running time of $\BaselineDKWS$ (resp. \DKWSBF{}, \DKWSPADS{}, \DKWSPINE{}, and \DKWSNP{}) increases by $71.59\%$ (resp. $82.61\%$, $138.97\%$, $81.38\%$, and $83.56\%$). c) \DKWSPINE{} consistently outperforms $\BaselineDKWS$, \DKWSBF{}, \DKWSPADS{} and \DKWSNP{} for all queries. Specifically, the results are presented in Fig.~\ref{fig:performance}\subref{fig:dkws-vary-tau-yago} to Fig.~\ref{fig:performance}\subref{fig:dkws-vary-tau-dbpedia}. In particular, \DKWSBF{} is $1.08$ times (resp. $9.18$, $4.37$, and $1.25$) times faster than $\BaselineDKWS$ on YAGO3 (resp. WebUK, DBLP, and DBpedia). \DKWSPADS{} is $1.33$ (resp. $11.80$, $4.76$, and $1.48$) times faster than $\BaselineDKWS$ on YAGO3 (resp. WebUK, DBLP, and DBpedia). \DKWSPADS{} is more efficient on $\tau$ since it can prune longer forward searches when $\tau$ increases. \DKWSNP{} is $3.02$ (resp. $19.0$, $6.10$, and $2.66$) times faster than $\BaselineDKWS$ on YAGO3 (resp. WebUK, DBLP, and DBpedia). \DKWSNP{} is more efficient since it pushed and notified tighter bounds early which was more efficient when $\tau$ was large. \DKWSPINE{} is $3.71$ (resp. $29.32$, $6.65$, and $16.2$) times faster than $\BaselineDKWS$ on YAGO3 (resp. WebUK, DBLP, and DBpedia).

\etitle{Impact of $k$.} We evaluated the scalability of $\DKWS$ by varying $k$. \vldbj{a) All algorithms take longer when $k$ gets larger since more matches are retrieved. b) All algorithms perform well with the increase of $k$. When $k$ increases from $5$ to $30$, the running time of $\BaselineDKWS$ (resp. \DKWSBF{}, \DKWSPADS{}, \DKWSPINE{}, and \DKWSNP{}) increases by $4.63\%$ (resp. $1.65\%$, $9.99\%$, $60.20\%$, and $47.22\%$). c) \DKWSNP{} outperforms $\BaselineDKWS$, \DKWSBF{}, \DKWSPADS{} and \DKWSPINE{} for all queries. Specifically,} the experiments are shown in Fig.~\ref{fig:performance}\subref{fig:dkws-vary-topk-webuk} to Fig.~\ref{fig:performance}\subref{fig:dkws-vary-topk-dblp}. On average, \DKWSBF{} is $1.58$ (resp. $14.96$, $4.07$, and $1.30$) times faster than $\BaselineDKWS$ on YAGO3 (resp. WebUK, DBLP, and DBpedia). \DKWSPADS{} is $1.67$ (resp. $15.97$, $4.39$, and $1.39$) times faster than $\BaselineDKWS$ on YAGO3 (resp. WebUK, DBLP, and DBpedia). \DKWSNP{} is $1.98$ (resp. $22.92$, $5.83$, and $6.37$) times faster than $\BaselineDKWS$ on YAGO3 (resp. WebUK, DBLP, and DBpedia). \DKWSPINE{} is $2.54$ (resp. $35.63$, $7.14$, and $19.66$) times faster than $\BaselineDKWS$ on YAGO3 (resp. WebUK, DBLP, and DBpedia).

\stitle{Exp-4: Communication costs.} We further investigated the communication cost in terms of the total message size. The communication costs on  WebUK and DBLP are reported in Fig.~\ref{fig:performance}\subref{fig:comm-webuk} and Fig.~\ref{fig:performance}\subref{fig:comm-DBLP}. The results on other datasets exhibit similar trends. We obtained the following findings. (a) The communication cost of \DKWSPADS{} is the same as that of \DKWSBF{} since \DKWSPADS{} only prunes the local traversals. \DKWSBF{} and \DKWSPADS{} ship $33.6\%$ (resp. $36\%$) of data transmitted by $\BaselineDKWS$ on WebUK (resp. DBLP). (b) \DKWSNP{} ships $29.8\%$ (resp. $30.4\%$) compared to that of $\BaselineDKWS$ on WebUK (resp. DBLP). This is because \DKWSNP{} yields tighter bounds and reduces unnecessary message exchange early. (c) \DKWSPINE{} ships $13.0\%$ (resp. $18\%$) compared to that of $\BaselineDKWS$ on WebUK (resp. DBLP). \DKWSPINE{} takes the advantage of preemptive execution of both $\bkws$ and $\fkws$, which reduces long or useless traversals. Consequently, the communication cost is reduced since the messages caused by such traversals have been avoided.

\noindent\TKDERF{\stitle{Exp-5: Impact of notification counter threshold.} We observed that on the four real-life datasets, setting the notification counter threshold to $2$ or $3$ resulted in a comparatively good performance. However, when the threshold exceeded $4$, there was no substantial difference in the performance improvement compared to when the notify-push paradigm was not used. This can be attributed to the fact that on these datasets, the number of times the local bounds were refined rarely exceeded $4$; thus, the push function was seldom invoked. The threshold of the notification counter in the coordinator can be determined by a simple experiment offline on the dataset. The details are presented in \cite{techreport}.}

\begin{figure}[tp]
\begin{center}
\includegraphics[width=0.4\textwidth]{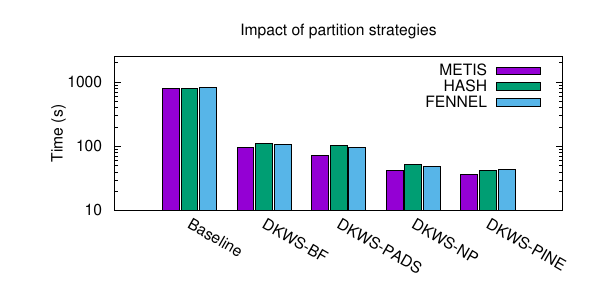}
\end{center}
\vspace{-3mm}
\caption{Impact of partition strategies (WebUK)}
\label{fig:partition}
\end{figure}

\noindent\TKDERF{\stitle{Exp-6: Impact of graph partition.} We evaluated the impact of different partition strategies, including METIS~\cite{karypis1995metis}, HASH~\cite{fan2017grape}, and FENNEL~\cite{tsourakakis2014fennel} in Fig.~\ref{fig:partition}. Among these strategies, all algorithms, except for the $\BaselineDKWS$, demonstrated faster performance under METIS partitioning. Considering the significance of efficiency, we selected METIS as the default partition strategy for our experiments, as mentioned earlier. Furthermore, we observed that METIS improved performance in \DKWSNP{} and \DKWSPINE{}. This can be attributed to the notify-push paradigm employed that helps alleviate the impact of load imbalances.}

\stitle{Exp-7: Comparison with a sequential algorithm.} We further compared our works with a sequential algorithm, \SKWSC~\cite{kacholia2005bidirectial}. The results are shown in Fig.~\ref{fig:performance}\subref{fig:dkws-baseline-yago} to Fig.~\ref{fig:performance}\subref{fig:dkws-baseline-dbpedia}. On average, \DKWSPINE{} is $82.58$ times faster than \SKWSC{}. This verifies that \DKWSPINE{} has exploited the efficiency of a distributed environment.

\section{Related work}\label{sec:related}

\stitle{Keyword search semantics.}  Recently, keyword search has attracted a lot of interest from both industry and research communities. Bhalotia et al.~\cite{bhalotia2002keyword} proposed keyword search on relational databases. He et al.~\cite{he2007blinks} proposed an index, called $\Blinks$ to reduce the search time. Kargar et al.~\cite{kargar2011keyword} proposed distance restrictions on the keyword nodes, \ie the distance between each pair of keyword nodes is smaller than $\tau$. \revise{Shi et al.~\cite{DBLP:conf/www/Shi0K20} proposed hub labelings to solve Group Steiner Trees (GST). Kargar et al.~\cite{efficient2020kargar} proposed an approximate algorithm to retrieve the GST on weighted graphs}. These studies optimize a specific keyword search semantic. Jiang et al.~\cite{Jiang2020TKDE} proposed a generic index for keyword search semantics running on a standalone machine. 

\stitle{Distributed systems.} Several distributed systems have been proposed for graphs. Popular graph systems include $\mathsf{Pregel}$~\cite{malewicz2010pregel}, $\mathsf{Giraph}$~\cite{avery2011giraph}, $\mathsf{GraphX}$ \cite{xin2013graphx}, $\mathsf{GraphLab}$~\cite{low2012distributed}, $\mathsf{PowerGrapah}$~\cite{gonzalez2012powergraph}, $\mathsf{Giraph}$++~\cite{tian2013think}, $\mathsf{Blogel}$~\cite{yan2014blogel}, $\mathsf{GPS}$~\cite{salihoglu2013gps}, $\grape$~\cite{fan2017parallel,fan2018think}, and $\mathsf{AAP}$~\cite{fan2020adaptive}. $\mathsf{Pregel}$~\cite{malewicz2010pregel} and $\mathsf{Giraph}$~\cite{avery2011giraph} are implemented with the vertex-centric programming model. A superstep executes a user-defined function at each vertex in parallel. $\mathsf{GraphX}$~\cite{xin2013graphx} is a component built on top of $\mathsf{Spark}$ for graphs which exposes a set of  operators (\eg subgraph, joinVertices, and aggregateMessages) as well as an optimized variant of the $\mathsf{Pregel}$~\cite{malewicz2010pregel}. $\mathsf{Blogel}$~\cite{yan2014blogel}, $\mathsf{Giraph}$++~\cite{tian2013think} and $\grape$~\cite{fan2017parallel} are implemented with the block-centric programming model. $\mathsf{AAP}$~\cite{fan2020adaptive} proposes an adaptive asynchronous parallel model for graph computations on ~\cite{fan2017parallel}. These systems are general-purpose. Keyword search algorithms have not been exploited. \TKDERF{For instance, $\DKWS$ can also be beneficial to existing systems. By integrating $\PINE$, the systems could make the query evaluation more fine-grained. By integrating the notify-push paradigm, $\DKWS$ allows each worker to broadcast local information to their peer workers which can alleviate the straggler problem.}

\stitle{Distributed $\kws$ algorithms.} \revise{Lu et al.~\cite{qin2014scalable} proposed a scalable algorithm for keyword search in MapReduce. However, the false matches were pruned at the last superstep, which may cause large messages. Yuan et al.~\cite{DBLP:journals/tkde/YuanLCYWS17} proposed a search strategy based on a compressed signature to avoid the exhaustive flooding search. \cite{DBLP:journals/tkde/YuanLCYWS17} sent all the local candidate matches to the coordinator at runtime which may require large messages and extra synchronization cost. $\DKWS$ differs from the above in the following aspects: (a) each worker computes the top-$k$ matches locally. $\DKWS$ sends the local matches to the coordinator when all of the workers terminate rather than sends massive local candidates matches;} and (b) $\DKWS$ exchanges the local upper bounds which prune some traversals early.

% !TeX root = main.tex

\section{Conclusions and future works}\label{sec:conclusions}

In this paper, we propose a distributed keyword search system called $\DKWS$. We derive new bounds for pruning some keyword searches that tackle the performance challenges of a general distributed system. We show that $\SKWS$, \TKDE{which can be used to express query algorithms for popular keyword semantics,} has a monotonic property that ensures the correct parallelization. \TKDE{We propose a notify-push paradigm allows asynchronously exchanging the upper bounds across the workers and the coordinators. We also propose a programming model $\PINE$ for $\DKWS$ which fits keyword search algorithms as they have distinguished $n$ phases, to allow preemptive searches to mitigate staleness in a distributed system}. We verify that $\DKWS$ significantly reduces the runtimes of distributed top-$k$ keyword searches.

In the future, we plan to implement $\PINE$ into the latest codebase of $\grape$. Moreover, we will extend $\DKWS$ to support {\em approximate} analysis for some keyword search semantics~\cite{kargar2011keyword,efficient2020kargar,DBLP:conf/www/Shi0K20}, some community detection algorithms~\cite{liao2022distributed} and some dense subgraph detection algorithms~\cite{jiang2022spade,ma2020efficient}.

\stitle{Acknowledgements.} This work is supported by HKRGC GRF 12203123, 12201119, 12200022, and 12202221, and C2004-21GF.

% \balance
\bibliographystyle{abbrv}
\bibliography{ref}

\begin{thebibliography}{10}

\bibitem{dblp}
{DBLP}.
\newblock {\sl https://dblp.org/}.

\bibitem{dbpedia}
{DBpedia}.
\newblock {\sl http://wiki.dbpedia.org/Datasets}.

\bibitem{webuk}
{WebUK}.
\newblock {\sl http://law.di.unimi.it/webdata/uk-union- 2006-06-2007-05}.

\bibitem{avery2011giraph}
C.~Avery.
\newblock Giraph: Large-scale graph processing infrastructure on hadoop.
\newblock {\em Proceedings of the Hadoop Summit. Santa Clara}, 11(3):5--9, 2011.

\bibitem{bhalotia2002keyword}
G.~Bhalotia, A.~Hulgeri, C.~Nakhe, S.~Chakrabarti, and S.~Sudarshan.
\newblock Keyword searching and browsing in databases using banks.
\newblock In {\em ICDE}, pages 431--440. IEEE, 2002.

\bibitem{coffman2012empirical}
J.~Coffman and A.~C. Weaver.
\newblock An empirical performance evaluation of relational keyword search techniques.
\newblock {\em IEEE Transactions on Knowledge and Data Engineering}, 26(1):30--42, 2012.

\bibitem{cohen2015all}
E.~Cohen.
\newblock All-distances sketches, revisited: Hip estimators for massive graphs analysis.
\newblock {\em IEEE Transactions on Knowledge and Data Engineering}, 27(9):2320--2334, 2015.

\bibitem{fan2018think}
W.~Fan, Y.~Cao, J.~Xu, W.~Yu, Y.~Wu, C.~Tian, J.~Jiang, and B.~Zhang.
\newblock From think parallel to think sequential.
\newblock {\em ACM SIGMOD Record}, 47(1):15--22, 2018.

\bibitem{fan2019dynamic}
W.~Fan, C.~Hu, M.~Liu, P.~Lu, Q.~Yin, and J.~Zhou.
\newblock Dynamic scaling for parallel graph computations.
\newblock {\em Proceedings of the VLDB Endowment}, 12(8):877--890, 2019.

\bibitem{fan2017incremental}
W.~Fan, C.~Hu, and C.~Tian.
\newblock Incremental graph computations: Doable and undoable.
\newblock In {\em SIGMOD}, pages 155--169. ACM, 2017.

\bibitem{fan2020application}
W.~Fan, R.~Jin, M.~Liu, P.~Lu, X.~Luo, R.~Xu, Q.~Yin, W.~Yu, and J.~Zhou.
\newblock Application driven graph partitioning.
\newblock In {\em Proceedings of the 2020 ACM SIGMOD International Conference on Management of Data}, pages 1765--1779, 2020.

\bibitem{fan2020adaptive}
W.~Fan, P.~Lu, W.~Yu, J.~Xu, Q.~Yin, X.~Luo, J.~Zhou, and R.~Jin.
\newblock Adaptive asynchronous parallelization of graph algorithms.
\newblock {\em ACM Transactions on Database Systems (TODS)}, 45(2):1--45, 2020.

\bibitem{fan2017grape}
W.~Fan, J.~Xu, Y.~Wu, W.~Yu, and J.~Jiang.
\newblock Grape: Parallelizing sequential graph computations.
\newblock {\em Proceedings of the VLDB Endowment}, 10(12):1889--1892, 2017.

\bibitem{fan2017parallel}
W.~Fan, J.~Xu, Y.~Wu, W.~Yu, J.~Jiang, Z.~Zheng, B.~Zhang, Y.~Cao, and C.~Tian.
\newblock Parallelizing sequential graph computations.
\newblock In {\em SIGMOD}, pages 495--510. ACM, 2017.

\bibitem{fang2016effective}
Y.~Fang, R.~Cheng, S.~Luo, and J.~Hu.
\newblock Effective community search for large attributed graphs.
\newblock {\em PVLDB}, 9(12):1233--1244, 2016.

\bibitem{gonzalez2012powergraph}
J.~E. Gonzalez, Y.~Low, H.~Gu, D.~Bickson, and C.~Guestrin.
\newblock Powergraph: Distributed graph-parallel computation on natural graphs.
\newblock In {\em Presented as part of the 10th $\{$USENIX$\}$ Symposium on Operating Systems Design and Implementation ($\{$OSDI$\}$ 12)}, pages 17--30, 2012.

\bibitem{he2007blinks}
H.~He, H.~Wang, J.~Yang, and P.~S. Yu.
\newblock Blinks: Ranked keyword searches on graphs.
\newblock In {\em SIGMOD}, pages 305--316, 2007.

\bibitem{techreport}
J.~Jiang, B.~Choi, X.~Huang, J.~Xu, and S.~S. Bhowmick.
\newblock Dkws: An efficient distributed system for keyword search on massive graphs.
\newblock {\footnotesize\sl https://www.comp.hkbu.edu.hk/\%7Ebchoi/DKWS.pdf}, 2023.

\bibitem{Jiang2020TKDE}
J.~Jiang, B.~Choi, J.~Xu, and S.~S. Bhowmick.
\newblock A generic ontology framework for indexing keyword search on massive graphs.
\newblock In {\em TKDE}, 2020.

\bibitem{Jiang2020PPKWSAE}
J.~Jiang, X.~Huang, B.~Choi, J.~Xu, S.~S. Bhowmick, and L.~Xu.
\newblock {PPKWS}: An efficient framework for keyword search on public-private networks.
\newblock In {\em ICDE}, pages 457--468, 2020.

\bibitem{jiang2022spade}
J.~Jiang, Y.~Li, B.~He, B.~Hooi, J.~Chen, and J.~K.~Z. Kang.
\newblock Spade: A real-time fraud detection framework on evolving graphs.
\newblock {\em Proceedings of the VLDB Endowment}, 16(3):461--469, 2022.

\bibitem{jiang2015exact}
M.~Jiang, A.~W.-C. Fu, and R.~C.-W. Wong.
\newblock Exact top-k nearest keyword search in large networks.
\newblock In {\em SIGMOD}, pages 393--404. ACM, 2015.

\bibitem{kacholia2005bidirectial}
V.~Kacholia, S.~Pandit, S.~Chakrabarti, S.~Sudarshan, R.~Desai, and H.~Karambelkar.
\newblock Bidirectional expansion for keyword search on graph databases.
\newblock In {\em PVLDB}, pages 505--516, 2005.

\bibitem{kargar2011keyword}
M.~Kargar and A.~An.
\newblock Keyword search in graphs: Finding r-cliques.
\newblock {\em PVLDB}, 4(10):681--692, 2011.

\bibitem{efficient2020kargar}
M.~Kargar, L.~Golab, D.~Srivastava, J.~Szlichta, and M.~Zihayat.
\newblock Effective keyword search over weighted graphs.
\newblock {\em IEEE Transactions on Knowledge and Data Engineering}, pages 1--1, 2020.

\bibitem{karypis1995metis}
G.~Karypis and V.~Kumar.
\newblock Metis--unstructured graph partitioning and sparse matrix ordering system, version 2.0.
\newblock 1995.

\bibitem{liao2022distributed}
X.~Liao, Q.~Liu, J.~Jiang, X.~Huang, J.~Xu, and B.~Choi.
\newblock Distributed d-core decomposition over large directed graphs.
\newblock {\em arXiv preprint arXiv:2202.05990}, 2022.

\bibitem{low2012distributed}
Y.~Low, D.~Bickson, J.~Gonzalez, C.~Guestrin, A.~Kyrola, and J.~M. Hellerstein.
\newblock Distributed graphlab: A framework for machine learning and data mining in the cloud.
\newblock {\em Proceedings of the VLDB Endowment}, 5(8):716--727, 2012.

\bibitem{ma2020efficient}
C.~Ma, Y.~Fang, R.~Cheng, L.~V. Lakshmanan, W.~Zhang, and X.~Lin.
\newblock Efficient algorithms for densest subgraph discovery on large directed graphs.
\newblock In {\em Proceedings of the 2020 ACM SIGMOD International Conference on Management of Data}, pages 1051--1066, 2020.

\bibitem{mahdisoltani2014yago3}
F.~Mahdisoltani, J.~Biega, and F.~Suchanek.
\newblock Yago3: A knowledge base from multilingual wikipedias.
\newblock In {\em Seventh Biennial Conference on Innovative Data Systems Research}, 2014.

\bibitem{malewicz2010pregel}
G.~Malewicz, M.~H. Austern, A.~J. Bik, J.~C. Dehnert, I.~Horn, N.~Leiser, and G.~Czajkowski.
\newblock Pregel: A system for large-scale graph processing.
\newblock In {\em Proceedings of the 2010 ACM SIGMOD International Conference on Management of Data}, pages 135--146. ACM, 2010.

\bibitem{michel2005klee}
S.~Michel, P.~Triantafillou, and G.~Weikum.
\newblock Klee: A framework for distributed top-k query algorithms.
\newblock In {\em Proceedings of the 31st International Conference on Very Large Data Bases}, pages 637--648, 2005.

\bibitem{pacaci2019experimental}
A.~Pacaci and M.~T. {\"O}zsu.
\newblock Experimental analysis of streaming algorithms for graph partitioning.
\newblock In {\em Proceedings of the 2019 International Conference on Management of Data}, pages 1375--1392, 2019.

\bibitem{qiao2013top}
M.~Qiao, L.~Qin, H.~Cheng, J.~X. Yu, and W.~Tian.
\newblock Top-k nearest keyword search on large graphs.
\newblock {\em PVLDB}, 6(10):901--912, 2013.

\bibitem{qin2014scalable}
L.~Qin, J.~X. Yu, L.~Chang, H.~Cheng, C.~Zhang, and X.~Lin.
\newblock Scalable big graph processing in mapreduce.
\newblock In {\em Proceedings of the 2014 ACM SIGMOD International Conference on Management of Data}, pages 827--838. ACM, 2014.

\bibitem{salihoglu2013gps}
S.~Salihoglu and J.~Widom.
\newblock Gps: A graph processing system.
\newblock In {\em Proceedings of the 25th International Conference on Scientific and Statistical Database Management}, page~22. ACM, 2013.

\bibitem{shi2016top}
J.~Shi, D.~Wu, and N.~Mamoulis.
\newblock Top-k relevant semantic place retrieval on spatial rdf data.
\newblock In {\em Proceedings of the 2016 International Conference on Management of Data}, pages 1977--1990, 2016.

\bibitem{DBLP:conf/www/Shi0K20}
Y.~Shi, G.~Cheng, and E.~Kharlamov.
\newblock Keyword search over knowledge graphs via static and dynamic hub labelings.
\newblock In Y.~Huang, I.~King, T.~Liu, and M.~van Steen, editors, {\em {WWW} '20: The Web Conference 2020, Taipei, Taiwan, April 20-24, 2020}, pages 235--245. {ACM} / {IW3C2}, 2020.

\bibitem{tian2013think}
Y.~Tian, A.~Balmin, S.~A. Corsten, S.~Tatikonda, and J.~McPherson.
\newblock From think like a vertex to think like a graph.
\newblock {\em Proceedings of the VLDB Endowment}, 7(3):193--204, 2013.

\bibitem{tian2008efficient}
Y.~Tian, R.~A. Hankins, and J.~M. Patel.
\newblock Efficient aggregation for graph summarization.
\newblock In {\em SIGMOD}, pages 567--580, 2008.

\bibitem{tsourakakis2014fennel}
C.~Tsourakakis, C.~Gkantsidis, B.~Radunovic, and M.~Vojnovic.
\newblock Fennel: Streaming graph partitioning for massive scale graphs.
\newblock In {\em Proceedings of the 7th ACM international conference on Web search and data mining}, pages 333--342, 2014.

\bibitem{valiant1990general}
L.~G. Valiant.
\newblock General purpose parallel architectures.
\newblock In {\em Algorithms and Complexity}, pages 943--971. Elsevier, 1990.

\bibitem{wang2010survey}
H.~Wang and C.~C. Aggarwal.
\newblock A survey of algorithms for keyword search on graph data.
\newblock In {\em Managing and mining graph data}, pages 249--273. Springer, 2010.

\bibitem{wu2013summarizing}
Y.~Wu, S.~Yang, M.~Srivatsa, A.~Iyengar, and X.~Yan.
\newblock Summarizing answer graphs induced by keyword queries.
\newblock {\em PVLDB}, 6(14):1774--1785, 2013.

\bibitem{xin2013graphx}
R.~S. Xin, J.~E. Gonzalez, M.~J. Franklin, and I.~Stoica.
\newblock Graphx: A resilient distributed graph system on spark.
\newblock In {\em First International Workshop on Graph Data Management Experiences and Systems}, page~2. ACM, 2013.

\bibitem{yan2014blogel}
D.~Yan, J.~Cheng, Y.~Lu, and W.~Ng.
\newblock Blogel: A block-centric framework for distributed computation on real-world graphs.
\newblock {\em Proceedings of the VLDB Endowment}, 7(14):1981--1992, 2014.

\bibitem{yang2021keyword}
J.~Yang, W.~Yao, and W.~Zhang.
\newblock Keyword search on large graphs: A survey.
\newblock {\em Data Science and Engineering}, 6(2):142--162, 2021.

\bibitem{yang2019efficient}
Y.~Yang, D.~Agrawal, H.~Jagadish, A.~K. Tung, and S.~Wu.
\newblock An efficient parallel keyword search engine on knowledge graphs.
\newblock In {\em 2019 IEEE 35th international conference on data engineering (ICDE)}, pages 338--349. IEEE, 2019.

\bibitem{yi2016autog}
P.~Yi, B.~Choi, S.~S. Bhowmick, and J.~Xu.
\newblock Autog: A visual query autocompletion framework for graph databases.
\newblock {\em PVLDB}, 9(13):1505--1508, 2016.

\bibitem{yu2010keyword}
J.~X. Yu, L.~Qin, and L.~Chang.
\newblock Keyword search in relational databases: A survey.
\newblock {\em IEEE Data Eng. Bull.}, 2010.

\bibitem{DBLP:journals/tkde/YuanLCYWS17}
Y.~Yuan, X.~Lian, L.~Chen, J.~X. Yu, G.~Wang, and Y.~Sun.
\newblock Keyword search over distributed graphs with compressed signature.
\newblock {\em {IEEE} Transactions on Knowledge and Data Engineering}, 29(6):1212--1225, 2017.

\bibitem{zheng2016semantic}
W.~Zheng, L.~Zou, W.~Peng, X.~Yan, S.~Song, and D.~Zhao.
\newblock Semantic sparql similarity search over rdf knowledge graphs.
\newblock {\em PVLDB}, 9(11):840--851, 2016.

\end{thebibliography}

\vspace{-3em}

\begin{IEEEbiography}[{\includegraphics[width=1in,height=1.2in,clip,keepaspectratio]{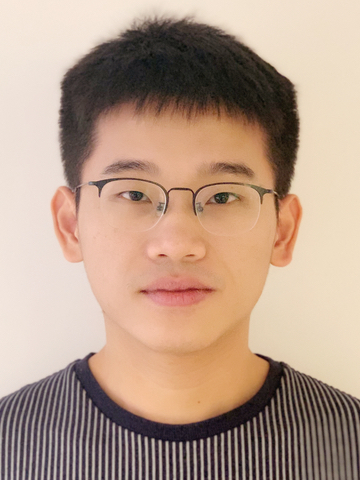}}]{Jiaxin~Jiang}
is a research fellow in the Institute of Data Science, National University of Singapore. He received his BEng degree in computer science and engineering from
 Shandong University in 2015 and PhD degree in computer science from Hong Kong Baptist University (HKBU) in 2020. His research interests include graph-structured databases, distributed graph computation and fraud detection.
\end{IEEEbiography}

\vspace{-3em}

\begin{IEEEbiography}[{\includegraphics[width=1in,height=1.2in,clip,keepaspectratio]{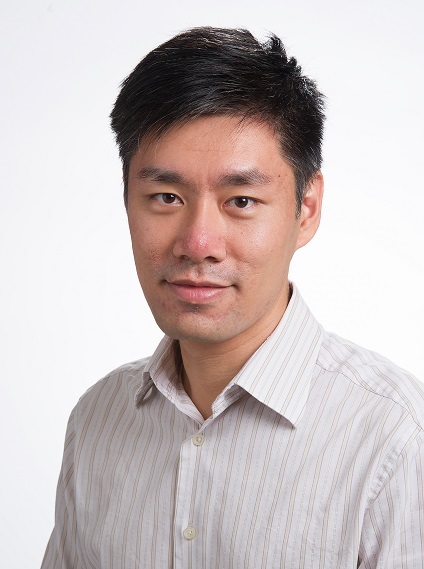}}]{Byron~Choi}
is a Professor in the Department of Computer Science at
the Hong Kong Baptist University. He received the bachelor of engineering degree
in computer engineering from the Hong Kong University of Science and Technology
(HKUST) in 1999 and the MSE and PhD degrees in computer and information science
from the University of Pennsylvania in 2002 and 2006, respectively.
\end{IEEEbiography}

\vspace{-3em}

\begin{IEEEbiography}[{\includegraphics[width=1in,height=1.2in,clip,keepaspectratio]{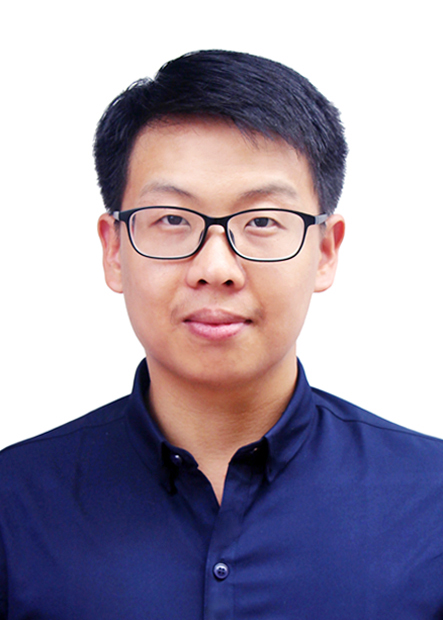}}]{Xin~Huang}
received the PhD degree from the Chinese University of Hong Kong (CUHK) in 2014. He is currently an Associate Professor at Hong Kong Baptist University. His research interests mainly focus on graph data management and mining.
\end{IEEEbiography}

\vspace{-3em}

\begin{IEEEbiography}[{\includegraphics[width=1in,height=1.2in,clip,keepaspectratio]{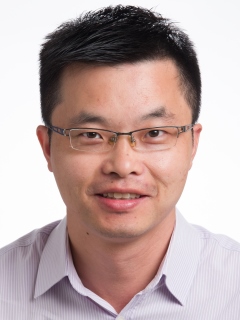}}]{Jianliang~Xu} 
is a Professor in the Department of Computer Science, Hong Kong
Baptist University (HKBU). He held visiting positions at Pennsylvania State
University and Fudan University.
% His current research interests include data
% management, database security \& privacy, and location-aware computing.
He has
published more than 150 technical papers in these areas, most of which appeared
in leading journals and conferences including SIGMOD, VLDB, ICDE, TODS, TKDE,
and VLDBJ.
% He has served as a program co-chair/vice chair for a number of major
% international conferences including IEEE ICDCS 2012, IEEE CPSNA 2015 and WAIM
% 2016. He is an Associate Editor of IEEE Transactions on Knowledge and Data
% Engineering (TKDE) and Proceedings of the VLDB Endowment (PVLDB 2018).
\end{IEEEbiography}

\vspace{-3em}

\begin{IEEEbiography}[{\includegraphics[width=1in,height=1.2in,clip,keepaspectratio]{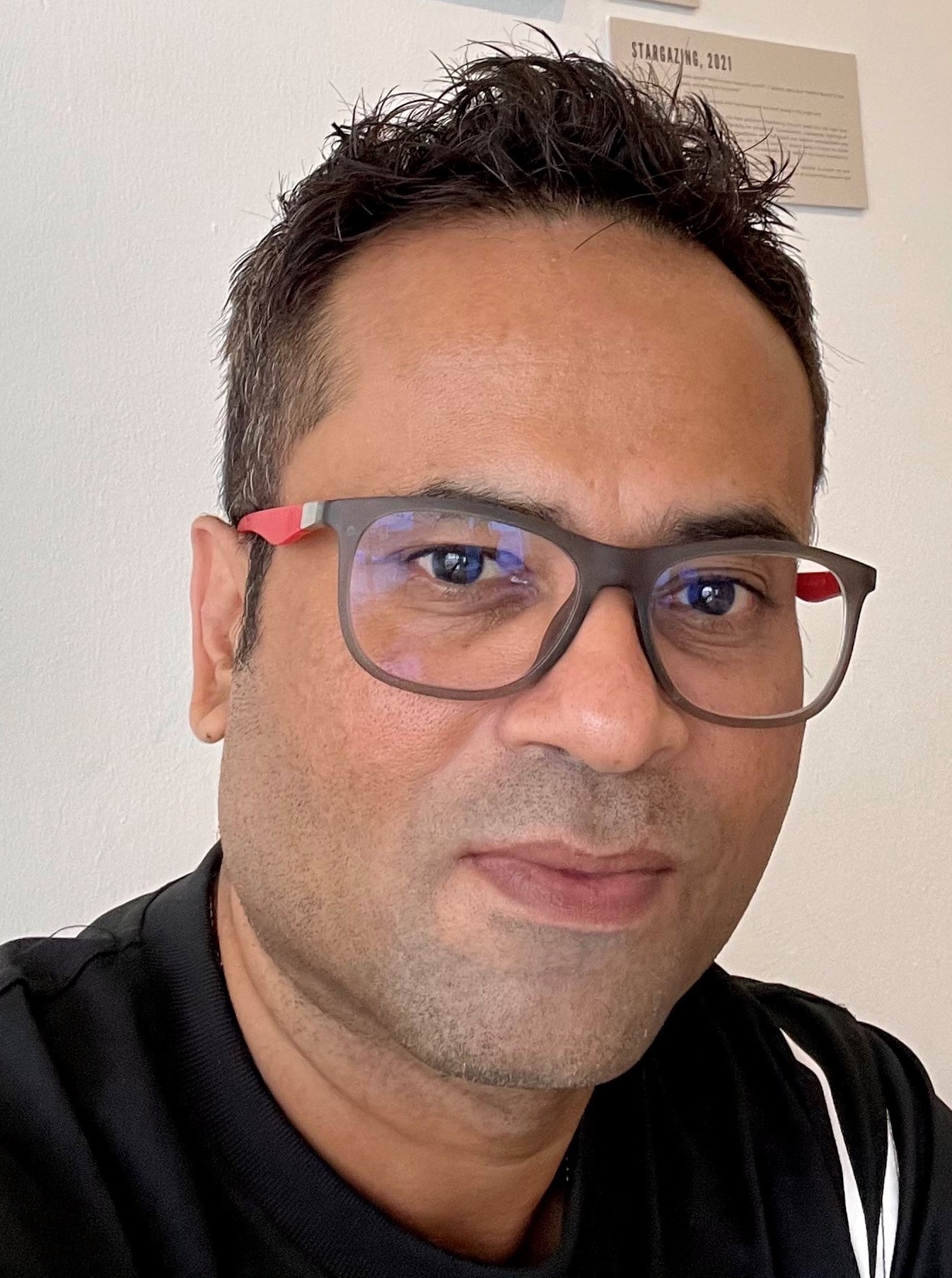}}]{Sourav~S.~Bhowmick}  is an Associate Professor in the School of Computer Science and Engineering (SCSE), Nanyang Technological University, Singapore. His core research expertise is in data management, human-data interaction, and data analytics. His research has appeared in premium venues such as ACM SIGMOD, VLDB, and VLDB Journal. He is co-recipient of Best Paper Awards in ACM CIKM 2004, ACM BCB 2011, and VLDB 2021. He is also co-recipient of the 2021 ACM SIGMOD Research Highlights Award. Sourav was inducted into Distinguished Members of the ACM in 2020.
\end{IEEEbiography}

\vfill

\clearpage
\appendices
\section{Appendix}

\subsection{Proof of Prop.~\ref{prop:s_completed}} \label{appendix-skws-completeness}

\eat{
\begin{manuallemma}{\ref{prop:s_completed}}
The node set visited by $\bkws$, $\VU$, has the following properties:
	\begin{enumerate}
		\item $\forall u\not\in \VU$, $\answer_u\not\in \answerset$; and
		\item $\forall \answer_u\in \answerset$, $u\in \VU$. 
	\end{enumerate}
\end{manuallemma}
}

\begin{proposition}
	The node set visited by $\bkws$, $\VU$, has the following properties:
	\begin{enumerate}
		\item $\forall u\not\in \VU$, $\answer_u\not\in \answerset$; and
		\item $\forall \answer_u\in \answerset$, $u\in \VU$. 
	\end{enumerate}
\end{proposition}
\begin{IEEEproof} 1) Suppose $u\not\in \VU$. We assume that the last vertex which is backward expanded for keyword $q_i$ as $v_i$. The distance between $v_i$ and the query keyword $q_i\in Q$ is $\dist(v_i,{q_i})$. It is worth noting that in Step 2 of the backward search, the vertex with the shortest distance to $O_{i}$ is chosen for expansion. 
	
	Since $u$ has not been visited, 
	\begin{equation}
	\dist(u,{q_i})\geq \dist(v_i,{q_i})
	\end{equation}
	
	Hence, 
	\begin{equation}
		\Score(u) = \Sigma \dist(u,{q_i}) \geq \Sigma \dist(v_i,{q_i}) >\prune
	\end{equation}
	The match rooted at $u$ is not among the top-$k$ matches. Therefore, $\answer_u\not\in \answerset$.

	2) We prove the second part of the proposition by contradiction. Suppose $\answer_u\in \answerset$. If $u\not\in\VU$, we have $\Score(u) > \prune$ based on the proof above. This contradicts with that the $\answer_u$ is among the top-$k$ matches. 
\end{IEEEproof}

\subsection{Proof of Prop.~\ref{prop:d_completed}} \label{appendix-completeness}
\begin{proposition}
    (\textbf{Completeness}) Suppose the top-$k$ matches to a keyword query is $\answerset$ and all the visited vertices $\VU$, we have:
    \begin{enumerate}
        \item $\forall u\not\in \VU$, $\answer_u\not\in \answerset$; and 
        \item $\forall \answer_u\in \answerset$, $u\in \VU$. 
    \end{enumerate}  
\end{proposition}

\begin{IEEEproof} 1) Suppose $u\not\in \VU$, and $u\in F_i.V$. The match rooted at $u$ is $\answer_u$. We consider the nearest query keyword $q_x\in Q$ of $u$, \ie $\dist(u,q_y) \geq \dist(u,q_x)$, where $q_y\in Q$. We denote the shortest path between $u$ and $q_x$ by $\Path(u,q_x)$.
    
    We consider the following cases: 
    
    \textbf{Case 1:} $\Path(u,q_x)$ is localized on $F_i$ completely, \ie $\Path(u,q_x)$ is a subgraph of fragment $F_i$. The proof is the same with that of Proposition~\ref{prop:s_completed}.

    We assume that the last vertex which to be backward expanded for keyword $q_y$ on $F_i$ as $v_y$. Since $u$ has not been visited, we have
	\begin{equation}
	\dist(u,q_x) \geq \dist(v_y,q_y)
    \end{equation}
    Hence, 
	\begin{equation}\label{equ:case1-start}
		\Score(u) = \sum\limits_{y\in [1,l]} \dist(u,q_y) \geq l\dist(u,q_x)  \geq \sum\limits_{y\in [1,l]} \dist(v_y,q_y)
    \end{equation}
    
    Based on the termination condition (Algo.~\ref{algo:inckws}), we have
    \begin{equation}\label{equ:case1-end}
        \sum\limits_{y\in [1,l]} \dist(v_y,q_y) > S_i
    \end{equation}

    Combining Equ~\ref{equ:case1-start} and Equ~\ref{equ:case1-end}, we derive
    \begin{equation}
        \Score(u) > S_i
    \end{equation}

    Hence, $\answer_u$ is not among the local top-$k$ matches on $F_i$. It is not among the global top-$k$ matches either in nature, \ie $\answer_u\not\in \answerset$.

    \textbf{Case 2:} $\Path(u,q_x)$ is not localized on $F_i$ completely, \ie the path spans through multiple fragments. We denote all the vertices which have been visited by the backward expansion of the query keyword $q_x$ on fragment $F_i$ by $\VU_{q_x,i} = \bigcup_{ s\in [1,R]}\Visit_{q_x,i}^s$. We consider the vertex $u$ and all the portal nodes on $\Path(u,q_x)$. Since $u$ is never visited, we denote the sequence of $u$ and the portal nodes by $\Portal=\{p_n,\ldots, p_{z+1}, p_z,\ldots, p_1\}$, where $p_n = u$, such that $p_z\in \VU_{q_x,j}$, $p_{z+1}\not\in \VU_{q_x,j}$, $p_z\in F_j.O$ and $p_{z+1}\in F_j.I$ or $p_{z+1}=u$, where $j\in [1, m]$. Intuitively, $p_{z+1}$ is the first portal node which is not expanded by $\bkws$ in the portal node sequence.
     
    The proof is similar to the \textbf{Case 1.} We assume that the last vertex to be backward expanded for keyword $q_y$ on $F_j$ as $v_y$. Since $p_{z+1}$ has not been visited by $\bkws$ of the query keyword $q_{x}$ on $F_j$, we have 

    \begin{equation}
        \dist(p_{z+1}, q_x) \geq \dist(v_y, q_y)
    \end{equation}

    Hence, 
    \begin{equation}\label{equ:case2-start}
        \footnotesize{
        \Score(u) = \sum\limits_{y\in [1,l]} \dist(u,{q_y}) \geq l\dist(u,{q_x}) \geq l\dist(p_{z+1},{q_x}) \geq \sum\limits_{y\in [1,l]} \dist(v_y,{q_y})
        }
    \end{equation}
    
    Based on the termination condition (Algo.~\ref{algo:inckws}), we have
    \begin{equation}\label{equ:case2-end}
        \sum\limits_{y\in [1,l]} \dist(v_y,{q_y}) > S_j
    \end{equation}

    Combining Equ~\ref{equ:case2-start} and Equ~\ref{equ:case2-end}, we derive
    \begin{equation}
        \Score(u) > S_j
    \end{equation}

    Hence, $\answer_u$ is not better than the local top-$k$ matches on $F_j$. It is not among the global top-$k$ matches, \ie $\answer_u\not\in \answerset$.

\end{IEEEproof}

\begin{theorem}\label{appendix-pine}
    Consider a $\PINE$ algorithm which consists of $n$ $\PEval$s (denoted by $\Psf_i$, where $i\in [1,n]$), $n$ $\IncEval$s (denoted by $\I_i$, where $i\in [1,n]$), and one $\Assemble$ (denoted by $\E$), and any partition strategy $\Partition$. If (a) $\Psf_i$ and $\I_i$ satisfy the monotonic condition, and (b) $\Psf_i$, $\I_i$ and $\E$ are correct \wrt $\Partition$, then $\DKWS$ with $\Psf_i$, $\I_i$ and $\E$ guarantee to terminate correctly.
\end{theorem}

\begin{IEEEproof}
        In each superstep, $\PINE$ performs $\I_i$ ($i\in [1,n]$) selected by the switch statement. Hence the monotonic condition of $\PINE$ is identical to that of $\I_i$.
\end{IEEEproof}

\subsection{Pseudocode of $\IncEval$ of $\fkws$}

\etitle{(2) Incremental computation} ($\IncEval$) for $\fkws$ (Algo.~\ref{algo:incfkws}). $\IncEval$ is derived from  $\fkws$ of $\SKWS$ with the following two modifications. 

\etitle{(2.1) Refinement propagation.} Firstly, $P_i$ receives the partial matches, $\answerf_u$, to the forward expansion \jiaxin{\em request} $\MF_u$ in previous supersteps from other fragments in $M_i^1$ via the portal nodes, where $u\in F_i.O$. If a shorter path between $u$ and $q$ is found crossing multiple fragments (\jiaxin{Line~\ref{incfkws:M1in}}), the forward match $\answerf_u[q]$ is refined (Line~\ref{incfkws:refine}). Then, $\IncEval$ propagates the distance refinement to the ancestor vertices. It also notifies the coordinator once the local upper bound is refined \jiaxin{in $\mathsf{forwardExpand}$ (Line~\ref{algo:incfkws_forwardExpand}).} 

\etitle{(2.2) Incremental forward expansion.} Secondly, upon receiving some forward expansion requests from other fragments, worker $P_i$ further forward expands to retrieve missing keywords on $F_i$ via the incoming portal nodes, $F_i.I$. Specifically, if $\MF_u^{\mathsf{in}}\in M_i^2$ is received and $u\not\in \VM$, $u$ is added into $\VM$ (Line~\ref{incfwks:unot}). Since the search requests come from different fragments, $\MF_u$ keeps the largest one for each keyword (Line~\ref{incfwks:umin}). If $u$ is forward expanded in previous iterations for query keyword $q$ or $\MF_u^{\mathsf{in}}[q]$ is smaller than $\MF_u[q]$, $\MF_u^{\mathsf{in}}[q]$ is skipped.

At the end of $\IncEval$, the partial matches found by forward expansions are grouped into $M_i^1$ and the remaining forward expansion requests are grouped into $M_i^2$, respectively, for fragment $F_i$ and sent to the corresponding fragments, which is the same as that of $\PEval$.

%%%%%%%%Answer Expanding
\begin{algorithm}[tb]
  \caption{{$\IncEval$} for $\fkws$}\label{algo:incfkws}
  \footnotesize
  \SetKwProg{Fn}{Function}{}{}
  \KwIn{$F_i(V, E, L)$, $Q=\{q_1,\ldots, q_l\}$, $\threshold$, message $M_i$}
  \KwOut{$Q(F_i \oplus M_i)$ consisting of current $\answer_u \in \answerset_i$} 

  \ForEach{\blue{$q\in Q$}}{
	\blue{init a priority queue $\Queue_q = \emptyset$ for $q$ to store the refinement} \\
	\ForEach{\blue{$\answer_u^{f,\mathsf{in}} \in M^1_i$}}{\label{incfkws:M1in}
		\If{\blue{$\answerf_u[q] > \answer_u^{f,\mathsf{in}}[q]$}}{
			\blue{$\answerf_{u}[q] = \answer_u^{f,\mathsf{in}}[q]$ \label{incfkws:refine}}\\
			\blue{$\Queue_q.\mathsf{insert}(\langle u,  \answerf_{u}[q]\rangle)$}
		}
	}
	\blue{$\mathsf{PropagrateUpdate}(\Queue_q, q)$}
  }
  
  \ForEach{\blue{$\MF_u^{\mathsf{in}} \in M^2_i$}}{
	\If{\blue{$u\not\in \VM$}}{ \label{incfwks:unot}
		\blue{$\VM.\mathsf{add}(u)$} 
	}\Else{
		\ForEach{\blue{$q\in \MF_u^{\mathsf{in}}$}}{\label{incfwks:umin}
			\blue{$\MF_u[q] = \max\{\MF_u[q], {\MF_u^{\mathsf{in}}}[q]\}$}
		}
	}
}

  \ForEach{$u\in \VM$}{\label{incfkws:order}
	$\mathsf{forwardExpand}(u,Q,\prune_i, \answerset_i)$ \label{algo:incfkws_forwardExpand}
  }
	\Fn{$\mathsf{PropagrateUpdate}(\Queue_q,q)$}{
		init a visited vertices set $\mathsf{Vis}=\emptyset$ \\

		\While{$\Queue_q$ \tnormal{is not empty}}{
			$\langle u,d\rangle = \Queue_q.\mathsf{top}()$ \\
			% $d = \Queue.\mathsf{top}().\distance$ \\
			$\mathsf{Vis}.\mathsf{add}(u)$ \\
			$\Queue_q.\mathsf{pop}()$\\ 
			\ForEach{$e = (u', u)\in E$ \tnormal{\textbf{and}} $u'\not\in \mathsf{Vis}$}{
				$d' = w(e) + d$ \\ 
				\If{$d'\leq \answerf_{u'}[q]$}{
					$\answerf_{u'}[q] = d'$\\
					$\Queue_q.\mathsf{insert}(\langle u',d'\rangle)$
				}

			}
		}	
	}
\etitle{Message segment:} \blue{$M^1_{i} = \{\answerf_u | u\in F_i.I\}$ and $M^2_{i} = \{\MF_u | u\in F_i.O\}$}
  \end{algorithm}

\section{Optimizations for $\DKWS$}\label{sec:opt}

\subsection{Backtrack graph for refinement propagation}

Refinement propagation is potentially costly when messages are large. If we conduct the refinement propagation from the portal nodes in $F.O$, one by one, the time complexity is bounded by $O(|Q||F.O|(|E| + |V|\log |V|))$. To address this issue, we propose backtrack pointers for $\fkws$  and extend them to the outgoing portal nodes to build a backtrack graph.

\stitle{Backtrack pointer.} To reduce duplicate forward traversals, $\fkws$ maintains a backtrack pointer $\invert$ for each visited vertex $v$ in the forward expansion of $u$. Specifically, let's assume the sequence of the shortest path that starts from $u$ and ends with $v$ is $[v_1,\ldots, v_n]$ such that $v_1=u$, $v_n=v$, $\invert_{v_{j+1}} = v_j$ and $q\in L(v_n)$. Once $v_n$ is expanded, $\SKWS$ refines the matches as follows: $\answer_{{v_j}}[q] = \answer_{v_{j+1}}[q] + w(v_j, v_{j+1})$ recursively, for all $j\in [1,n-1]$.

\stitle{Backtrack graph.} Given a query keyword $q\in Q$, a backtrack graph  $G_{\invert}^q = (\invertV^q, \invertE^q)$ stores all  the paths between the vertices in $\VM$ and the outgoing portal nodes $F.O$ when conducting the forward expansion to retrieve the missing keyword $q$. $\DKWS$ propagates the refinement in {\em one batch} of the outgoing portal nodes for one query keyword rather than {\em vertex by vertex} from all the outgoing portal nodes. {\em The time complexity on fragment $F$ is reduced to} $O(|Q|(|\invertE^{q}|+|\invertV^{q}|\log|\invertV^{q}|))$.

\etitle{Construction of backtrack graph.} We consider a vertex $u\in \VM$. For any vertex $v\in F.O$ that is forward expanded during the forward expansion starting from $u$ for retrieving query keyword $q$, the sequence of the shortest path between $u$ and $v$, $[v_1,\ldots, v_n]$ (such that $v_1=u$, $v_n=v$), is added into $G_{\invert}^{q}$. That is, $\DKWS$ inserts  the edge $(v_j,v_{j+1})$ ($j\in [1,n-1]$) to $E_{\invert}^{q}$, implemented by backtrack pointers.

\begin{proposition}
    The size of $G_{\invert}^{q}$ is bounded by $O(|V|+|E|)$.
\end{proposition}

\eat{
\begin{proof}
     As we introduced above, $\forall v\in \invertV^{i}$, $v\in V$. Moreover, $\forall e=(v_j,v_{j+1})\in \invertE^{i}$, $e\in E$. Hence, $G_{\invert}^{i}$ is a subgraph of $F$. The size is bounded by $O(|V|+|E|)$. 
\end{proof}
}

\subsection{Indexing for pruning false matches}

As proposed in Sec.~\ref{sec:skws}, $\PRADS$ and $\KPADS$  prune some partial matches locally. Here, we extend $\PRADS$ and $\KPADS$ to propose $\BPADS$ that skips some forward expansions that span across multiple fragments.

%$\PRADS$ and $\KPADS$ are used to prune the partial answers during the forward expansion. However,

\stitle{Boundary-{$\PRADS$} ($\BPADS$).} For each fragment $F$, we build a sketch for all the out-portal nodes $F.O$, denoted by $\BPADS(F)$. $\BPADS(F)$ is built by merging $\PRADS^{\mathsf{out}}$ of out-portal nodes, \ie {$\PRADS^{\mathsf{out}}(v)$}, where $v\in F.O$. Formally, given a center $(w_j,d_j) \in \PRADS^{\mathsf{out}}(v)$, $(w_j,d_j)\in \BPADS(F)$ {\em iff}~~$\forall (w_j,d_j')\in \PRADS^{\mathsf{out}}(v')$ and $v'\in F.O$, $d_i'\geq d_i$.

\etitle{Pruning by indexes.} Consider a vertex $u\in \VM$ that needs to be forward expanded. The forward expansion starting from $u$ can be skipped if there is either no {\em local path} or no path {\em across multiple fragments} between $u$ and $q$ whose length is smaller than  $\MF_u[q]$: a) pruning forward expansion locally. We denote the lower bound of the shortest distance between $u$ and each missing keyword $q\in Q$ by $\Ld(u,q)$. If $\Ld(u,q)$ $> \MF_u[q]$, then there is no local path between $u$ and $q$ whose length is smaller than $\MF_u[q]$ in $F$.   $\Ld(u,q)$ is derived by Eq~\ref{equ:lowerbound} with $\PRADS^{\mathsf{out}}(u)$ and $\KPADS^{\mathsf{out}}(q)$ (Line~\ref{algo:checkuq} of Algo.~\ref{algo:pevalfkws}); and b) pruning forward expansion across multiple fragments. We denote the lower bound of the shortest distance between $u$ and the nearest out-portal node  by $\Ld(u,F.O)$. If $\Ld(u,F.O)$ is larger than $\MF_u[q]$, then no path across multiple fragments between $u$ and $q$ which is smaller than $\MF_u[q]$ could be found. $\Ld(u,F.O)$ is derived by $\PRADS^{\mathsf{out}}(u)$ and $\BPADS$ (Line~\ref{algo:checkup} of Algo.~\ref{algo:pevalfkws}). If $\Ld(u,q)$ and $\Ld(u,F.O)$ are both larger than $\MF_u[q]$, the forward expansion starting from $u$ can simply be skipped, since such expansion cannot yield any matches.

\etitle{Complexities.} This optimization does not change the time complexity of $\fkws$. A vertex $u\in \VM$ that needs to be forward expanded may be pruned in $O(\ln |V|)$ time. Given a vertex $u$, the size of $\PRADS(u)$ is $O(\ln |V|)$. Given a keyword $q$, the size of $\KPADS(q)$ is $O(|V|)$. The space complexity of $\BPADS$ is $O(|F.O|\ln|V|)$.

\subsection{Forward expansion order for  $\fkws$}\label{sec:opt-bfkws}

A vertex may be expanded multiple times in different forward expansions. To reduce this, we propose forward expansion orders, \ie Lines~\ref{algo:pevalfkws_forward_start}-\ref{algo:pevalfkws_forward_end_f} of Algo.~\ref{algo:pevalfkws} and Line~\ref{incfkws:order} of Algo.~\ref{algo:incfkws}. $\DKWS$ stores $\VM$ in a priority queue of {\em the match score $\Score(u)$ in descending order}. Intuitively, the larger $\Score(u)$ is, the easier $\Score(u)$  exceeds the upper bound $\prune$. As a consequence, the smaller region is expanded for $u$. Consider a forward expansion that starts from another vertex $u'\in \VM$, where $\Score(u') < \Score(u)$. If the expansion meets $u$, the following expansion starting from $u$ could be skipped, since $\dist(u,q)$ has already been computed and stored at $\answer_u[q]$ (Line~\ref{algo:found} of Algo.~\ref{algo:pevalfkws}). And the distance between $u'$ and $q$ via $u$ is computed by $\dist(u',u) + \answer_u[q]$.

\stitle{Exp-5: Effectiveness of the optimizations.} We performed a set of experiments to investigate the effectiveness of the proposed optimizations on  WebUK and DBLP. The results on other datasets exhibit similar trends and are hence not shown.

\begin{figure}[tp]
	\centering
	\begin{subfigure}[b]{0.24\textwidth}
		\centering
		\includegraphics[width=\textwidth]{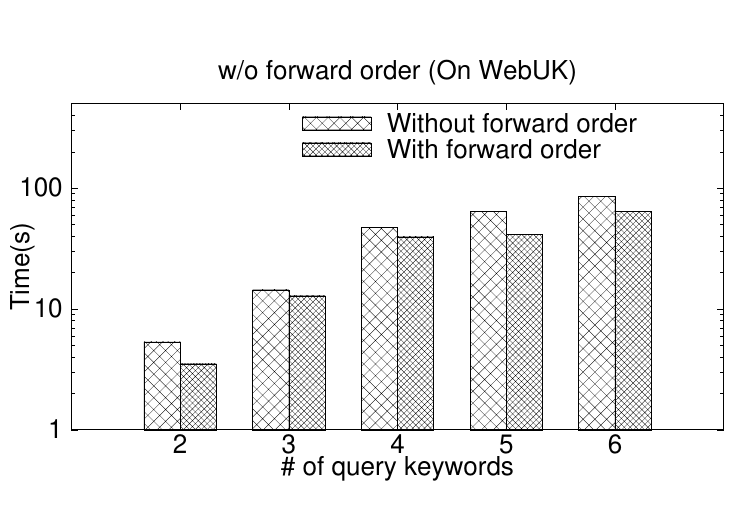}
	\end{subfigure}%\hspace{-5em}
	\begin{subfigure}[b]{0.24\textwidth}
		\centering
		\includegraphics[width=\textwidth]{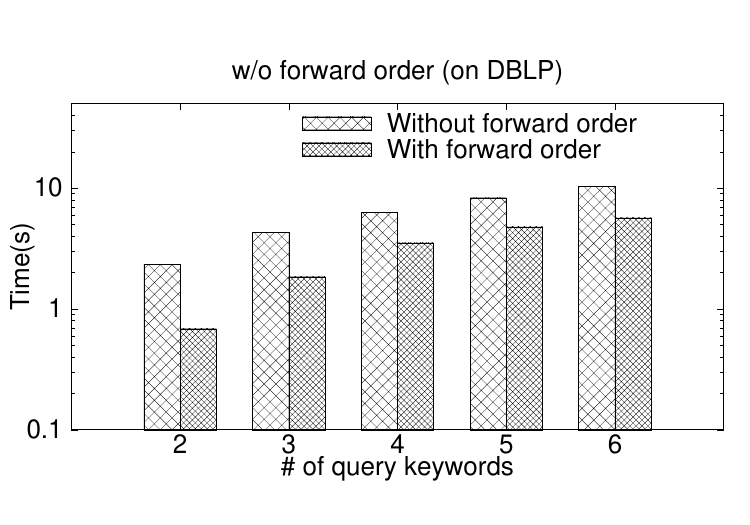}
	\end{subfigure}
	\caption{Performance of the forward expansion order}
	\label{fig:opt-order}
\end{figure}

\etitle{Effectiveness of expansion order.} We turned the expansion order of $\fkws$ on and off. The results are reported in Fig.~\ref{fig:opt-order}. With the optimization of expansion order, \DKWSPINE{} is $1.13$ (resp. $1.86$) times faster on WebUK (resp. DBLP). The improvement with expansion order is more significant on DBLP than that on WebUK since shorter forward expansions are computed early and some longer forward expansions can be computed from the shorter ones.

\begin{figure}[tp]
	\centering
	\begin{subfigure}[b]{0.24\textwidth}
		\centering
		\includegraphics[width=\textwidth]{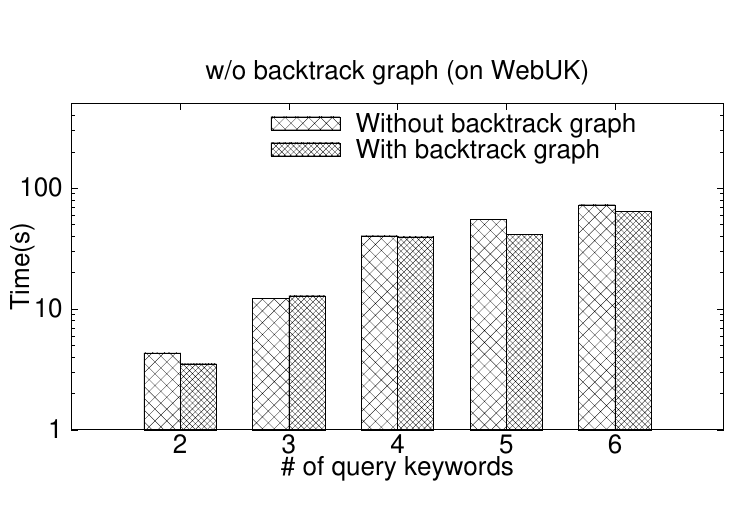}
	\end{subfigure}
	%\hfill
	\begin{subfigure}[b]{0.24\textwidth}
		\centering
		\includegraphics[width=\textwidth]{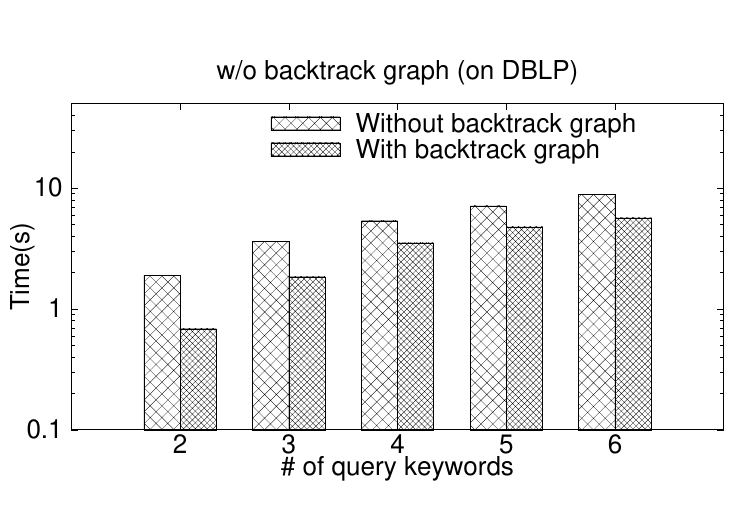}
	\end{subfigure}
	\caption{Performance of the backtrack graph}
	\label{fig:opt-bt}
\end{figure}

\etitle{Effectiveness of backtrack graph.} We also investigated the effectiveness of the backtrack graph by turning the optimization on and off. The results are reported in Fig.~\ref{fig:opt-bt}. With the optimization of the backtrack graph, \DKWSPINE{} is $1.34$ (resp. $2.22$) times faster on WebUK (resp. DBLP) on average. The improvement of this optimization is more significant on DBLP than that on WebUK since more refinement propagation on DBLP is shared.

\subsection{Supplementary experiment}

\begin{figure}[tp]
\begin{center}
\includegraphics[width=0.4\textwidth]{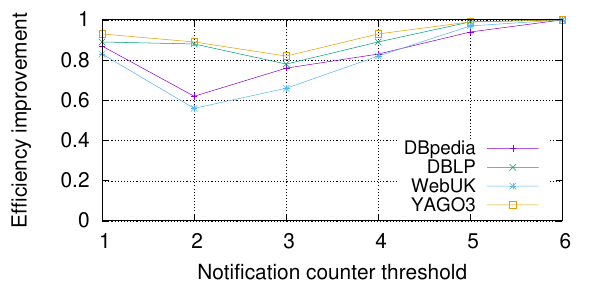}
\end{center}
\caption{Impact of notification counter threshold}
\label{fig:ncthreshold}
\end{figure}

\noindent\TKDERF{
\stitle{Impact of Notification Counter.} As depicted in Fig.~\ref{fig:ncthreshold}, we normalized the execution time by the time taken when the notify-push paradigm was not activated. We observed that on the four real-life datasets, setting the notification counter to $2$ or $3$ resulted in a comparatively good performance. However, when the notification counter exceeded 4, there was no substantial difference in the performance acceleration compared to when the notify-push paradigm was not used. In the context of keyword search semantics, this observed behavior can be traced back to the fact that, within these datasets, the discrepancy in the number of times each worker refines the bounds rarely surpasses four times; thus, the push function was seldom invoked. We also noted that, on larger graphs, setting the threshold to 2 performed better than setting it to 3. This is because the rapid synchronization of global bounds on larger graphs provides more significant performance acceleration.
}

\subsection{Time complexity of $\bkws$}\label{sec:complexity_bkws}

$\bkws$ takes $O(|Q|(|E|+|V|\log |V|))$, where $|Q|$ is the number of query keywords. To derive the complexity, we add a dummy vertex $v^{d}_i$ for each query keyword $q_i\in Q$.  Also, we add a set of dummy edges to connect $v^{d}_i$ and the vertices in the search origin $V_{q_i}$. We denote the graph with the dummy vertex $v^{d}_i$ and edges by $G^d_i$. The complexity of $\bkws$ for each query keyword on $G$ is identical to that of  Dijkstra's algorithm starting at $v^{d}_i$ on $G^d_i$, \ie $O(|E|+|V|\log |V|)$. Hence, the complexity of $\bkws$ is bounded by   $O(|Q|(|E|+|V|\log |V|))$.

\subsection{PageRank-based All Distance Sketches ($\PRADS$)}\label{sec:prads}

In this subsection, we  review {$\ADS$} and then propose our index. It is known that $\ADS{}$ is small in size, accurate, and efficient in answering shortest distance queries. Our main idea is to use PageRank to determine the chance of a node to be included in the sketch (\ie  the index).

\stitle{All-Distances sketches ($\ADS$).} Recall that in~\cite{cohen2015all}, given a graph $G=(V,E)$, each vertex $v$ is associated with a sketch, which is a set of vertices and their corresponding shortest distances from $v$. To select the vertices in $V$ and put them as the centers in the sketch of $v$,  each vertex is initially assigned a {\em random} value in $[0,1]$. If a vertex $u\in V$ has the $k$-th largest value among the vertices which have been traversed from  $v$ in the Dijkstra order, then $u$ is added to the sketch of $v$. $k$ is a user-defined parameter set by user. A larger $k$ results in larger and more accurate sketches. The shortest distance between $u$ and $v$ can be estimated by the intersection set of $\ADS$($u$) and $\ADS$($v$) (a.k.a. the common centers).

A drawback of {$\ADS$} is that it does not consider the relative importance of the vertices when generating the sketch. The vertices with high PageRanks, which roughly estimates the importance of the vertices in a graph, should be added to the sketch to cover the shortest paths. On the contrary, the vertices with low PageRanks are unlikely to be on many shortest paths and should not be added to the sketch.

\etitle{PageRank}. We employ any efficient algorithms to obtain the PageRank of the vertices of a graph $G$. We use a function $pr$: $V \rightarrow$ [0,1] to denote the PageRank of a vertex $v$ by $pr$($v$).

\etitle{Dijkstra rank.} We recall that we can efficiently obtain the Dijkstra rank of a vertex $v$ w.r.t a source vertex $s$ as follows. We run the Dijkstra's algorithm starting at $s$ and obtain the order of the visited nodes $[v_1,v_2,\ldots, v_l]$. The Dijkstra rank of $v_i$ w.r.t $s$ is $i$, denoted as $\pi(s,v_i) = i$.

\stitle{PageRank based all-distances sketches ($\PRADS$)}. Given a Dijkstra rank $\pi$, the PageRank, a vertex $v$, and a threshold $k$,   the {$\PRADS$} of $v$ is defined as follows:
\begin{equation}\label{fma:pads}
\mathsf{\PRADS}(v) = \{(u, d(v,u))\;|\;pr(u) \geq k(v,u)\}, 
\end{equation}
where $k(v,u)$ is the $k$-th largest PageRank among the nodes from $v$ to $u$ according to $\pi$.

\begin{algorithm}[tb]
	\caption{$\PRADS$ $\;$ construction}\label{algo:prads}
	\footnotesize
	\SetKwProg{Fn}{Function}{}{}
	\KwIn{Graph $G=(V,E)$}
	\KwOut{$\PRADS$}
	compute the PageRank $pr$ of the vertices in $G$\\
	initialize $\PRADS$($v$) = \{($v$, 0)\} for each vertex $v\in V$\\
	sorted the vertices $V$ by the descending order of $pr(v)$\\
	\For{$v\in V$}{
		\For{$u$ \textnormal{in the Dijkstra's traversal}}{
			\uIf{$|\{(w,d)\in \PRADS(u)~|~d \leq d(v,u) \}| < k$}{ \label{algo:prads:bound}
				add $(v,d(v,u))$ into $\PRADS(u)$
			}
			\uElse{
				continue the traversal on the next vertex \\
			}
		}
	}
	
	\Return $\PRADS$
\end{algorithm}

\begin{figure}[t]
	\begin{center}
	\includegraphics[width=5.5cm]{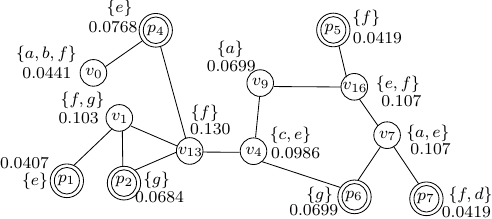}
		\end{center}
	\caption{A graph (fragment) and the PageRank}
	\label{fig:publicgraph}
\end{figure}

\begin{example}
	$(${$\PRADS$} construction$)$ Consider the graph $G$ in Fig.~\ref{fig:publicgraph}. Assume $k=1$. We compute the PageRank values for all the vertices in the graph, as shown below the vertices'  labels. \jiaxin{$v_{13}$ covers $41$ out of $156$ shortest paths in the graph $G$ in total, which is the largest among all the vertices. This shows that the node having a large PageRank value, $pr(v_{13})=0.130$, can be an effective center.} To determine the {$\PRADS$} of $v_1$, we run the Dijkstra's algorithm by taking $v_1$ as the source vertex to obtain the Dijkstra ranked list $[v_1, p_1, p_2, v_{13}, v_4, \ldots, p_7]$. Since the PageRank value of $v_{13}$ is the highest among the first four vertices in the ranked list, $v_{13}$ is added to $\PRADS(v_{1})$ with its distance to $v_1$. Similarly, $v_{1}$ is added to $\PRADS(v_{1})$. 
\end{example} 

\etitle{Shortest distance estimation.} Given a shortest distance query $(u,v)$ and the $\PRADS$, $\hat d(u,v)$ is computed by the intersection of $\PRADS(u)$ and $\PRADS(v)$ as follows: 
\begin{equation}\label{equ:pradsestimate}
\hat d(u,v) = \min\{(d_1 + d_2)\},
\end{equation} 
where $(w, d_1) \in \PRADS (u), (w, d_2) \in \PRADS(v)$.

\etitle{Spae complexity.} The expected size of $\PRADS$($v$) is $O(k\ln n)$, where $n$ is the number of nodes reachable from $v$, which is bounded by $O(k\ln |V|)$. (The analysis of \cite{cohen2015all} can be applied to $\PRADS$.)

\etitle{Time complexity.} The time complexity  is bounded by $O$($k|E|\ln|V|$).

Consider the graph $G$ in Fig.~\ref{fig:publicgraph}. We set $k=1$ and compute its {$\ADS$} shown in Tab.~\ref{tab:ads} and its {$\PRADS$} shown in Tab.~\ref{tab:pads}. We can see that there are two advantages of $\PRADS$. First, the size of {$\PRADS$} is significantly smaller than that of {$\ADS$}. Second, the {$\PRADS$}'s estimation is much more accurate than that of {$\ADS$}.

%to put them side by side to see the saving in  size 
\begin{table}[t]
\centering
\caption{An {$\ADS$} label for the graph in Fig.~\ref{fig:publicgraph}}
\label{tab:ads}
%\begin{tiny}
\tiny
\begin{tabular}{|p{0.15\linewidth}|p{0.68\linewidth}|}
    \hline
    {\bf Vertex ID} & {\bf ADS} \\
\hline
$v_0$ & $\{(v_0,0),(p_4,1),(v_1,3),(p_1,4),(p_7,6)\}$ \\
$p_4$ & $\{(p_4,0),(v_1,2),(p_1,3),(p_7,5)\}$ \\
$v_{13}$ & $\{(v_{13},0),(p_4,1),(v_1,1),(p_1,2),(p_7,4)\}$ \\
$v_1$ & $\{(v_1,0),(p_1,1),(p_7,5)\}$ \\
$p_1$ & $\{(p_1,0),(p_7,6)\}$ \\
$p_2$ & $\{(p_2,0),(v_1,1),(p_1,2),(p_7,5)\}$ \\
$v_4$ & $\{(v_4,0),(v_{13},1),(v_9,1),(p_4,2),(v_1,2),(p_1,3),(p_7,3)\}$ \\
$v_9$ & $\{(v_9,0),(p_4,3),(v_1,3),(p_7,3)\}$ \\
$p_6$ & $\{(p_6,0),(v_4,1),(v_{13},2),(v_9,2),(p_7,2)\}$ \\
$v_{16}$ & $\{(v_{16},0),(v_9,1),(p_7,2)\}$ \\
$v_7$ & $\{(v_7,0),(v_{16},1),(p_7,1)\}$ \\
$p_5$ & $\{(p_5,0),(v_9,2),(p_7,3)\}$ \\
$p_7$ & $\{(p_7,0)\}$ \\
\hline
%    $\dist(v_1,v_2)$ & The shortest between $v_1$ and $v_2$ in the original graph \DataGraph\\ \hline
%    $\desc(v,G)$ & The descendants of $v$ in graph $G$\\ \hline
%    $\support(l,G)$ & The occurrence  of the label $l$ in graph $G$\\ \hline
%    $\DT(G,\Generalization)$ & The distortion of the generalization $C$ on graph $G$ \\ \hline
%    $\degree(v)$ & The degree of vertex $v$. \\ \hline
 \end{tabular}  
%\end{tiny}
\end{table}

\begin{table}[t]
\centering
\caption{The {$\PRADS$} label for the graph in Fig.~\ref{fig:publicgraph}}
\label{tab:pads}
%\begin{tiny}
\tiny
\begin{tabular}{|p{0.15\linewidth}|p{0.43\linewidth}|}
    \hline
    {\bf Vertex ID} & {\bf PADS} \\
\hline

$v_0$ & $\{(v_0,0),(p_4,1),(v_{13},2)\}$ \\
$p_4$ & $\{(p_4,0),(v_{13},1)\}$ \\
$v_{13}$ & $\{(v_{13},0)\}$ \\
$v_1$ & $\{(v_1,0),(v_{13},1)\}$ \\
$p_1$ & $\{(p_1,0),(v_1,1),(v_{13},2)\}$ \\
$p_2$ & $\{(p_2,0),(v_1,1),(v_{13},1)\}$ \\
$v_4$ & $\{(v_4,0),(v_{13},1)\}$ \\
$v_9$ & $\{(v_9,0),(v_4,1),(v_{16},1),(v_{13},2)\}$ \\
$p_6$ & $\{(p_6,0),(v_4,1),(v_7,1),(v_{13},2)\}$ \\
$v_{16}$ & $\{(v_{16},0),(v_7,1),(v_{13},3)\}$ \\
$v_7$ & $\{(v_7,0),(v_{16},1),(v_{13},3)\}$ \\
$p_5$ & $\{(p_5,0),(v_{16},1),(v_7,2),(v_{13},4)\}$ \\
$p_7$ & $\{(p_7,0),(v_7,1),(v_{16},2),(v_{13},4)\}$ \\

\hline
%    $\dist(v_1,v_2)$ & The shortest between $v_1$ and $v_2$ in the original graph \DataGraph\\ \hline
%    $\desc(v,G)$ & The descendants of $v$ in graph $G$\\ \hline
%    $\support(l,G)$ & The occurrence  of the label $l$ in graph $G$\\ \hline
%    $\DT(G,\Generalization)$ & The distortion of the generalization $C$ on graph $G$ \\ \hline
%    $\degree(v)$ & The degree of vertex $v$. \\ \hline
 \end{tabular}  
%\end{tiny}
\end{table}

\begin{example}
	$($Shortest distance estimation.$)$ Consider the graph $G$ in Fig.~\ref{fig:publicgraph} and its {$\PRADS$}s in Tab.~\ref{tab:pads}. Given two vertices $v_9$ and $v_7$, there are two common centers $v_{16}$ and $v_{13}$ in $\PRADS(v_9)$ and $\PRADS(v_7)$. The shortest distance is estimated by Eq~\ref{equ:pradsestimate}, \ie $\hat{d}(v_9, v_7) =2$ (\ie $0$\% error). By $\ADS$, $\hat{d}(v_9, v_7) =4$ is returned (\ie $100$\% error). More specifically, we compare the estimation accuracy of {$\ADS$} and {$\PRADS$} between all pairs of the vertices in Fig.~\ref{fig:publicgraph}. The average error of {$\PRADS$} (resp. {$\ADS$}) is around $3\%$ (resp. $38\%$).
\end{example}
It is worth noting that {$\PRADS$} exhibits the theoretical guarantee of the shortest path estimation stated below.

\begin{lemma}\label{lemma:uvbound}
The distance between two vertices $u$ and $v$ is estimated using Eq~\ref{equ:pradsestimate} with an approximation factor  $(2c-1)$, where $c=\lceil \frac{\ln |V|}{\ln k}\rceil$ with a constant probability, \ie $\hat{d}(u,v) \leq (2c-1)d(u,v)$.
\end{lemma}
\begin{IEEEproof}
Let $d=d(u,v)$. Let $N_{i}(u)$ denote the neighbors of vertices $u$ within $id$ hops. For simple exposition, we denote the intersection and union of $N_{i}(u)$ and $N_{i}(v)$ as $I_i = N_{i}(u)\cap N_{i}(v)$ and $U_i = N_{i}(u)\cup N_{i}(v)$, respectively. It is worth noting that $I_i\subseteq U_i \subseteq I_{i+1}$. Consider the ratio of $\frac{|I_i|}{|U_i|}$ and a ratio threshold $\frac{m}{k}$. Given the vertices with $k$ largest $pr$ values in $U_i$, if one of them (say $w$) hits $I_i$, $w$ belongs to both $\PRADS(v)$ and $\PRADS(u)$. The real distance $d$ can be estimated within $2id$. The probability \jiaxin{of {\em at least one of the vertices, which has the k largest PageRank values in $U_i$, hits the $I_i$}} is $1 - (1 - \frac{\alpha}{k})^k \approx 1- e^{-\alpha}$. Since there are $n$ vertices in graph $G$ at most, $|U_i| \leq n$. Hence, there exists $i\leq \log_{k/\alpha}n$.
\end{IEEEproof}

\subsection{PageRank-based Keyword Distance Sketches ($\KPADS$)} \label{subsec:KPRADS}
We denote the {\em shortest distance between a vertex} $v$ and {\em a keyword} $t$ by $d(v,t)$, where $d(v,t)=\min\{d(v,u) | t \in L(u), u \in V\}$. To estimate the distance between a given vertex and keyword, we propose {$\KPADS$}, which is constructed by $\PRADS$-merging: Given any two vertices $u$ and $u'$ where $t\in L(u)$ and $t\in L(u')$, there may exist common centers in $\PRADS$($u$) and $\PRADS$($u'$). Hence, we only keep the smallest one among $\hat{d}(v,u')$ and $\hat{d}(v,u)$, since both of them are the upper bound of $d(v,t)$. 

\stitle{Keyword-{$\PRADS$} ($\KPADS$).} For each keyword $t\in \Sigma$, we build a sketch $\KPADS(t)$. $\KPADS(t)$ can be built by merging $\PRADS$ of those vertices that contain $t$, \ie {$\PRADS$($v$)} where $t\in L(v)$. More formally, given a center $(w_i,d_i) \in \PRADS(v)$, $(w_i,d_i)\in \KPADS(t)$ {\em iff}~~$\forall (w_i,d_i')\in \PRADS(v')$ and $t\in L(v')$, $d_i'\geq d_i$.

\etitle{Shortest keyword-vertex distance estimation.} Given a vertex $v$ and a keyword $t$, the shortest distance $\hat{d}(v,t)$ can be computed as follows: 
\begin{equation}\label{equ:kvde}
\footnotesize
\hat d(v,t) = \min\{(d_1 + d_2) | \\
(w, d_1) \in \PRADS (v)\; \textnormal{\small and}\;  (w, d_2) \in \KPADS (t) \}
\end{equation}

\begin{table}[tb]
	\centering
	\caption{The {$\KPADS$} label for the graph in Fig.~\ref{fig:publicgraph}}
	\label{tab:kpads}
	%\begin{tiny}
	\footnotesize
	\tiny 
	\begin{tabular}{|p{0.07\linewidth}|p{0.80\linewidth}|}
		\hline
		{\bf Terms} & {\bf KPADS} \\
		\hline
$a$ & $\{(v_9,0), (v_4,1), (p_4,1), (v_7,0), (v_{13},2), (v_{16},1), (v_0,0)\}$ \\
$b$ & $\{(v_0,0), (v_{13},2), (p_4,1)\}$ \\
$c$ & $\{(v_{13},1), (v_4,0)\}$ \\
$d$ & $\{(v_{13},4), (v_7,1), (p_7,0), (v_{16},2)\}$ \\
$e$ & $\{(v_{13},1), (v_4,0), (v_1,1), (v_7,0), (p_4,0), (v_{16},0), (p_1,0)\}$ \\
$f$ & $\{(p_5,0), (v_1,0), (v_{13},0), (p_4,1), (v_7,1), (v_{16},0), (v_0,0), (p_7,0)\}$ \\
$g$ & $\{(p_6,0), (v_1,0), (v_4,1), (v_{13},1), (v_7,1), (p_2,0)\}$ \\
		\hline
		%    $\dist(v_1,v_2)$ & The shortest between $v_1$ and $v_2$ in the original graph \DataGraph\\ \hline
		%    $\desc(v,G)$ & The descendants of $v$ in graph $G$\\ \hline
		%    $\support(l,G)$ & The occurrence  of the label $l$ in graph $G$\\ \hline
		%    $\DT(G,\Generalization)$ & The distortion of the generalization $C$ on graph $G$ \\ \hline
		%    $\degree(v)$ & The degree of vertex $v$. \\ \hline
	\end{tabular}  
	%\end{tiny}
\end{table}

%% Remove a naive example for the limite space
\eat{We present the {\textnormal \KPADS} construction by taking keyword $a$ as an example. There are three vertices $v_0$, $v_7$ and $v_9$ containing the keyword $a$ in $G$. Since $v_{13}$ is the common center in the {\textnormal \PRADS} of these three vertices, it will be merged and added into $\textnormal \KPADS(a)$ with the smallest distance $\min\{d(v_0, v_{13}), d(v_7, v_{13}), d(v_9, v_{13})\}$, \ie ($v_{13}, 2$) will be added into $\textnormal \KPADS(a)$.}

\begin{example}
Consider the graph $G$ in Fig.~\ref{fig:publicgraph} and its $\PRADS$ in Tab.~\ref{tab:pads}. The $\KPADS{}$ is shown in Tab.~\ref{tab:kpads}. Consider the shortest distance between $a$ and $p_4$. The distance can be estimated by the intersection of $\KPADS$($a$) and $\PRADS$($p_4$). There are two common centers, $p_4$ and $v_{13}$. $\hat{d}(a,p_4)=1$ is returned by the common center $p_4$.
\end{example}

\end{document}